\documentclass[onecolumn,superscriptaddress,showpacs,nofootinbib]{revtex4}

 \usepackage{graphicx}

\usepackage{bm,color}

\begin{document}

\title{Derivation of a generalized Schr\"odinger equation from the theory of scale relativity}

\author{Pierre-Henri Chavanis}
\affiliation{Laboratoire de Physique Th\'eorique, 
Universit\'e Paul Sabatier, 118 route de Narbonne 31062 Toulouse, France}


\begin{abstract}
Using Nottale's theory of scale relativity relying on a fractal
space-time, we derive a generalized Schr\"odinger equation taking into account
the interaction of the system with the external environment. This equation
describes the irreversible evolution of the system towards a static
quantum state. We first interpret the scale-covariant equation of dynamics
stemming from Nottale's theory as a hydrodynamic viscous Burgers
equation for a potential flow involving a complex
velocity field and an imaginary viscosity.  We show that the Schr\"odinger
equation
can be directly obtained from this equation by performing a  Cole-Hopf
transformation equivalent to the WKB transformation.  We then introduce a
friction force proportional and opposite to the complex velocity in the
scale-covariant equation of dynamics in a way that preserves the local
conservation of the normalization condition. We find that the resulting
generalized Schr\"odinger equation, or the corresponding fluid equations
obtained from the Madelung transformation, involve not only a damping term but
also an effective thermal term. The friction coefficient and the temperature
are related to the real and imaginary parts of the  complex friction coefficient
in the scale-covariant equation of dynamics. This may be viewed as a form of
fluctuation-dissipation theorem. We show that our generalized Schr\"odinger
equation satisfies an $H$-theorem for the quantum Boltzmann free energy. As a
result, the probability distribution relaxes towards an equilibrium state
which can be viewed as a Boltzmann distribution including a quantum potential.
We propose to apply this generalized Schr\"odinger equation to dark matter
halos in the Universe, possibly  made of self-gravitating Bose-Einstein
condensates.

\end{abstract}



\maketitle


\section{Introduction}
\label{sec_introduction}

The Schr\"odinger equation is the cornerstone of quantum
mechanics. Schr\"odinger introduced this equation in
1926 from an ingenious procedure \cite{schrodinger2,schrodinger4,schrodingerPR}
obtained by combining the de Broglie \cite{brogliethese} relations and the
standard
wave equation (second order in time) with a space dependent phase
velocity.\footnote{A short account of the early development of quantum
mechanics is given in the Introduction of \cite{chavmatos}. See in particular
Appendix F of that paper for a brief presentation of the
historical derivation of  Schr\"odinger's equation.} The Schr\"odinger equation
can also be obtained from a correspondence principle
\cite{schrodingerCORR,broglie1,broglie2,dirac} by
writing the energy of the particle as  $E={\bf p}^2/2m+m\Phi$  and replacing
the 
variables $E$ and ${\bf p}$ by the operators $i\hbar\partial_t$ and
$-i\hbar\nabla$ acting on the wave function $\psi$. This is how the
Schr\"odinger equation is usually introduced in standard textbooks of quantum
mechanics. The Schr\"odinger equation has proven to be extremely successful. Its
time-independent solution leads
to a fundamental eigenvalue problem $\hat{H}\psi=E\psi$ which determines the
quantification of the energy.\footnote{
Schr\"odinger initially obtained this eigenvalue equation from a variational
principle
\cite{schrodinger1,schrodingerCORR} before introducing 
his time-dependent equation (see \cite{chavmatos} for a short review).} For the
Coulomb potential,
Schr\"odinger recovered the energy spectrum of the hydrogen atom
heuristically obtained by Bohr \cite{bohr1,bohr2}. For the harmonic oscillator
and for the rotator, Schr\"odinger recovered the results
previously obtained by Heisenberg \cite{heisenberg}
from his more abstract matrix mechanics. More generally, Schr\"odinger
\cite{schrodingerCORR} showed the equivalence between his wave mechanics and the
Heisenberg-Born-Jordan quantum mechanics \cite{heisenberg,bornjordan,bhj}. On
the other hand, by introducing a relation $\psi=e^{i{\cal
S}/\hbar}$ between the wave function $\psi$ and the complex action ${\cal S}$,
Wentzel
\cite{wentzel}, Brillouin \cite{brillouin} and Kramers \cite{kramers} showed
that
the Schr\"odinger equation reduces to the Hamilton-Jacobi equation in 
the semi-classical limit $\hbar\rightarrow 0$. This is the so-called WKB
approximation which allows
one to study the semi-classical regime of a quantum system. One recovers the
classical mechanics from the wave mechanics in the limit $\hbar\rightarrow 0$ in the same manner
that one recovers geometric optics from the theory of undulatory optics when the
wavelength $\lambda\rightarrow 0$. Despite all its successes, the  
Schr\"odinger equation remains very mysterious (in part because of the complex
nature of the wave function) and it is fair to say that this equation has been
essentially {\it postulated} rather than being {\it derived} from first
principles.

The physical interpretation of the Schr\"odinger equation when applied to the
hydrogen atom posed itself since the start. Schr\"odinger \cite{schrodinger4}
introduced a density $\rho=|\psi|^2$
and a current and derived a local conservation equation for the density. 
This gives an interpretation to the
wave function in the sense that $|\psi|^2({\bf r},t)$ characterizes the
``presence'' of the particle at some point. Schr\"odinger thought that the wave
function represents a particle that is spread out, most of the particle being
where the modulus of the wave function $|\psi|^2$ is large. For example,
according to Schr\"odinger's view, the charge of the electron is not
concentrated in a point, but is spread out through the whole space, proportional
to the quantity $|\psi|^2$. Alternatively, Born
\cite{bornf,bornnature,born} developed a probabilistic interpretation of the
wave
function.\footnote{As reported by
Bhaumik
\cite{bhaumik}, Born was strongly influenced by Einstein in his interpretation
as he stated in his Nobel lecture: ``Again an idea of Einstein's gave me the
lead. He had tried to make the duality of particles - light quanta or photons -
and waves comprehensible by interpreting the square of the optical wave
amplitudes as probability density for the occurrence of photons. This concept
could at once be carried over to the $\psi$-function:
$|\psi|^2$ ought to represent the probability
density for electrons (or other particles).''} Born proposed that the magnitude
of the wavefunction $\psi({\bf r},t)$ does not tell us how much of the particle
is at position ${\bf r}$ at time $t$, but rather the  probability that the
particle is at ${\bf r}$ at time $t$. This gives an interpretation
to the wave function in the sense that $|\psi|^2$ represents the probability of
presence of
the electron at some point. Therefore, quantum mechanics essentially provides a
probabilistic description of Nature. However, the origin of this probabilistic
description divided the researchers. In the {\it complementarity school} led by
Bohr, the quantum theory is complete. It must be accepted as it is, and one
should not try to understand the origin of probabilities from a more fundamental
principle. This interpretation of quantum mechanics was criticized by the {\it
statistical school} associated with Einstein who argued that the quantum
theory is incomplete  and that one must understand the origin of probabilities
from a causal subdynamics (``God does not play dice'').
Therefore, the interpretation of quantum mechanics gave rise to intense debates between the view of Bohr who considers that
the wave function defines a physical state completely and the view of Einstein
who did not accept this view and preferred a more stochastic and statistical
interpretation. In brief, Bohr argues that the physical state of an
individual system is completely specified by a wave function that determines
only the probabilities of actual results that can be obtained in a statistical
ensemble of similar experiments. Alternatively, Einstein believes that, even at the quantum
level, there must exist dynamical variables determining (as in classical
physics) the actual behavior of each individual system, and not merely its
probable behavior. Following this point of view, there have been
attempts to explain quantum phenomena within the framework of classical statistical
mechanics. Several researchers tried to found quantum
mechanics in terms of more familiar stochastic models and probability theory in
order to gain a better understanding.  Without attempting to be
exhaustive, we give below a short account of different attempts to justify the
postulates of quantum mechanics, and derive the Schr\"odinger equation, in terms
of classical stochastic theories.

In a very intriguing paper, Madelung \cite{madelung}
showed that the complex, linear, Schr\"odinger equation could be transformed
into a pair of real, nonlinear, hydrodynamic equations similar to an equation of
continuity and an Euler equation involving a quantum potential arising from the
finite value of $\hbar$. As a result, the Schr\"odinger equation can be
represented by the picture of an
irrotational flow of a compressible perfect fluid with a particular quantum
potential (also equivalent to an anisotropic pressure tensor). The quantum
potential, or quantum force, accounts for the Heisenberg uncertainty principle.
Because
of this quantum term, the particle's motion does not exactly follow the laws of classical mechanics. The Madelung transformation suggests
the possibility of a semiclassical description of quantum systems through the fluid
dynamical viewpoint. Initially, very little attention was paid to this
approach\footnote{Madelung's hydrodynamic approach
was further developed by Kennard \cite{kennard} in 1927. However, apart from
that work, the paper of Madelung was very little quoted.} and
Pauli expressed the opinion that it is not
interesting (see the comment in Ref. \cite{spiegel}).

At about the same period, de Broglie
\cite{broglie1927a,broglie1927b,broglie1927c} (see also London \cite{london})
developed a relativistic hydrodynamic representation of the Klein-Gordon (KG)
equation (he was apparently not aware of the work of Madelung).
He derived a relativistic Euler equation that contains a Lorentz
invariant
quantum potential. This is the relativistic version of the
classical Madelung quantum potential. De Broglie interpreted the
quantum force as a force of internal
tensions existing around the corpuscles. He also interpreted
the continuity equation as a conservation equation for a
density transported by a velocity. The aim of de Broglie
was to provide a
causal and objective interpretation of
wave mechanics, in accordance with the wish expressed many times by Einstein,
and in contrast to the purely probabilistic interpretation of quantum mechanics
put forward by Born, Bohr, and Heisenberg. This is what he called the pilot-wave
theory because the particle is guided by
the wave $\psi$. The pilot-wave theory of de Broglie
was criticized by Pauli during the October
1927 Solvay Physics Conference (see the comment in
\cite{solvay}), and de Broglie abandoned
it.

An interpretation of the Schr\"odinger equation in terms of particle trajectories was proposed by
Bohm \cite{bohm}  in 1952. In this interpretation, the evolution of each individual
system is determined by definite laws analogous to (but not identical with) the
classical equations of motion. Bohm interpreted the deviation from the Newtonian
equations of motion as being due to a quantum-mechanical potential associated
with the wave function. This quantum potential, which depends on the density and
density gradients of the system, suggests an interpretation of the quantum
theory
in terms of ``hidden'' variables, in the same sense that in macroscopic physics
the coordinates and momenta of individual atoms are hidden variables which in a
large scale system manifest themselves only as statistical averages.  Actually,
Bohm rediscovered the quantum potential previously introduced Madelung and de
Broglie in 1926.\footnote{Bohm \cite{bohm}  mentions that the work of de Broglie
was drawn to his attention after he completed his paper. On the other hand, the
paper of Madelung was mentioned in a {\it Note added in proofs}.} 
The quantum potential plays a
central role in the deterministic
interpretation of quantum mechanics discussed by Bohm.
This renewal
of interest for a causal interpretation of quantum mechanics stimulated de
Broglie
to return to the problem again and undertake a fresh examination of his old
ideas \cite{ndb1,ndb2,ndb3}.

The interpretation of quantum mechanics in terms of a classical
statistical or hydrodynamical picture (based on the Madelung-de Broglie transformation) was
further discussed by Takabayasi \cite{takabayasi1,takabayasi2}.\footnote{Takabayasi also criticized certain
aspects of Bohm's interpretation.}
This formulation is based on the procedure to transform the equation for a
state vector (the wave function) into an ensemble of classical motions. There is
a formal equivalence between the picture of waves and that of an ensemble of
trajectories. This suggests an alternative formulation of quantum mechanics in
terms of a classical picture. The quantum potential, which depends on the
probability itself,  leads to a statistical blurring of the classical
trajectory. Diffusion of wave packets, interferences, tunnel effect etc. may  be
interpreted in terms of this quantum force as previously emphasized by Bohm
\cite{bohm}. In this
approach, the trajectories
deviate, due to the quantum force, from purely classical ones, generally showing
very complicated fluctuations in non-stationary cases. According to Takabayasi, it
might be conceivable to deduce the fluctuations of the particles' trajectories
due to the quantum potential through a mechanism similar to that of
Brownian motion; that is, we may regard the fluctuations as produced
by the random action from the outside medium. In this viewpoint, we must
postulate a virtual medium (an aether). Therefore, ``hidden
variables'' are introduced not only as the simultaneously defined particle
position and velocity as in the original statistical picture, but also as the
freedom of the medium.\footnote{Similar interpretations of the Schr\"odinger equation have been
developed by other authors. For example, Bohm and Vigier \cite{bv} introduced the notion of
random fluctuations arising from the interaction with a subquantum medium. On
the other hand, F\"urth \cite{furth}, F\'enyes \cite{fenyes}, Weizel
\cite{weizel}, Kershaw \cite{kershaw} and Comisar \cite{comisar}  tried to
describe the motion of
quantum particles in terms of a Markoff process by analogy with Brownian
motion. They pointed out the formal analogy between the Schr\"odinger
equation and
the Fokker-Planck equation and introduced an imaginary
diffusion coefficient $D=i\hbar/2m$ (this formula first
appeared
in the paper of F\"urth \cite{furth}). In the work of Weizel \cite{weizel}, the
random aspects of the motion of a quantum particle are due to the interaction
with hypothetical particles that he called zerons.} However, 
despite some resemblances between quantum mechanical motion and diffusion
phenomena as the result of the formal similarity between the Schr\"odinger
equation and the diffusion equation, Takabayasi
\cite{takabayasi1,takabayasi2} emphasized that the nature of the stochastic
process in the two cases is very different. In the quantum theory,  the
trajectories may have  very complicated fluctuations but these fluctuations are
not at random since, for each individual trajectory, they are completely
determined by the density $\rho=|\psi|^2$ appearing in the quantum potential. By
contrast, in Brownian theory, the particle experiences a fluctuating force that
is uncorrelated at every successive time and independent on the probability
distribution $\rho$.\footnote{It is interesting to note that, in the
context of generalized thermodynamics and nonlinear Fokker-Planck
(NFP) equations, the
random force acting on a particle is allowed to depend on the probability
distribution
$\rho({\bf r},t)$ (see, e.g., \cite{nfp,entropy} for a review on NFP
equations).}

In a remarkable paper, Nelson \cite{nelson} proposed a derivation
of the Schr\"odinger equation from Newtonian mechanics using a stochastic
theory based on Fokker-Planck equations.\footnote{Many authors have pointed out the formal equivalence
between the Schr\"odinger and the Fokker-Planck equation  with {\it imaginary}
time (see footnote 7) but the approach of Nelson is different.}
The hypothesis is that every particle of mass $m$ is subject to a
Brownian motion with a diffusion coefficient ${\cal D}=\hbar/2m$ (where $\hbar$ is
the Planck constant) and no friction. In ordinary Brownian motion, friction
plays an important role. In Nelson's theory, quantum particles experience no
friction in order to preserve Galilean invariance. As a result, the quantum
Brownian motion is non dissipative and reversible while the usual Brownian motion is
dissipative and irreversible (this is in agreement with the fact that the Schr\"odinger equation conserves the energy while the Fokker-Planck equation dissipates the free energy). In this stochastic formulation of quantum mechanics,  the
random motion represents quantum fluctuations due to a sub-quantum medium (a
sort of aether). Nelson
introduced a mean forward velocity ${\bf u}_{+}$ and a mean backward velocity
${\bf u}_{-}$, associated with forward and backward Fokker-Planck equations,
defined a mean acceleration, and derived a pair of coupled hydrodynamic
equations for the
current velocity ${\bf u}=({\bf u}_{+}+{\bf u}_{-})/2$ and the osmotic velocity
${\bf u}_Q=({\bf u}_{+}-{\bf u}_{-})/2$. Using the Madelung
transformation,\footnote{Nelson \cite{nelson} does not explicitly
refers to Madelung's paper.} he showed that
these equations are equivalent to the Schr\"odinger equation!

More recently, Nottale \cite{nottale} developed a theory of scale
relativity and managed to derive the Schr\"odinger equation from Newton's law by
using a principle of scale covariance. Nottale gives up the concept of differentiability
of space-time and postulates that space-time is fractal. In a fractal
space-time the trajectories of the quantum particles are continuous but nowhere
differentiable. The velocity is the sum of a differentiable part and a
non-differentiable (fractal) part. The new component is an explicit
scale-dependent fractal fluctuation. The fractal dimension of the Brownian
motion is $D_{\rm F}=2$ \cite{feynman}. The
theory of scale relativity extends Einstein's theory of relativity to scale
transformations of resolutions. In this approach, the particles have a
stochastic motion that is due to the fractal nature of the space-time itself. The fractal (non-differentiable) nature of the trajectories leads to introducing  twin velocities ${\bf u}_{+}$ and ${\bf u}_{-}$ whose meaning is
different from the one given in the stochastic quantum mechanics
of Nelson.\footnote{The theory of Nottale differs fundamentally from the one of
Nelson even if there exist close analogies (see Appendix
\ref{sec_ri}).} From these twin velocities, one can form a
classical velocity ${\bf u}=({\bf u}_{+}+{\bf u}_{-})/2$ and a quantum velocity
${\bf u}_Q=({\bf u}_{+}-{\bf u}_{-})/2$. These real velocities can be
combined together into a complex velocity ${\bf U}={\bf u}-i{\bf u}_Q$ and this
duality is viewed as the fundamental origin of the complex nature of the wave
function $\psi$ in quantum mechanics (the wave function is related to the
complex velocity by ${\bf U}=-i(\hbar/m)\nabla\ln\psi$).

The Schr\"odinger equation applies to systems that are isolated from the surrounding.  As a
consequence, the energy is conserved and the evolution is reversible. However,
in reality, a system is never totally isolated from the surrounding. It is
therefore of interest to try to generalize the Schr\"odinger equation by taking
into account the interaction of the system with the external environment. This
 gives rise to dissipative effects and irreversibility. How can one introduce
dissipative effects into the Schr\"odinger equation? This apparently cannot be
done from the correspondence principle. This shows the limit of this heuristic
approach. By contrast, it is possible to generalize the approach of Nottale
\cite{nottale} by including a damping force in the fundamental equation of
dynamics (Newton's law). However, there is a little difficulty. If we naively
introduce a friction force $-\gamma {\bf U}$  in the scale-covariant equation of
dynamics [see Eq. (\ref{pb1})], we obtain a damped Schr\"odinger
equation  [see Eq. (\ref{pb5})] that does not conserve the normalization
condition locally. Therefore, we slightly
generalize the scale-covariant equation of dynamics by writing the friction
force as $-{\rm Re} (\gamma {\bf U})$ [see Eq. (\ref{qb1})] in order to obtain
a damped Schr\"odinger equation [see Eqs.  (\ref{qb5}) and (\ref{qb10})] that
conserves the normalization condition locally. Interestingly, when we perform
the Madelung transformation on this equation, we obtain a quantum Euler equation
 [see Eq. (\ref{tmad6})] that includes not only a friction term $-\xi {\bf u}$
(as expected)
but also a pressure term associated with an effective isothermal equation
of state $P=\rho k_B T/m$ [see Eq. (\ref{tmad7})]. The friction coefficient
$\xi$ and the temperature $T$ are related to the real and imaginary parts  of
the complex friction $\gamma$ [see Eq. (\ref{qb11})]. The emergence of both
friction and temperature is
reminiscent of the fluctuation-dissipation theorem in statistical mechanics but
it takes here a novel form. It arises from the complex nature of quantum
mechanics which is intrinsically due to the non-differentiability of
space-time. Indeed, $\xi$ and $2k_B T/\hbar$ may be viewed as 
twin friction coefficients forming a complex friction coefficient
$\gamma=\xi+i2k_B T/\hbar$ similarly to the twin velocities ${\bf u}$ and ${\bf
u}_Q$ forming a complex velocity ${\bf U}={\bf u}-i{\bf u}_Q$.

The paper is organized as follows. In Sec. \ref{sec_derivation}, we recall the
scale-covariant equation of dynamics [see Eq. (\ref{s0})] obtained from
Nottale's theory of scale relativity and emphasize its resemblance with a
complex viscous Burgers equation for a potential flow in hydrodynamics [see Eq.
(\ref{s1})]. Then,
making the complex Cole-Hopf transformation [see Eq. (\ref{s3b})], and noting
its resemblance with the WKB transformation [see Eq. (\ref{s4})], we derive
the Schr\"odinger equation [see Eq. (\ref{s9})]. This direct derivation
considerably simplifies and clarifies the derivation of Nottale \cite{nottale}.
It shows that the Schr\"odinger equation is formally equivalent to the complex
viscous Burgers equation in the same sense that the diffusion equation is
equivalent to the real viscous Burgers equation. Having obtained the
Schr\"odinger equation, we recall the Madelung transformation and show that the
Schr\"odinger equation is formally equivalent to real hydrodynamic equations
formed by a continuity equation and an Euler equation including a quantum
potential [see Eqs. (\ref{tmad4c})-(\ref{ntmad8})]. In Sec. \ref{sec_diss},  we
start from a scale-covariant equation of dynamics including a friction force
that preserves the local conservation of the normalization condition
[see Eq. (\ref{qb1})] and derive a generalized Schr\"odinger equation. We
obtain a nonlinear Schr\"odinger equation  [see Eq.  (\ref{qb10})] exhibiting
{\it simultaneously} dissipative effects and (effective) thermal effects.
Using the Madelung transformation, we show that our generalized
Schr\"odinger equation is equivalent to quantum damped isothermal Euler
equations [see Eqs. (\ref{tmad4})-(\ref{tmad7})]. In the strong friction limit
$\xi\rightarrow +\infty$, we obtain a quantum Smoluchowski equation [see Eq.
(\ref{tmad11})]. In Sec. \ref{sec_effth}, we develop an effective
thermodynamic formalism. We show that our generalized Schr\"odinger equation
satisfies an $H$-theorem for a quantum free energy associated with the Boltzmann
entropy. As a result, the probability distribution relaxes towards an
equilibrium state which can be viewed as a Boltzmann distribution including a
quantum potential. This is at variance
with the Schr\"odinger equation and the quantum Euler equation (without
friction) that admit
steady states but that do not relax towards such states from a generic initial
condition. This is because the Schr\"odinger equation is reversible 
while our generalized Schr\"odinger equation is irreversible. In Sec.
\ref{sec_virial}, we establish the virial theorem associated with our
generalized Schr\"odinger equation. In Sec. \ref{sec_conn}, we
discuss the connection between our generalized Schr\"odinger equation and the
nonlinear Schr\"odinger equations obtained previously by Kostin \cite{kostin}
and Bialynicki-Birula and Mycielski \cite{bbm} who introduced {\it individually}
dissipative effects and nonlinear effects equivalent to thermal effects. In a
sense, our approach based on the theory of scale relativity unifies these two
equations into  a single one and shows that dissipative effects and thermal
effects arise from a single complex friction coefficient [see Eq.
(\ref{qb11})]. In Sec. \ref{sec_ags}, we consider another
dissipative Schr\"odinger equation. In the Conclusion, we propose to apply our
results to self-gravitating Bose-Einstein
condensates (BECs) possibly describing dark matter halos in cosmology (this
model is developed in detail
in \cite{dgp}).
The Appendices contain additional material that is used
throughout the paper. In this paper, we
do not go into the details of the interpretation of the theory of scale
relativity that are explained in the book of Nottale \cite{nottale}. Rather, we
use the principle of scale covariance as a mathematical tool to obtain
generalized Schr\"odinger equations. We note that frictional effects cannot be
directly obtained from the correspondence principle. Therefore, the formalism of
scale-relativity seems to be more general and more constructive from this point
of view.

\section{Derivation of the Schr\"odinger equation}
\label{sec_derivation}

In this section, we recall the derivation of the Schr\"odinger equation from the
theory of scale relativity. We propose a more direct derivation than the one
given by Nottale \cite{nottale}. We interpret the scale-covariant equation of
dynamics stemming from Nottale's theory as a hydrodynamic viscous Burgers
equation for a potential flow involving a complex velocity field and an
imaginary
viscosity.  We show that the Schr\"odinger equation can be directly obtained
from this equation by performing the complex Cole-Hopf transformation
equivalent to the WKB transformation. Therefore, the Schr\"odinger equation is
formally equivalent to the complex viscous Burgers equation in the same sense
that the diffusion equation is equivalent to the real viscous Burgers equation
(see Appendix \ref{sec_c}).

\subsection{Basic tools of scale relativity}
\label{sec_basic}

A fundamental consequence of the non-differentiable nature of space-time,
breaking the symmetry $dt\leftrightarrow -dt$,  is the two-valuedness character
of the
velocity which is at the origin of complex numbers in quantum mechanics. Indeed,
 when the flow ${\bf r}(t,dt)$ is not differentiable, the derivative $d{\bf
r}/dt$ is not defined (contrary to standard mechanics) and one has to introduce
two velocities ${\bf u}_+({\bf r}(t),t)$ and  ${\bf u}_-({\bf r}(t),t)$ defined
from $t-dt$ to $t$ for ${\bf u}_-$ and from $t$ to $t+dt$ for ${\bf u}_+$. The
elementary displacement $d{\bf r}_{\pm}$ for both processes  is the sum of a
differential part $d{\bf r}_{\pm}={\bf u}_{\pm}\, dt$ and a non-differentiable
part which is a scale-dependent fractal fluctuation $d{\bf b}_{\pm}$. This
fractal fluctuation is described by a stochastic variable which, by definition,
is of zero mean $\langle d{\bf b}_{\pm}\rangle ={\bf 0}$. It can
be shown that quantum mechanics has a fractal dimension $D_{\rm F}=2$
\cite{feynman} similar  to that of Brownian motion or more generally of Markov
processes. Therefore, we can write
\begin{equation}
\label{b1}
d{\bf r}_{\pm}={\bf u}_{\pm}\, dt+d{\bf b}_{\pm},
\end{equation}
with
\begin{equation}
\label{b2}
\langle d{\bf b}_{\pm}\rangle ={\bf 0},\qquad \langle db_{\pm i} db_{\pm j}\rangle =\pm 2{\cal D}\delta_{ij}dt,
\end{equation}
where ${\cal D}$ is a a sort of ``quantum diffusion coefficient'' measuring the
covariance of the noise.\footnote{In Eq. (\ref{b2}), we consider that $dt>0$ for
the $(+)$ process and
$dt<0$ for the $(-)$ process so that $\pm 2{\cal D}\delta_{ij}dt$ is always
positive \cite{nottale}.} In other
words, ${\cal D}$ characterizes the amplitude of the fractal fluctuations.
 We
can also introduce two classical derivative operators $d_+/dt$ and $d_-/dt$
which yield the twin classical velocities when they are applied to the position
vector ${\bf r}$, namely
\begin{equation}
\label{b3}
\frac{d_+{\bf r}}{dt}={\bf u}_+,\qquad \frac{d_-{\bf r}}{dt}={\bf u}_-.
\end{equation}
It proves convenient in the formalism to replace the twin velocities $({\bf u}_+,{\bf u}_-)$ by the couple $({\bf u}, {\bf u}_Q)$ where
\begin{equation}
\label{b4}
{\bf u}=\frac{{\bf u}_++{\bf u}_-}{2},\qquad {\bf u}_Q=\frac{{\bf u}_+-{\bf u}_-}{2}.
\end{equation}
With these two velocities, we can form a complex velocity
\begin{equation}
\label{b5}
{\bf U}={\bf u}-i{\bf u}_Q.
\end{equation}
We note that it does not correspond to the velocity of a point-like particle but
as that of the fluid of geodesic itself \cite{nottale}.
In the classical limit, we have ${\bf u}_+={\bf u}_-={\bf u}$ (differentiable
trajectories), so the real 
part ${\bf u}$ can be identified  with the classical velocity ${\bf u}$ and the
imaginary part ${\bf u}_Q$ vanishes. More generally, the velocity ${\bf u}$ is
interpreted as the classical velocity and the velocity  ${\bf u}_Q$ as the
quantum velocity. The quantum velocity is at the origin of the complex number
$i$ in the equations of quantum mechanics. We can also define a complex
derivative operator from the classical (differential) parts as
\begin{equation}
\label{b6}
\frac{D}{Dt}=\frac{d_++d_-}{2dt}-i \frac{d_+-d_-}{2dt}
\end{equation}
in terms of which
\begin{equation}
\label{b7}
\frac{D{\bf r}}{Dt}={\bf U}.
\end{equation}

The total derivative with respect to time of a function $f({\bf r}(t),t)$ of fractal dimension $D_{\rm F}=2$ writes
\begin{equation}
\label{b8}
\frac{df}{dt}=\frac{\partial f}{\partial t}+\nabla f\cdot \frac{d{\bf r}}{dt}+\frac{1}{2}\sum_{i,j}\frac{\partial^2 f}{\partial x_i\partial x_j}\frac{dx_i dx_j}{dt}.
\end{equation}
Using Eq. (\ref{b2}), we
find that the classical (differentiable) part of this expression is
\begin{equation}
\label{b9}
\frac{d_{\pm}f}{dt}=\frac{\partial f}{\partial t}+{\bf u}_{\pm}\cdot \nabla f\pm {\cal D}\Delta f.
\end{equation}
Substituting Eq. (\ref{b9}) into Eq. (\ref{b6}), we obtain the expression of the
complex time derivative operator
\begin{equation}
\label{b10}
\frac{D}{Dt}=\frac{\partial}{\partial t}+{\bf U}\cdot \nabla-i {\cal D}\Delta.
\end{equation}
Now, the fundamental postulate of Nottale's theory of scale relativity is that the equations of quantum mechanics (non-differentiable trajectories) can be obtained from the equations of classical mechanics (differentiable trajectories) by replacing the standard velocity ${\bf u}$ by the complex velocity ${\bf U}$ and the standard time derivative $d/dt$ by the complex time derivative $D/Dt$. In other words, $D/Dt$ plays the role of a ``covariant derivative operator'' in terms of which the fundamental equations of physics keep the same form in the classical and quantum regimes. This is similar to the principle of covariance in Einstein's theory of relativity according to which the form of the equations of physics should be conserved under all transformations of coordinates.

\subsection{The Schr\"odinger equation}
\label{sec_s}

We now use the principle of covariance to derive the Schr\"odinger equation from Newton's law. Using the covariance principle, we write the fundamental equation of dynamics as
\begin{equation}
\label{s0}
\frac{D{\bf U}}{Dt}=-\nabla\Phi,
\end{equation}
where ${\bf F}=-\nabla\Phi$ is the force by unit of mass exerted on the
particle.  This equation can also be obtained
from an action principle based on
a complex action ${\cal S}$. Indeed, it corresponds to the Euler-Lagrange
equation associated with the stationarity of the complex action: $\delta {\cal
S}=0$. Using the expression (\ref{b10}) of the complex time derivative operator,
the foregoing equation can be rewritten as
\begin{equation}
\label{s1}
\frac{\partial {\bf U}}{\partial t}+({\bf U}\cdot \nabla){\bf U}=i {\cal D}\Delta{\bf U}-\nabla\Phi.
\end{equation}
This equation is similar to the viscous Burgers equation of fluid mechanics,
except that in the present case the velocity field ${\bf U}({\bf r},t)$ is
complex and the
viscosity $\nu=i{\cal D}$ is imaginary. Therefore, quantum mechanics may be
interpreted as a generalized hydrodynamics involving a complex velocity field
and an imaginary viscosity. In the  Lagrangian formalism
\cite{nottale}, the
complex impulse ${\bf P}=m{\bf U}$ and the complex energy ${\cal E}$ write
\begin{equation}
{\bf P}=\nabla {\cal S},\qquad {\cal E}=-\frac{\partial {\cal S}}{\partial t},
\end{equation}
where $m$ is the mass of the particle and ${\cal S}({\bf r},t)$ is the  complex
action. The complex velocity field is potential since it can be
written as the gradient of a function:
\begin{equation}
\label{s2}
{\bf U}=\frac{\nabla {\cal S}}{m}.
\end{equation}
As a consequence of Eq. (\ref{s2}), the flow
is irrotational: $\nabla\times {\bf U}={\bf 0}$. Using the well-known identities
of fluid mechanics $({\bf U}\cdot \nabla){\bf U}=\nabla ({{\bf U}^2}/{2})-{\bf
U}\times (\nabla\times {\bf U})$ and $\Delta{\bf U}=\nabla(\nabla\cdot {\bf
U})-\nabla\times(\nabla\times{\bf U})$  which reduce to $({\bf U}\cdot
\nabla){\bf U}=\nabla ({{\bf U}^2}/{2})$ and $\Delta{\bf U}=\nabla (\nabla\cdot
{\bf U})$ for an irrotational flow, and using the identity $\nabla\cdot {\bf
U}=\Delta{\cal S}/m$ resulting from Eq. (\ref{s2}), we find that Eq. (\ref{s1})
is equivalent to
\begin{equation}
\label{s3}
\frac{\partial {\cal S}}{\partial t}+\frac{1}{2m}(\nabla {\cal S})^2-i{\cal D}\Delta{\cal S}+m\Phi=0.
\end{equation}
Equation (\ref{s3}) can be viewed as a quantum Hamilton-Jacobi equation 
for a complex action, or as a Bernoulli equation  for a complex potential. We
define the wave function $\psi({\bf r},t)$ through the complex Cole-Hopf
transformation
\begin{equation}
\label{s3b}
{\cal S}=-2im{\cal D}\ln\psi.
\end{equation}
Since macroscopic bodies do not feel the
non-differentiable nature of space-time, the quantum
diffusion coefficient ${\cal D}$ must depend on the mass $m$ of the particle
and must tend to zero when $m\rightarrow +\infty$. We shall assume that
${\cal D}$ is inversely proportional to  $m$ and
set\footnote{This relation first appeared in Nelson's stochastic interpretation
of quantum mechanics \cite{nelson} so we shall call it the Nelson relation.}
\begin{equation}
\label{s8}
{\cal D}=\frac{\hbar}{2m}.
\end{equation}
The constant $\hbar$ has the dimension of an action. We shall see that it can
be identified with the Planck constant. With the relation (\ref{s8}), Eq.
(\ref{s3b}) can be rewritten as
\begin{equation}
\label{s5}
{\cal S}=-i\hbar\ln\psi,
\end{equation}
where the mass $m$ of the particle does not appear anymore (this is a strong
argument in favor of Eq. (\ref{s8})). Equation (\ref{s5}) can be rewritten in
the WKB form 
\begin{equation}
\label{s4}
\psi=e^{i{\cal S}/\hbar},
\end{equation}
relating the wavefunction to the complex action. Therefore, the complex 
Cole-Hopf transformation (\ref{s3b}) is equivalent to the WKB
transformation (\ref{s4}) provided that the relation (\ref{s8})
holds.\footnote{As shown in
Appendix \ref{sec_qdc}, it is only when the relation (\ref{s8}) is satisfied
in Eq. (\ref{s4}) 
that the integral  $\int |\psi|^2\, d{\bf r}$ is conserved, allowing us to
identify $|\psi|^2$ with a probability density (see Eqs. (\ref{s10}) and
(\ref{s11})).} In
particular, the factor $2$ in Eq. (\ref{s8}) has  the same origin as the one
arising in
the Cole-Hopf
transformation (\ref{s3b}). 
Substituting Eq. (\ref{s5}) into Eq. (\ref{s3}), and using the identity
\begin{equation}
\label{s6}
\Delta(\ln\psi)=\frac{\Delta\psi}{\psi}-\frac{1}{\psi^2}(\nabla\psi)^2,
\end{equation}
we obtain the Schr\"odinger equation
\begin{equation}
\label{s9}
i\hbar\frac{\partial\psi}{\partial t}=-\frac{\hbar^2}{2m}\Delta\psi+m\Phi\psi.
\end{equation}
This justifies interpreting $\hbar$ as the Planck constant. This also justifies
Eq. (\ref{s8}) {\it a posteriori}. From the
Schr\"odinger equation (\ref{s9}), we can easily derive the identity
\begin{eqnarray}
\label{stress1}
\frac{\partial |\psi|^2}{\partial t}+\nabla\cdot {\bf j}=0,
\end{eqnarray}
where
\begin{eqnarray}
\label{stress2}
{\bf j}=\frac{\hbar}{2im}\left (\psi^*\nabla\psi-\psi\nabla\psi^*\right )
\end{eqnarray}
is a current. Since the Schr\"odinger equation (\ref{s9}) conserves the integral  $\int |\psi|^2({\bf r},t)\, d{\bf r}$, we can interpret
\begin{equation}
\label{s10}
\rho({\bf r},t)=|\psi|^2({\bf r},t)
\end{equation}
as the probability density of finding the particle at position ${\bf r}$ at time
$t$ (Born's postulate).\footnote{See Sec. \ref{sec_ni} for an interpretation
of this result in the framework of Nottale's theory of scale relativity.} This
interpretation yields the normalization condition
\begin{equation}
\label{s11}
\int |\psi|^2({\bf r},t)\, d{\bf r}=1.
\end{equation}
Equation (\ref{stress1}) shows  that the conservation of the
normalization condition is locally satisfied.

{\it Remark:} By making the complex Cole-Hopf transformation (\ref{s3b}),
equivalent to the WKB transformation (\ref{s4}), we have shown that the complex
viscous Burgers equation (\ref{s1}) is equivalent to the Schr\"odinger equation
(\ref{s9}) in the same sense that the real viscous Burgers equation is
equivalent to the diffusion equation (see Appendix \ref{sec_c}). It is also
possible to proceed the other way round. Starting from the
Schr\"odinger equation (\ref{s9}) and making the WKB transformation (\ref{s4}),
we obtain the complex quantum Hamilton-Jacobi equation (\ref{s3}). This equation
was known by the founders of quantum mechanics (see the Introduction of
\cite{chavmatos}). If we define a complex velocity field according to Eq.
(\ref{s2}), we obtain the complex viscous Burgers equation (\ref{s1}) with an
imaginary
viscosity given by $\nu=i\hbar/2m$. Therefore, the
Schr\"odinger equation is {\it equivalent} to a hydrodynamic equation involving
a  complex velocity and an imaginary viscosity. To our knowledge, this complex
hydrodynamic equation was ``missed'' by the founders of quantum mechanics and
first appeared in the work of Nottale \cite{nottale} at the very start of his
theory of scale relativity. It could have been obtained much earlier by
directly starting from the Schr\"odinger equation. Of course, the interest of
Nottale's
theory is to justify this equation from first principles and then to derive the 
Schr\"odinger equation from it.

{\it Remark:} For $\hbar=0$, Eq. (\ref{s3}) reduces to the classical
Hamilton-Jacobi equation 
\begin{equation}
[{S}]\equiv \frac{\partial {S}}{\partial
t}+\frac{1}{2m}(\nabla {S})^2+m\Phi=0.
\end{equation}
The complex quantum Hamilton-Jacobi equation  (\ref{s3}) can be
written as
\begin{equation}
[{\cal S}]=i\frac{\hbar}{2m}\Delta{\cal S}.
\end{equation}
This equation connects the classical mechanics to the wave mechanics. It could
be used as postulate, similar to the correspondance principle, to derive the
Schr\"odinger equation (\ref{s9}). Actually, this is how De Donder \cite{dd2}
 derived the KG equation in the relativistic context (see footnote 7 in
\cite{chavmatos}).

{\it Remark:} In the previous calculations, $\hbar$ needs not be the Planck
constant. Indeed, Nottale's formalism may apply to situations different from
quantum mechanics \cite{nottale}. In any case, the ``constant''  $\hbar$ (that
Nottale writes ${\cal S}_0$) must be related to ${\cal D}$ according to Eq.
(\ref{s8}), see footnote 14. Therefore, the general Schr\"odinger equation can
be written as
\begin{equation}
i{\cal D}\frac{\partial\psi}{\partial
t}=-{\cal D}^2\Delta\psi+\frac{1}{2}\Phi\psi,
\end{equation}
where ${\cal D}$ depends on the system under consideration. It is interesting
to note that, under that form, the mass $m$ of the particle does not appear
anymore, as is obvious from the original equation (\ref{s0}).

\subsection{The Madelung transformation}
\label{sec_mag}

Using the Madelung \cite{madelung} transformation, the Schr\"odinger equation (\ref{s9}) can be written  in the form of real hydrodynamic equations. We write the wavefunction as
\begin{equation}
\label{ntmad2}
\psi({\bf r},t)=\sqrt{\rho({\bf r},t)} e^{iS({\bf r},t)/\hbar},
\end{equation}
where $\rho$ is the probability density for the particle position and $S$ is the
real action. They can be expressed in terms of the wave function as
\begin{equation}
\label{ntmad2b}
\rho=|\psi|^2,\qquad S=-i\frac{\hbar}{2}\ln\left (\frac{\psi}{\psi^*}\right ).
\end{equation}
Following Madelung, we introduce the real potential velocity field
\begin{equation}
\label{ntmad3}
{\bf u}=\frac{\nabla S}{m}.
\end{equation}
The flow defined in this way is irrotational since $\nabla\times {\bf u}={\bf
0}$. We also define the energy
\begin{equation}
\label{ntmad3q}
E=-\frac{\partial S}{\partial t}.
\end{equation}
Substituting Eq. (\ref{ntmad2})  into Eq. (\ref{s9}), separating real and imaginary parts, and using Eq. (\ref{ntmad3}), we obtain the hydrodynamic equations
\begin{equation}
\label{tmad4c}
\frac{\partial\rho}{\partial t}+\nabla\cdot (\rho {\bf u})=0,
\end{equation}
\begin{equation}
\label{ntmad5}
\frac{\partial S}{\partial t}+\frac{1}{2m}(\nabla S)^2+m\Phi+Q=0,
\end{equation}
\begin{equation}
\label{ntmad6}
\frac{\partial {\bf u}}{\partial t}+({\bf u}\cdot \nabla){\bf u}=-\nabla\Phi-\frac{1}{m}\nabla Q,
\end{equation}
where \begin{equation}
\label{ntmad8}
Q=-\frac{\hbar^2}{2m}\frac{\Delta \sqrt{\rho}}{\sqrt{\rho}}
=-\frac{\hbar^2}{4m}\left\lbrack
\frac{\Delta\rho}{\rho}-\frac{1}{2}\frac{(\nabla\rho)^2}{\rho^2}
\right\rbrack=-\frac{\hbar^2}{8m}({\nabla\ln\rho})^2-\frac{\hbar^2}{4m}
\Delta(\ln\rho)
\end{equation}
is the quantum potential (we have used the identity (\ref{s6}) to obtain the
last equality). Equation (\ref{tmad4c}) is similar to the continuity equation in
hydrodynamics. It accounts for the local conservation of the normalization
condition.\footnote{Taking the gradient of the action in Eq. (\ref{ntmad2b}) and
comparing its expression with Eq. (\ref{stress2}), we find that ${\bf j}=\rho
{\bf u}$. As a result, the continuity equation (\ref{tmad4c}) is equivalent to
Eq. (\ref{stress1}).} It arises from the imaginary part of the Schr\"odinger
equation in the Madelung transformation. Equation (\ref{ntmad5}) is similar to
the Hamilton-Jacobi equation or to the Bernoulli equation with an additional
quantum potential $Q$. It arises from the real part of the Schr\"odinger
equation in the Madelung transformation. Equation (\ref{ntmad6}) has a form
similar to the Euler equation of hydrodynamics with an additional quantum force
$-\nabla Q$. It is obtained by taking the gradient of Eq. (\ref{ntmad5}) and
using the well-known  identity $({\bf u}\cdot \nabla){\bf u}=\nabla ({{\bf
u}^2}/{2})-{\bf u}\times (\nabla\times {\bf u})$ which reduces to $({\bf u}\cdot
\nabla){\bf u}=\nabla ({{\bf u}^2}/{2})$ for an irrotational flow.  Equations
(\ref{tmad4c})-(\ref{ntmad6}) are called the quantum Euler equations.

\subsection{The quantum force}

The quantum potential (\ref{ntmad8}) first appeared in the paper of  Madelung
\cite{madelung} (see also his less well-known paper \cite{madelungearly}) and
was rediscovered by Bohm \cite{bohm}. For that reason, it is sometimes called
``the Bohm potential''.\footnote{A relativistic version of the
quantum potential appeared in the works 
of de Broglie \cite{broglie1927a,broglie1927b,broglie1927c} and
London \cite{london} who developed a hydrodynamic representation of the KG
equation independently of Madelung \cite{madelung}.}  The quantum force by unit
of mass writes
\begin{equation}
\label{ntmad13a}
{\bf F}_Q=-\frac{1}{m}\nabla Q.
\end{equation}
We note the identity
\begin{equation}
\label{mad19}
(F_Q)_i=-\frac{1}{m}\partial_i Q= -\frac{1}{\rho}\partial_j P_{ij},
\end{equation}
where $P_{ij}$ is the quantum stress (or pressure) tensor defined by
\begin{equation}
\label{mad20}
P_{ij}^{(1)}=-\frac{\hbar^2}{4m^2}\rho\, \partial_i\partial_j\ln\rho\qquad {\rm or}\qquad
P_{ij}^{(2)}=\frac{\hbar^2}{4m^2}\left (\frac{1}{\rho}\partial_i\rho\partial_j\rho-\delta_{ij}\Delta\rho\right ).
\end{equation}
This tensor is manifestly symmetric: $P_{ij}=P_{ji}$.  The identity
(\ref{mad19}) shows that the 
quantum force $-\nabla Q$ is equivalent to the force produced by
an  anisotropic pressure tensor $P_{ij}$. The tensors defined by Eq.
(\ref{mad20}) are related to each other by
\begin{equation}
\label{tmad16}
P_{ij}^{(1)}=P_{ij}^{(2)}+\frac{\hbar^2}{4m^2}(\delta_{ij}\Delta\rho-\partial_i\partial_j\rho).
\end{equation}
They differ by a tensor
$\chi_{ij}=\delta_{ij}\Delta\rho-\partial_i\partial_j\rho$ satisfying 
$\partial_j\chi_{ij}=0$. Contracting the indices, we obtain
\begin{equation}
\label{mad20b}
P_{ii}^{(1)}=-\frac{\hbar^2}{4m^2}\rho\, \Delta\ln\rho,\qquad P_{ii}^{(2)}=\frac{\hbar^2}{4m^2}\left \lbrack\frac{(\nabla\rho)^2}{\rho}-d\Delta\rho\right \rbrack,
\end{equation}
and the relation
\begin{equation}
\label{mad21c}
P_{ii}^{(1)}=P_{ii}^{(2)}+(d-1)\frac{\hbar^2}{4m^2}\Delta\rho.
\end{equation}
According to Eq. (\ref{mad19}) we note that
\begin{equation}
\label{zqf}
\langle {\bf F}_Q\rangle=\int \rho {\bf F}_Q\, d{\bf r}={\bf 0}
\end{equation}
so there is no resultant of the quantum force.  This property is at the basis of
the  Ehrenfest theorem \cite{ehrenfest} (see Appendix \ref{sec_eh}).

{\it Remark:} We provide below another derivation of Eq. (\ref{zqf}). Using
Eqs. (\ref{ntmad8}) and (\ref{ntmad13a}), the average value of the quantum force
is
\begin{equation}
\label{ide1}
\langle {\bf F}_Q\rangle =\int \rho {\bf F}_Q\, d{\bf r}=-\frac{1}{m}\int \rho
\nabla Q\, d{\bf r}=\frac{1}{m}\int Q\nabla \rho \, d{\bf
r}=-\frac{\hbar^2}{2m^2}\int \frac{\Delta\sqrt{\rho}}{\sqrt{\rho}}\nabla \rho \,
d{\bf r}=-\frac{\hbar^2}{m^2}\int {\Delta\sqrt{\rho}}\nabla \sqrt{\rho} \, d{\bf
r}.
\end{equation}
Since
\begin{equation}
\label{ide2}
\int {\Delta\sqrt{\rho}}\nabla \sqrt{\rho} \, d{\bf r}=-\int
{\nabla\sqrt{\rho}}\Delta \sqrt{\rho} \, d{\bf r}={\bf 0},
\end{equation}
obtained by integration by parts, we recover Eq. (\ref{zqf}).

\subsection{The conservation of the average energy}

Using Eqs. (\ref{ntmad3}) and  (\ref{ntmad3q}), the quantum
Hamilton-Jacobi equation (\ref{ntmad5}) can be written
as
\begin{equation}
\label{new1}
E({\bf r},t)=\frac{1}{2}m{\bf u}^2+m\Phi+Q.
\end{equation}
The energy $E({\bf r},t)$ is the sum of the kinetic energy, the classical
potential and the
quantum potential. This is the quantum generalization of the classical equation
of mechanics $E=m{\bf u}^2/2+m\Phi$. We note that the Euler equation
(\ref{ntmad6}) can be written as
\begin{equation}
\label{new1b}
\frac{\partial {\bf u}}{\partial t}=-\frac{\nabla E}{m}.
\end{equation}
The average energy is
\begin{equation}
\label{dif1q}
\langle E\rangle=\int\rho E \, d{\bf r}=\int \rho m\frac{{\bf u}^2}{2}\,d{\bf
r} +\int \rho m\Phi\, d{\bf r}+\int \rho Q\, d{\bf r}.
\end{equation}
We can show that it is conserved (see Appendix \ref{sec_cote}):
\begin{equation}
\label{dote}
\dot{\langle E\rangle}=0.
\end{equation}
The energies defined by Eqs. (\ref{ntmad3q})
and (\ref{new1}) first appeared in the papers of Madelung
\cite{madelung,madelungearly}. He also mentioned the conservation of the average
energy (\ref{dif1q}).

\subsection{Stationary quantum states}
\label{sec_tigpa}

The time-independent solutions of a quantum system are special solutions
$\rho({\bf r})$ and ${\bf u}({\bf r})$ of the hydrodynamic equations
(\ref{tmad4c})-(\ref{ntmad8}). Two types of stationary solutions must be
considered.

\subsubsection{Static states}

For static states, $\rho({\bf r})$ is independent on time and ${\bf u}({\bf
r})={\bf 0}$. The continuity equation (\ref{tmad4c}) is automatically
satisfied and the quantum Euler equation (\ref{ntmad6}) reduces to the condition
of quantum hydrostatic equilibrium
\begin{equation}
\label{sta2}
\nabla\Phi+\frac{1}{m}\nabla Q={\bf 0}.
\end{equation}
Therefore, a static state results from the balance between the external force
and the quantum force. On the other hand, according to Eq. (\ref{ntmad3}),
$\nabla S={\bf 0}$ so that $S$ depends only on time: $S=S(t)$. From Eqs.
(\ref{ntmad3q}) and (\ref{ntmad5}), we obtain
\begin{equation}
\label{sta1}
E=-\frac{d S}{d t}=m\Phi+Q.
\end{equation}
Since the second term depends only on time and the third 
term depends only on ${\bf r}$, $E$ is necessarily a constant. We
then find that
$S=-Et$. Taking the gradient of Eq. (\ref{sta1}), we recover the condition
of quantum hydrostatic equilibrium (\ref{sta2}). On the other hand, since
$S=-Et$, the wave function (\ref{ntmad2}) can be written as
\begin{equation}
\label{sta3}
\psi({\bf r},t)=\phi({\bf r})e^{-i E t/\hbar},
\end{equation}
where $\phi({\bf r})=\sqrt{\rho({\bf r})}$ is real. Therefore, we can define a quantum static state by Eq. (\ref{sta3}). Substituting Eq. (\ref{sta3}) into Eq. (\ref{s9}), we obtain the time-independent Schr\"odinger equation
\begin{eqnarray}
\label{sta4}
-\frac{\hbar^2}{2m}\Delta\phi+m\Phi\phi=E\phi.
\end{eqnarray}
Equation (\ref{sta4})  is an eigenvalue equation for the wave function $\phi({\bf r})$ where the eigenvalue $E$ is the energy. It is equivalent to the previous equations expressed in terms of $\rho$. Indeed, dividing Eq. (\ref{sta4}) by $\phi$ and using $\rho=\phi^2$, we get
\begin{equation}
\label{sta5}
-\frac{\hbar^2}{2m}\frac{\Delta \sqrt{\rho}}{\sqrt{\rho}}+m\Phi=E \qquad {\rm
i.e.}\qquad Q+m\Phi=E.
\end{equation}
As shown above, this equation can be directly derived from the 
quantum Hamilton-Jacobi equation (\ref{ntmad5}) by setting  $S=-Et$. It can
also be obtained from Eq. (\ref{new1}) by taking $E({\bf r},t)=E$ and ${\bf
u}={\bf 0}$.

\subsubsection{Dynamic states}

For dynamic states, $\rho({\bf r})$ and ${\bf u}({\bf r})$ are independent on
time and ${\bf u}({\bf r})\neq {\bf 0}$.  The continuity equation (\ref{tmad4c})
and the quantum Euler equation (\ref{ntmad6}) reduce to
\begin{equation}
\label{sta8}
\nabla\cdot (\rho {\bf u})=0,
\end{equation}
\begin{equation}
\label{sta9}
({\bf u}\cdot \nabla){\bf u}=-\nabla\Phi-\frac{1}{m}\nabla Q.
\end{equation}
Therefore, a dynamic state results from the balance between the inertia, the
external force and the quantum force. On the other hand, according to Eq.
(\ref{ntmad3}), $\nabla S$ does not depend on time. Therefore, $S$ must be of
the form $S=S_1({\bf r})+S_2(t)$. From Eqs. (\ref{ntmad3q}) and (\ref{ntmad5}),
we obtain
\begin{equation}
\label{sta7}
E=-\frac{\partial S}{\partial t}=\frac{1}{2}m {\bf u}^2+m\Phi+Q.
\end{equation}
Since the second term depends only on time and the third 
term depends only on ${\bf r}$, $E$ is necessarily a constant. We then find that
$S_2=-Et$ so that $S=S_1({\bf r})-Et$. Taking the gradient of Eq.
(\ref{sta7}), we obtain 
\begin{equation}
\label{sta9qw}
\nabla
\cdot \left ( \frac{{\bf u}^2}{2}+\Phi+\frac{Q}{m} \right )={\bf 0},
\end{equation}
which is equivalent to Eq. (\ref{sta9}). On the other hand, since
$S=S_1({\bf r})-Et$, the wave function (\ref{ntmad2}) can be written as
\begin{equation}
\label{sta10}
\psi({\bf r},t)=\Psi({\bf r})e^{-i E t/\hbar},
\end{equation}
where $\Psi({\bf r})=\phi({\bf r})e^{i S_1({\bf r})/\hbar}$ is complex and $\phi({\bf r})=\sqrt{\rho({\bf r})}$ is real. Substituting Eq. (\ref{sta10}) into Eq. (\ref{s9}), we obtain the time-independent Schr\"odinger equation
\begin{eqnarray}
\label{sta11}
-\frac{\hbar^2}{2m}\Delta\Psi+m\Phi\Psi=E\Psi,
\end{eqnarray}
where $\Psi$ is complex. Equation (\ref{sta11}) is equivalent to the previous
equations expressed in terms of $\rho$ and ${\bf u}$.

\subsection{Traditional interpretation of the Madelung transformation}
\label{sec_interpretation}

In the hydrodynamic representation of the Schr\"odinger equation developed 
by Madelung \cite{madelung}, Bohm \cite{bohm}, and
Takabayasi \cite{takabayasi1,takabayasi2}, Eqs.  (\ref{tmad4c})-(\ref{ntmad6})
are regarded
as representing an ensemble of particle trajectories which belong to a single
velocity potential $S$ by the relation of Eq. (\ref{ntmad3}).\footnote{This is
what de Broglie \cite{broglie1927a,broglie1927b,broglie1927c} 
called the pilot-wave theory in his relativistic approach based on the KG
equation since the velocity is determined by the wave $\psi$.} Introducing the
material derivative $d/dt=\partial/\partial t+{\bf u}\cdot\nabla$, the quantum
Euler equation (\ref{ntmad6}) can be rewritten as
\begin{equation}
\label{int1}
m\frac{d{\bf u}}{dt}=-m\nabla\Phi-\nabla Q,
\end{equation}
where ${\bf u}(t)={\bf u}({\bf r}(t),t)=d{\bf r}/dt$ is 
the Lagrangian  velocity of a fluid particle.  Equation (\ref{int1}) is
similar to
the classical equation of motion of a particle. However, in the present case,
the particle experiences an extra quantum force $-\nabla Q$. This quantum force
is extremely complex since it depends on the state of the system itself via its
density $\rho$. This is what leads to the peculiar effects of quantum mechanics
such as diffusion of wave packets, interferences, tunnel effect etc. As
Bohm \cite{bohm} states, the quantum potential corresponds to the interaction
of the $\psi$
field with its own particle.

If we define the energy of a quantum particle by\footnote{This corresponds to
Eq. (\ref{new1}) with $E(t)=E({\bf r}(t),t)$ in the Lagrangian description.}
\begin{equation}
\label{int2}
E(t)=\frac{1}{2}m{\bf u}^2+m\Phi+Q,
\end{equation}
we get
\begin{equation}
\label{int3}
\dot {E}=\frac{\partial Q}{\partial t}
\end{equation}
since the quantum potential $Q({\bf r},t)$  depends explicitly on time.
Therefore, the energy of a quantum  particle is not conserved.
  As a result, the trajectory of a quantum particle exhibits a
complicated fluctuation. This is responsible for the blurring of the classical
trajectory. However, if we consider the ensemble of particle trajectories,
the average energy $\langle E\rangle$ given by Eq. (\ref{dif1q}) 
is conserved (see Appendix \ref{sec_cote}).

Because of the quantum potential, the particle can 
reach an equilibrium state in which ${\bf u}={\bf 0}$. Bohm \cite{bohm} writes
that ``the
absence of motion is possible because the applied force, $-m\nabla\Phi$, is
balanced by the quantum-mechanical force, $-\nabla Q$, produced by the
Schr\"odinger $\psi$-field  acting on its own particle. There is, however, a
statistical ensemble of possible positions of the particle, with a probability
density, $P({\bf r})=|\psi({\bf r})|^2$.'' In the hydrodynamic
language, this corresponds to the condition of quantum hydrostatic equilibrium
\begin{equation}
\label{int5}
m\nabla\Phi-\nabla Q={\bf 0}.
\end{equation}
For example, in the case of the hydrogen atom, Eq. (\ref{int5}) describes the
balance between the attractive Coulomb force and  the repulsion due to the
quantum force. This fluid dynamic formulation offers a simple physical
explanation of the stability of matter because of quantum mechanics. Without the
quantum potential, the
electron would fall on the nucleus. In this sense, the hydrodynamic picture
leads to a conception completely different from the original Bohr
model. The tunnel effect
can also be explained in terms of the quantum potential. Bohm
\cite{bohm} mentions that the
existence of $Q$ partially compensates for the large value of $\Phi$ and enables
the particle to ``ride over the barrier''. In the wave interpretation of the
Schr\"odinger equation, the effect of the quantum potential is played by the
Heisenberg uncertainty principle.

{\it Remark:} We may note certain analogies between the motion of a quantum
particle and the notion of phase mixing and violent relaxation introduced by
Lynden-Bell \cite{lb} in the context of collisionless stellar systems described
by the Vlasov-Poisson equations. For
example, the energy of an individual star in a protogalaxy is not conserved
because of the
fluctuations of the gravitational field [compare Eq. (1) in
\cite{lb} with Eq. (\ref{int3})] but the total energy of the system is
conserved. On the other hand, as shown in \cite{csr}, collective effects in a
stellar system produce
a sort of pressure force [see Eq. (5.11) in \cite{csr}] that can balance the
gravitational force and lead to a
quasi-stationary state (QSS) on a coarse-grained scale. This pressure
force plays a role similar to the quantum force, which can also be interpreted
as a pressure force according to Eq. (\ref{mad19}).

\subsection{Interpretation of the Madelung transformation in the theory of
scale relativity}
\label{sec_ni}

Historically, the Madelung transformation is 
based on the Schr\"odinger equation (\ref{s9}) which is at the basis of the
wave mechanics initiated by de Broglie \cite{brogliethese}. In that
context, the fundamental
object is the wave function $\psi({\bf r},t)$.
According to
the Born postulate, the square norm of the wave function $\rho({\bf
r},t)=|\psi|^2$ is interpreted as the probability density of finding the 
particle at ${\bf r}$ at time $t$. Starting from the wave function,
Madelung \cite{madelung} defined a velocity field by Eq. (\ref{ntmad3}), in a
rather {\it ad hoc} manner, and showed that the
Schr\"odinger equation is equivalent to the hydrodynamic equations
(\ref{tmad4c})-(\ref{ntmad6}). However, this is essentially a formal analogy. In
the framework of conventional quantum mechanics, this analogy with hydrodynamics
remains relatively mysterious.\footnote{It can be given a more physical
interpretation in the context of BECs as discussed in the Conclusion.} One
important new feature of this hydrodynamic representation is the quantum
potential (\ref{ntmad8}) that accounts for all quantum phenomena.

In the theory of scale relativity,  the
interpretation is very
different\footnote{We briefly summarize the main points of the discussion of
Nottale \cite{nottale}.} because the fundamental object is the complex velocity
${\bf U}$ of the fractal geodesics and the complex hydrodynamic equation
(\ref{s1}) of these
geodesics from which the Schr\"odinger equation (\ref{s9}) is derived. 
Now, we expect the fluid of geodesics to be more concentrated at some places and
less at others as does a real fluid. To find the probability density of presence
of the paths we can
remark that Eqs. (\ref{s1}) and (\ref{s9})  are equivalent to the real fluid
equations (\ref{tmad4c})-(\ref{ntmad6}) [see Appendix \ref{sec_gne}]. In that
context, Eq.  (\ref{ntmad3}) is not an {\it ad hoc} definition but it
corresponds to the
classical velocity (the real part of ${\bf U}$). On the other hand, in the
theory of scale relativity, the
quantum potential is a manifestation of the geometry of spacetime, namely, of
its non-differentiability and fractality, in similarity with  the Newtonian
potential being a manifestation of the curvature of spacetime in Einstein's
theory of general relativity. Then, Eqs.
(\ref{tmad4c})-(\ref{ntmad6}), which are equivalent to Eqs. (\ref{s1})  and
(\ref{s9}), describe a fluid of fractal geodesics in a
non-differentiable space-time.
They have therefore a clear physical interpretation. As a result $\rho({\bf
r},t)=|\psi|^2$ represents the density of the geodesic fluid, and the
probability
density for the quantum ``particle'' to be found at a given position
must
be proportional to  $\rho({\bf
r},t)$ (the constant of proportionality can be set equal to
unity). As a result, the Born postulate is naturally justified in the theory of
scale relativity. 

In a sense, Madelung \cite{madelung} and Nottale \cite{nottale} proceed in
opposite directions. Madelung starts from the wave equation of Schr\"odinger and
shows that it is equivalent to fluid equations. Inversely, Nottale's theory is
directly based on fluid equations [see Eq. (\ref{s1})] and he shows that they
are equivalent to the Schr\"odinger equation.

It is interesting to come back to the manner how the density $\rho({\bf
r},t)$ arises in the theories of Born, Nelson, and Nottale. In the original
interpretation of quantum mechanics by Born \cite{bornf,bornnature,born}, the
relation 
$\rho({\bf
r},t)=|\psi|^2$ is introduced {\it at the end} as an independent postulate that
gives an interpretation to the Schr\"odinger equation: $\rho({\bf
r},t)=|\psi|^2$ determines the probability of presence of a particle.  We note
that the  Schr\"odinger equation (\ref{s9}) is obtained without having to
specify $\rho({\bf
r},t)$ (see the original derivations of the Schr\"odinger equation summarized in
Appendix F of \cite{chavmatos}).  In the stochastic interpretation
of Nelson \cite{nelson},  $\rho({\bf
r},t)$ is introduced {\it from the start} by analogy with a diffusion process.
Finally, in Nottale's theory of
scale relativity \cite{nottale}, the relation $\rho=|\psi|^2$  that
represents the density of the geodesic fluid, or the probability of presence of
a particle, is not a new postulate. It is fixed {\it at the end} by Eqs. 
(\ref{tmad4c})-(\ref{ntmad6}) that are equivalent to the equation of
fractal geodesics (\ref{s1}). We note that the Schr\"odinger equation
(\ref{s9}) is obtained without having
to specify $\rho({\bf r},t)$.

\subsection{Complex and real hydrodynamic equations}

As we have seen previously, the Schr\"odinger 
equation is formally equivalent to hydrodynamic equations. These
hydrodynamic equations can be either complex or real. 

By making the complex
Cole-Hopf transformation (\ref{s3b}), equivalent to the WKB transformation
(\ref{s4}), the Schr\"odinger equation (\ref{s9}) can be transformed into the
complex viscous Burgers equation (\ref{s1}). This is a single complex equation
for the complex velocity ${\bf U}$ with an imaginary viscosity $\nu=i\hbar/2m$
and without quantum potential.

By making the
Madelung transformation (\ref{ntmad2})-(\ref{ntmad3}), the
Schr\"odinger equation (\ref{s9}) can be transformed into a set of two real
hydrodynamic
equations, the
continuity equation (\ref{tmad4c}) for the density $\rho$ and the quantum Euler
equation (\ref{ntmad6}) for the real velocity ${\bf u}$. The Euler equation
involves a quantum potential $Q$ but no viscosity.

As shown in Appendix \ref{sec_ri}, the hydrodynamic equations 
(\ref{tmad4c}) and (\ref{ntmad6}) correspond to the real and imaginary
parts of the complex viscous Burgers equation (\ref{s1}). Furthermore, the
viscous term $i {\cal D}\Delta{\bf U}$ in the complex viscous Burgers equation
(\ref{s1}) can be interpreted as a complex quantum force (see Appendix
\ref{sec_cqq}).

{\it Remark:} Nottale \cite{nottale} does not interpret Eq. (\ref{s1}) as a
complex Burgers equation with an imaginary viscosity but rather as an equation
of geodesics. Similarly, he does not interpret Eq. (\ref{s3b}) as a Cole-Hopf
transformation. We believe that our ``hydrodynamic interpretation'' clarifies
the derivation of the Schr\"odinger equation by making a close analogy with
fluid mechanics (see Appendix \ref{sec_c}).

\section{Derivation of a generalized Schr\"odinger equation}
\label{sec_diss}

In this section, we use the formalism of scale relativity to derive a generalized Schr\"odinger equation that includes dissipative effects. As we shall see, our procedure automatically yields a term accounting for (effective) thermal effects. The two terms are related by a sort of fluctuation-dissipation theorem.

\subsection{The generalized Schr\"odinger equation}

In order to incorporate dissipative effects into the Schr\"odinger equation, 
it would seem natural to introduce a damping term $-\gamma {\bf U}$ in the
fundamental equation of dynamics (\ref{s0}). However, if we proceed too naively,
we obtain a generalized Schr\"odinger equation that does not satisfy the local
conservation of the normalization condition (see Appendix \ref{sec_pb}). In
order to conserve the normalization condition locally, while taking dissipative
effects into account, we write the fundamental equation of dynamics under the
form
\begin{equation}
\label{qb1}
\frac{D{\bf U}}{Dt}=-\nabla\Phi-{\rm Re} (\gamma {\bf U}),
\end{equation}
where $\gamma$ is a complex friction coefficient. As shown in Sec. \ref{sec_mag}, the continuity equation (\ref{tmad4c}) is obtained by performing the Madelung transformation and taking the imaginary part of the Schr\"odinger equation. Therefore, by projecting the damping term $-\gamma {\bf U}$ on the real axis, we are guaranteed to obtain a generalized Schr\"odinger equation  that satisfies the local conservation of the normalization condition since the continuity equation will not be affected by the damping term.  Using the expression (\ref{b10}) of the covariant derivative, Eq. (\ref{qb1}) can be rewritten as a damped complex viscous Burgers equation of the form
\begin{equation}
\label{qb2}
\frac{\partial {\bf U}}{\partial t}+({\bf U}\cdot \nabla){\bf U}=i {\cal D}\Delta{\bf U}-\nabla\Phi-{\rm Re} (\gamma {\bf U}).
\end{equation}
Using Eq. (\ref{s2}), and proceeding as in Sec. \ref{sec_s}, the corresponding complex Hamilton-Jacobi equation is
\begin{equation}
\label{qb3}
\frac{\partial {\cal S}}{\partial t}+\frac{1}{2m}(\nabla {\cal S})^2-i{\cal D}\Delta{\cal S}+m\Phi+V(t)+{\rm Re} (\gamma {\cal S})=0,
\end{equation}
where $V(t)$ is a ``constant'' of integration depending on time that has a
nonzero value when dissipative effects are taken into account (see below).
Introducing the wave function from Eq. (\ref{s5}) and repeating the derivation
of Sec. \ref{sec_s} with the additional damping term, we obtain the generalized
Schr\"odinger equation
\begin{equation}
\label{qb5}
i\hbar\frac{\partial\psi}{\partial t}=-\frac{\hbar^2}{2m}\Delta\psi+m\Phi\psi+V\psi+\hbar \, {\rm Im}(\gamma\ln\psi)\psi.
\end{equation}
Writing $\gamma=\gamma_R+i\gamma_I$, where $\gamma_R$ is the classical friction
coefficient and $\gamma_I$ is the
quantum friction coefficient, and using  the identity
\begin{equation}
\label{qb6}
{\rm Im}(\gamma\ln\psi)=\gamma_I\ln|\psi|-\frac{1}{2}i \gamma_R\ln\left
(\frac{\psi}{\psi^*}\right ),
\end{equation}
we can rewrite Eq. (\ref{qb5}) in the equivalent form
\begin{equation}
\label{qb7}
i\hbar\frac{\partial\psi}{\partial t}=-\frac{\hbar^2}{2m}\Delta\psi
+m\Phi\psi+V\psi+\hbar \gamma_I\ln|\psi|\, \psi-i\frac{\hbar}{2}\gamma_R\ln\left
(\frac{\psi}{\psi^*}\right )\, \psi.
\end{equation}
Introducing the notations
\begin{equation}
\label{qb8}
\gamma_R=\xi,\qquad \gamma_I=\frac{2k_B T}{\hbar},
\end{equation}
the generalized Schr\"odinger equation (\ref{qb7}) takes the form
\begin{equation}
\label{qb9}
i\hbar\frac{\partial\psi}{\partial t}=-\frac{\hbar^2}{2m}\Delta\psi+m\Phi\psi+V\psi+2k_B T\ln|\psi|\, \psi-i\frac{\hbar}{2}\xi\ln\left (\frac{\psi}{\psi^*}\right )\, \psi.
\end{equation}
Using the hydrodynamic representation of the
generalized Schr\"odinger equation (\ref{qb9}), we will see in Sec.
\ref{sec_tmad} that $\xi$ plays
the role of an ordinary friction coefficient while $T$ plays the role of an
effective temperature. The friction coefficient $\xi$ must be positive in
order to guarantee that the generalized Schr\"odinger equation (\ref{qb9})
relaxes towards a stable equilibrium state (see Sec. \ref{sec_ht}). By
contrast, since
the temperature is effective, it can be positive or negative. Finally, we choose
the function $V(t)$ so that the average value of the friction term proportional
to $\xi$ is equal to zero. This
gives
\begin{equation}
\label{qb10q}
V(t)=i\frac{\hbar}{2}\xi\left\langle \ln\left (\frac{\psi}{\psi^*}\right )\right\rangle,
\end{equation}
where $\langle X\rangle=\int \rho X\, d{\bf r}$. We finally obtain the generalized Schr\"odinger equation
\begin{equation}
\label{qb10}
i\hbar\frac{\partial\psi}{\partial t}=-\frac{\hbar^2}{2m}\Delta\psi+m\Phi\psi+2k_B T\ln|\psi|\, \psi-i\frac{\hbar}{2}\xi\left\lbrack \ln\left (\frac{\psi}{\psi^*}\right )-\left\langle \ln\left (\frac{\psi}{\psi^*}\right )\right\rangle\right\rbrack \psi.
\end{equation}
We note that the temperature term  in the generalized Schr\"odinger equation can
be written as $k_B T\ln\rho\, \psi$ and the dissipative term as $\xi(S-\langle
S\rangle)\psi$. It is interesting to note that the complex nature of the
friction coefficient
\begin{equation}
\label{qb11}
\gamma=\xi+i\frac{2k_B T}{\hbar}
\end{equation}
leads to a generalized Schr\"odinger equation exhibiting {\it simultaneously} a
friction term (as expected) 
and an effective temperature term (unexpected). They correspond to the
real and imaginary parts of $\gamma$. This may be viewed as a new form of
fluctuation-dissipation theorem (see Appendix \ref{sec_el}).

{\it Remark:} Equation (\ref{qb10}) can be interpreted as a dissipative
Gross-Pitaevskii (GP)
equation with a logarithmic nonlinearity. This is a particular case of the
generalized GP equations considered in \cite{dgp}. General properties of these
equations are established in \cite{dgp}. In the present paper, we give the 
main properties of the dissipative logarithmic Schr\"odinger equation
(\ref{qb10}) and refer to  \cite{dgp} for more details,
derivations, and generalizations.

\subsection{The Madelung transformation}
\label{sec_tmad}

Using the Madelung transformation, and proceeding as in Sec. \ref{sec_mag}, we
find that 
the generalized Schr\"odinger equation (\ref{qb10}) is equivalent to the
hydrodynamic equations
\begin{equation}
\label{tmad4}
\frac{\partial\rho}{\partial t}+\nabla\cdot (\rho {\bf u})=0,
\end{equation}
\begin{equation}
\label{tmad5}
\frac{\partial S}{\partial t}+\frac{1}{2m}(\nabla S)^2+m\Phi+Q+k_B T\ln\rho+\xi (S-\langle S\rangle)=0,
\end{equation}
\begin{equation}
\label{tmad6}
\frac{\partial {\bf u}}{\partial t}+({\bf u}\cdot \nabla){\bf
u}=-\frac{1}{\rho}\nabla P-\nabla\Phi-\frac{1}{m}\nabla Q-\xi {\bf u},
\end{equation}
\begin{equation}
\label{tmad7}
P=\rho \frac{k_B T}{m},
\end{equation}
where $Q$ is the quantum potential defined in Eq. (\ref{ntmad8}). 
The first equation is the continuity equation, the second equation is the
quantum Hamilton-Jacobi equation (or the quantum Bernoulli equation), and the
third equation is the quantum Euler equation. It involves a damping term and a
pressure term with an isothermal equation of state. For that reason, it is
called the quantum damped isothermal Euler equation \cite{sspcosmo}. Using the
continuity
equation (\ref{tmad4}), it can can be rewritten as
\begin{eqnarray}
\label{tmad9}
\frac{\partial}{\partial t}(\rho {\bf u})+\nabla(\rho {\bf u}\otimes {\bf u})
=-\nabla P-\rho\nabla\Phi-\frac{\rho}{m}\nabla Q-\xi\rho {\bf u}.
\end{eqnarray}
When the quantum potential is neglected, we recover the classical damped
isothermal Euler equation. For dissipationless systems
($\xi=0$), they reduce to the quantum and classical isothermal Euler equations.
On the other hand, in the overdamped limit $\xi\rightarrow +\infty$, we can
formally neglect the inertial term in Eq. (\ref{tmad6}) so that
\begin{equation}
\label{tmad10}
\xi{\bf u}\simeq -\frac{1}{\rho}\nabla P-\nabla\Phi-\frac{1}{m}\nabla Q.
\end{equation}
Substituting this relation into the continuity equation (\ref{tmad4}), we 
obtain the quantum  Smoluchowski  equation \cite{sspcosmo}:
\begin{equation}
\label{tmad11}
\xi\frac{\partial\rho}{\partial t}=\nabla\cdot\left (\nabla
P+\rho\nabla\Phi+\frac{\rho}{m}\nabla Q\right ).
\end{equation}
When the quantum potential is neglected, we obtain the
classical Smoluchowski
equation with a diffusion coefficient $D=k_B T/\xi m$ (see
Appendix \ref{sec_el}).\footnote{The classical Smoluchowski
equation \cite{smoluchowski} was introduced in the context of the Brownian
theory
initiated by Einstein \cite{einstein}. It describes the evolution of the
probability density of an overdamped
Brownian particle moving in a potential $\Phi$ and experiencing a random force.
In that case, it can be interpreted as a Fokker-Planck equation
\cite{fokker,planck,risken}. It is interesting to note that the Smoluchowski
equation
can also be
obtained from the generalized Schr\"odinger equation (\ref{qb10}) in a strong
friction limit. However, we emphasize
that (besides the presence of the quantum potential) its physical interpretation
is completely different. For example, the
diffusive term in the Smoluchowski equation for Brownian particles comes from
stochastic processes (it is due to a random force) while the diffusive term in
the Smoluchowski equation derived from the generalized Schr\"odinger equation
(\ref{qb10}) comes from the logarithmic nonlinearity. On the other hand, the
quantum damped Euler equation (\ref{tmad6}) can be rigorously derived from
the generalized Schr\"odinger equation (\ref{qb10}) while the classical damped
Euler equation is not rigorously justified for
Brownian particles except in the strong friction limit $\xi\rightarrow +\infty$
where it reduces to the classical Smoluchowski equation  (see the discussion in
\cite{just1,just2}).} Finally, if
we neglect the advection term $\nabla(\rho {\bf u}\otimes {\bf u})$ in Eq.
(\ref{tmad9}), but retain the term $\partial (\rho{\bf u})/\partial t$, and
combine the
resulting
equation with the continuity equation (\ref{tmad4}), we obtain the quantum
telegraphic  equation
\begin{equation}
\label{tmad12}
\frac{\partial^2\rho}{\partial t^2}+\xi\frac{\partial\rho}{\partial
t}=\nabla\cdot\left (\nabla P+\rho\nabla\Phi+\frac{\rho}{m}\nabla Q\right )
\end{equation}
which can be seen as a generalization of the quantum Smoluchowski equation
(\ref{tmad11}) taking inertial (or memory) effects into account.
When the quantum potential is neglected, we obtain the
classical telegraphic equation.

\subsection{The energy}

If we define the energy by
\begin{equation}
\label{enxia}
E=-\left (\frac{\partial S}{\partial t}\right )_{\xi=0}
\end{equation}
and use the
Hamilton-Jacobi equation (\ref{tmad5}), we
obtain
\begin{equation}
\label{enxi}
E({\bf r},t)=\frac{1}{2}m{\bf u}^2+m\Phi+Q+k_B T\ln\rho.
\end{equation}
This is the sum of the kinetic energy, the classical potential, the
quantum potential, and the enthalpy \cite{dgp}. We note that the quantum damped 
isothermal Euler equation (\ref{tmad6}) can be written as
\begin{equation}
\label{tmad6q}
\frac{\partial {\bf u}}{\partial t}=-\frac{\nabla E}{m}-\xi {\bf u}.
\end{equation}

\subsection{Time-independent solutions}
\label{sec_tigp}

If we consider a wave function of the form
\begin{equation}
\label{tigp1b}
\psi({\bf r},t)=\phi({\bf r})e^{-i E t/\hbar},
\end{equation}
where $\phi({\bf r})=\sqrt{\rho({\bf r})}$ is real, and substitute Eq. (\ref{tigp1b}) into Eq. (\ref{qb10}), we obtain the generalized time-independent Schr\"odinger equation
\begin{eqnarray}
\label{tigp2}
-\frac{\hbar^2}{2m}\Delta\phi+m\Phi\phi+2k_B T(\ln\phi)\phi=E\phi.
\end{eqnarray}
Equation (\ref{tigp2})  is an eigenvalue equation for the wave function $\phi({\bf r})$ where the eigenvalue $E$ is the energy. Dividing Eq. (\ref{tigp2}) by $\phi$ and using $\rho=\phi^2$, we get
\begin{equation}
\label{tigp3}
-\frac{\hbar^2}{2m}\frac{\Delta \sqrt{\rho}}{\sqrt{\rho}}+m\Phi+k_B
T\ln\rho=E\qquad {\rm i.e.}\qquad Q+m\Phi+k_B T\ln\rho=E.
\end{equation}
Using the last equality of Eq. (\ref{ntmad8}), we
can rewrite  Eq. (\ref{tigp3}) into the form
\begin{equation}
\label{tigp4b}
-\frac{\hbar^2}{8m}(\nabla\ln\rho)^2-\frac{\hbar^2}{4m}\Delta\ln\rho+m\Phi+k_B
T\ln\rho=E.
\end{equation}
Equation (\ref{tigp3}) can also be derived from the damped quantum
Hamilton-Jacobi equation (\ref{tmad5}) by setting  $S=-Et$ or from Eq.
(\ref{enxi}) by setting $E({\bf r},t)=E$ and ${\bf
u}={\bf 0}$. We note that dissipative effects do
not alter the time-independent solutions of the generalized Schr\"odinger
equation because $S=-Et$ is uniform so that $\xi(S-\langle S\rangle)=0$.

{\it Remark:} If we take $V=0$ in the generalized Schr\"odinger equation (\ref{qb9}), we get a Hamilton-Jacobi equation of the form
\begin{equation}
\label{tigp5}
\frac{\partial S}{\partial t}+\frac{1}{2m}(\nabla S)^2+m\Phi+Q+k_B T\ln\rho+\xi S=0.
\end{equation}
We note that Eqs. (\ref{tmad4}) and (\ref{tmad6}) are not modified. If we look
for a static solution of the generalized Schr\"odinger equation such that
$S$ depends only on time (this implies ${\bf u}={\bf 0}$) while $\rho$ depends
only on position, the Hamilton-Jacobi equation (\ref{tigp5}) can be
written as
\begin{equation}
\label{tigp6}
-\frac{dS}{dt}-\xi S=m\Phi+Q+k_B T\ln\rho=E_0.
\end{equation}
Since the first term depends only on time and the second term depends only on
position, they must be individually equal to a constant $E_0$. The second part
of Eq. (\ref{tigp6}) returns Eq. (\ref{tigp3}) where $E$ is replaced by $E_0$.
The first part of Eq. (\ref{tigp6}) can be solved for $S(t)$ leading to
\begin{equation}
\label{tigp7}
S(t)=\frac{E_0}{\xi}(e^{-\xi t}-1),
\end{equation}
where the constant of integrating has been determined such that Eq. (\ref{tigp7}) reduces to $S\sim -E_0 t$ when $\xi\rightarrow 0$ or when $t\rightarrow 0$. We note that $S(t)\rightarrow -E_0/\xi$ when $t\rightarrow +\infty$. Defining the energy by $E(t)=-dS/dt$, we obtain
\begin{equation}
\label{tigp8}
E(t)=E_0 e^{-\xi t}.
\end{equation}
This equation shows that $E_0$ represents the initial energy. Furthermore, it
shows that the energy $E(t)$ is 
dissipated in the static state and that it decays to zero for $t\rightarrow
+\infty$.

\subsection{Hydrostatic equilibrium}
\label{sec_he}

The time-independent Schr\"odinger equation (\ref{tigp3}) 
can also be obtained from the damped quantum isothermal Euler equation
(\ref{tmad6}) since it is equivalent to the generalized Schr\"odinger equation
(\ref{qb10}). The equilibrium state  of the damped quantum isothermal Euler
equation (\ref{tmad6}), obtained by setting $\partial_t\rho=0$ and ${\bf u}={\bf
0}$, satisfies 
\begin{equation}
\label{he1}
\nabla P+\rho\nabla\Phi+\frac{\rho}{m}\nabla Q={\bf 0} \qquad {\rm i.e.}\qquad
\frac{k_B T}{m}\nabla \rho+\rho\nabla\Phi-\frac{\hbar^2\rho}{2m^2}\nabla \left
(\frac{\Delta\sqrt{\rho}}{\sqrt{\rho}}\right )={\bf 0}.
\end{equation}
This equation generalizes the usual condition 
of hydrostatic equilibrium by incorporating the contribution of the quantum
potential. From the hydrodynamic representation, we clearly understand why
frictional effects do not influence the equilibrium state since they vanish when
${\bf u}={\bf 0}$. Equation (\ref{he1}) describes the (possible)
balance between the repulsion (for $T>0$) or the attraction (for $T<0$)
due to the pressure $P$,
the repulsion or the attraction due to the potential $\Phi$, and the repulsion
due to the quantum potential $Q$ arising from the Heisenberg uncertainty
principle. This equation is equivalent to Eq. (\ref{tigp3}). Indeed,
dividing Eq. (\ref{he1})  by $\rho$ and
integrating the resulting equation with the aid of Eq. (\ref{tmad7}), we obtain
Eq. (\ref{tigp3}) where the eigenenergy  $E$ appears as a
constant of integration.

\section{Effective thermodynamics}
\label{sec_effth}

In this section, we show that the generalized Schr\"odinger equation (\ref{qb10}) is consistent with a notion of effective thermodynamics based on the Boltzmann entropy.

\subsection{The free energy}
\label{sec_ef}

The free energy associated with the generalized Schr\"odinger equation (\ref{qb10}), or equivalently with the damped quantum isothermal Euler equations (\ref{tmad4})-(\ref{tmad7}),  can be written as
\begin{eqnarray}
\label{ef1}
F=\Theta_c+\Theta_Q+W+U.
\end{eqnarray}
The first two terms in Eq. (\ref{ef1}) correspond to the total kinetic energy
$\Theta$ [see Eq. (\ref{ef2})]. Using the Madelung
transformation, it can be decomposed
into  the  classical kinetic energy 
\begin{eqnarray}
\label{ef3}
\Theta_c=\int\rho m \frac{{\bf u}^2}{2}\, d{\bf r}
\end{eqnarray}
and the quantum kinetic energy\footnote{This functional was introduced by von
Weizs\"acker
\cite{wei}. It is related to the Fisher \cite{fisher} entropy
$S_F=\int (\nabla\rho)^2/\rho\, d{\bf r}$ by the
relation
$\Theta_Q=(\hbar^2/8m)S_F$.
Actually, the functional (\ref{ef5}) was already introduced by Madelung
\cite{madelung,madelungearly} under the equivalent form
$\Theta_Q=-({\hbar^2}/{2m})\int \sqrt{\rho}\Delta\sqrt{\rho}\, d{\bf
r}$ [see Eq. (\ref{ef5b})].} 
\begin{eqnarray}
\label{ef5}
\Theta_Q=\frac{\hbar^2}{8m}\int \frac{(\nabla\rho)^2}{\rho}\, d{\bf r}.
\end{eqnarray}
Using Eq. (\ref{ntmad8}), integrating by parts, and assuming that the boundary
term can be neglected, we get
\begin{eqnarray}
\label{ef5b}
\int \rho Q\, d{\bf r}=-\frac{\hbar^2}{2m}\int \sqrt{\rho}\Delta\sqrt{\rho}\, d{\bf r}
=\frac{\hbar^2}{2m}\int (\nabla\sqrt{\rho})^2\, d{\bf r}=\frac{\hbar^2}{8m}\int
\frac{(\nabla\rho)^2}{\rho}\, d{\bf r}.
\end{eqnarray}
Therefore, the quantum kinetic
energy can be rewritten as
\begin{equation}
\label{ef4}
\Theta_Q=\int \rho Q\, d{\bf r}.
\end{equation}
It can be interpreted  as a potential energy associated with the quantum
potential $Q$.\footnote{This is not obvious since $Q$ is a function of the
density (it is {\it not} an external potential).} 
The third term in Eq. (\ref{ef1}) is the potential energy
\begin{eqnarray}
\label{ef6}
W=\int\rho m \Phi\, d{\bf r}.
\end{eqnarray}
The fourth term in Eq. (\ref{ef1}) is the internal energy
\begin{eqnarray}
\label{ef1e}
U=m\int\rho\int^{\rho}\frac{P(\rho')}{{\rho'}^2}\, d\rho'\, d{\bf r}.
\end{eqnarray}
The density of internal energy $\rho m u$ satisfies the first law
of thermodynamic $du=-Pd(1/\rho)$. The internal energy takes into account the
contribution of the nonlinear
potential in the generalized Schr\"odinger equation. For the logarithmic
potential, leading to the effective isothermal equation of state (\ref{tmad7}),
we get
\begin{eqnarray}
\label{ef1eb}
U=k_B T\int \rho(\ln\rho-1)\, d{\bf r}.
\end{eqnarray}
We note that
\begin{eqnarray}
\label{ef1ec}
U=-T S
\end{eqnarray}
where
\begin{eqnarray}
\label{ef7}
S=-k_B\int \rho(\ln\rho-1)\, d{\bf r}
\end{eqnarray}
is the Boltzmann entropy. Regrouping all these results, the free energy can be explicitly written as
\begin{eqnarray}
\label{ef8}
F=\int\rho m \frac{{\bf u}^2}{2}\, d{\bf r}+\int \rho Q\, d{\bf r}+\int\rho m \Phi\, d{\bf r}+k_B T\int \rho(\ln\rho-1)\, d{\bf r}.
\end{eqnarray}
Introducing the energy
\begin{eqnarray}
\label{ef1b}
E_*=\Theta_c+\Theta_Q+W=\int\rho m \frac{{\bf u}^2}{2}\, d{\bf r}+\int \rho Q\, d{\bf r}+\int\rho m \Phi\, d{\bf r}
\end{eqnarray}
that includes the classical kinetic energy $\Theta_c$, 
the quantum kinetic energy $\Theta_Q$ and the potential energy $W$, the free
energy can be written in the standard form
\begin{eqnarray}
\label{ef1bol}
F=E_*-T S.
\end{eqnarray}
It coincides with the Boltzmann free energy. The free energy associated with the
quantum Smoluchowski equation
(\ref{tmad11}) is
\begin{eqnarray}
\label{ef12}
F=\Theta_Q+W+U=\Theta_Q+W-TS
\end{eqnarray}
since the classical kinetic energy $\Theta_c$, which is of order $O(\xi^{-2})$,
can be neglected in the overdamped limit $\xi\rightarrow +\infty$.  

\subsection{Difference between the average energy and the free energy}
\label{sec_dif}

The average value of the energy $E({\bf r},t)$ defined by Eq.
(\ref{enxi}) is
\begin{equation}
\label{dif1}
\langle E\rangle=\int\rho E\, d{\bf r}=\int \rho m\frac{{\bf
u}^2}{2}\,d{\bf
r} +\int \rho m\Phi\, d{\bf r}+\int \rho Q\, d{\bf r}+k_B T\int \rho \ln\rho\,
d{\bf r}=\Theta_c+W+\Theta_Q+U+k_B T.
\end{equation}
It coincides with the average value of the energy operator (see
Appendix \ref{sec_ae}). Comparing Eqs. (\ref{ef8}) and (\ref{dif1}), we find
that
\begin{eqnarray}
\label{dif2}
F=\langle E\rangle-k_B T.
\end{eqnarray}
Therefore
\begin{equation}
\label{ae12}
F\neq \langle E\rangle.
\end{equation}
It is only in the case of the linear Schr\"odinger equation ($T=0$)
that
$F=\langle E\rangle$. For the logarithmic Schr\"odinger equation, $F$ and
$\langle E\rangle$ differ by a constant $k_B T$. Since they just differ by a
constant, they will satisfy the same properties. For more general nonlinear
Schr\"odinger equations,  $F$ and
$\langle E\rangle$  differ by a more complicated functional and can behave very
differently \cite{dgp}.

\subsection{The $H$-theorem}
\label{sec_ht}

It is shown in Appendix D of \cite{dgp} that the time
derivative of the free energy (\ref{ef1}) satisfies the identity
\begin{eqnarray}
\label{ef11}
\dot F=-\xi\int \rho m {\bf u}^2\, d{\bf r}\qquad {\rm i.e.}\qquad \dot
F=-2\xi\Theta_c.
\end{eqnarray}
We have to consider two situations:

(i) For dissipationless systems ($\xi=0$), 
Eq. (\ref{ef11}) shows that the generalized Schr\"odinger  equation
(\ref{qb10}), or the quantum isothermal Euler equations
(\ref{tmad4})-(\ref{tmad7}), conserve the free energy ($\dot F=0$).\footnote{For
dissipationless systems, $F$ is called the total energy $E_{\rm tot}$ of the
system, not the free energy. However, for convenience, we shall always refer to
$F$ as the free energy.}  In that case, it can be shown from general
arguments \cite{holm} that a minimum of free
energy at fixed mass determines a steady state of the  generalized
Schr\"odinger equation, or of the quantum isothermal Euler equations, that is
formally nonlinearly dynamically stable.

(ii) For dissipative systems ($\xi>0$), Eq. (\ref{ef11}) shows that 
the generalized Schr\"odinger equation (\ref{qb10}), or the 
quantum damped isothermal Euler equations (\ref{tmad4})-(\ref{tmad7}), decrease
the
free energy ($\dot F\le 0$). When $\dot F=0$,  Eq. (\ref{ef11}) implies that
${\bf u}={\bf 0}$. From the Euler equation (\ref{tmad6}), we obtain the
condition of hydrostatic equilibrium (\ref{he1}). Therefore, Eq. (\ref{ef11})
forms an $H$-theorem for the generalized Schr\"odinger  equation or for the
damped quantum isothermal Euler equations: $\dot F\le 0$ and $\dot F=0$ if, and
only if, the system is at equilibrium. In that case, $F$ is called a
Lyapunov functional. From Lyapunov's direct method, we can
show that the system will relax, for $t\rightarrow +\infty$, towards an
equilibrium state that is a (local) minimum of free energy at fixed mass. Maxima
or saddle points of free energy are unstable. If several local minima of free
energy exist, the selection will depend on the initial condition and on a notion
of basin of attraction.

The free energy associated with the quantum Smoluchowski equation
(\ref{tmad11}) is given by Eq. (\ref{ef12}). Its time
derivative satisfies the
identity\footnote{For the classical diffusion
equation (corresponding to Eq. (\ref{tmad11}) with $\Phi=Q=E_*=0$, $P=\rho k_B
T/m$, $D=k_B T/\xi m$ and
$F_B=-TS_B$ where $S_B$ is the Boltzmann entropy) we find from 
Eq. (\ref{ef13}) that $\dot S_B=D k_B S_F$ where $S_F$ is the Fisher entropy.}
\begin{eqnarray}
\label{ef13}
\dot F=-\frac{1}{\xi}\int \frac{m}{\rho}\left (\nabla
P+\rho\nabla\Phi+\frac{\rho}{m}\nabla Q\right )^2\, d{\bf r}.
\end{eqnarray}
This equation can be obtained from the quantum 
Smoluchowski equation (see Appendix D of \cite{dgp}). It
can also be  directly obtained from Eq. (\ref{ef11}) by using Eq. (\ref{tmad10})
which is valid in the strong friction limit $\xi\rightarrow +\infty$. When $\dot
F=0$, Eq. (\ref{ef13})
implies that the term in parenthesis vanishes, leading to the condition of
hydrostatic equilibrium (\ref{he1}).  Therefore, Eq. (\ref{ef13}) forms an
$H$-theorem for the quantum Smoluchowski equation.

{\it Remark:} Since the dissipative ($\xi\neq 0$) Schr\"odinger equation
(\ref{qb10}), the damped
quantum isothermal Euler equations (\ref{tmad4})-(\ref{tmad7}), and the quantum
Smoluchowski equation
(\ref{tmad11}) are relaxation equations, they  can be
used as {\it numerical algorithms} to compute stable equilibrium states of the
conservative ($\xi=0$)  Schr\"odinger equation, or
quantum isothermal Euler equations. This can be very useful on a
practical point of view because it
is generally not easy to solve the time-independent equations (\ref{tigp3}) and
(\ref{he1}) directly and be
sure that
the solution is stable.\footnote{See Appendix E of \cite{nfp}, and \cite{vp},
for
numerical algorithms in the form of relaxation equations that can be used to
construct stable steady states of the
Vlasov-Poisson and 2D Euler-Poisson equations.}

\subsection{The equilibrium state}
\label{sec_es}

According to the previous discussion, a stable equilibrium state  of the
 generalized Schr\"odinger   equation (\ref{qb10}), or equivalently of the
damped quantum isothermal Euler equations 
(\ref{tmad4})-(\ref{tmad7}), is the
solution of the minimization problem
\begin{eqnarray}
\label{ef14}
F(M,T)=\min_{\rho,{\bf u}} \left\lbrace F[\rho,{\bf u}]\quad \biggl|\quad
\int\rho\, d{\bf r}=1\right\rbrace.
\end{eqnarray}
A
critical point of free energy satisfying the normalization condition  is
determined by the variational principle
\begin{eqnarray}
\label{ef14zb}
\delta F-\mu\delta \left (\int\rho\, d{\bf r}\right )=0,
\end{eqnarray}
where $\mu$ is a Lagrange multiplier taking into account the normalization
constraint. Using the results of Appendix C of \cite{dgp},
this variational problem gives ${\bf u}={\bf 0}$ (the equilibrium state is
static) and the condition
\begin{eqnarray}
\label{ef14b}
Q+m\Phi+k_B T\ln\rho=\mu.
\end{eqnarray}
Taking the gradient of Eq. (\ref{ef14b}) and using Eq. (\ref{tmad7}), 
we recover the condition of hydrostatic equilibrium (\ref{he1}). Equation
(\ref{ef14b}) is equivalent to the time-independent Schr\"odinger equation
(\ref{tigp3}) provided that we make the identification
\begin{eqnarray}
\label{ef14c}
\mu=E.
\end{eqnarray}
This shows that the Lagrange multiplier $\mu$
(chemical potential) in the variational problem associated with Eq. (\ref{ef14})
can be identified with the eigenenergy $E$.  Inversely, the eigenenergy $E$ may
be
interpreted as a chemical potential $\mu$.  Equation (\ref{ef14b}) can be
written
as
\begin{eqnarray}
\label{ef14bb}
\rho=e^{-\beta(m\Phi+Q-\mu)}.
\end{eqnarray}
This equation shows that the equilibrium state of the generalized Schr\"odinger 
 equation (\ref{qb10}), or equivalently of the damped quantum isothermal Euler
equations (\ref{tmad4})-(\ref{tmad7}), is given by a generalized Boltzmann
distribution including the contribution of the quantum potential (this is
actually a differential equation). If we neglect
the contribution of the quantum potential ($Q=0$), we recover the usual
Boltzmann distribution.
Finally, considering the second order variations of free energy, we find that the equilibrium state is stable if, and only if,
\begin{eqnarray}
\label{ef16}
\delta^2 F= k_B T\int \frac{(\delta\rho)^2}{2\rho}\, d{\bf r}+\frac{\hbar^2}{8m}\int\frac{1}{\rho}\left\lbrack \left (\frac{\Delta\rho}{\rho}-\frac{(\nabla\rho)^2}{\rho^2}\right ){(\delta\rho)^2}+{(\nabla\delta\rho)^2}\right\rbrack\, d{\bf r}>0
\end{eqnarray}
for all perturbations that conserve the normalization condition: $\int \delta\rho\, d{\bf r}=0$. This inequality can also be written as
\begin{eqnarray}
\label{ef16b}
\delta^2 F= k_B T\int \frac{(\delta\rho)^2}{2\rho}\, d{\bf r}
+\frac{\hbar^2}{8m}\int  \left \lbrack \nabla \left (\frac{\delta\rho}{\sqrt{\rho}}\right )\right\rbrack^2\, d{\bf r}+\frac{\hbar^2}{8m}\int \frac{\Delta\sqrt{\rho}}{\rho^{3/2}}(\delta\rho)^2\, d{\bf r}>0.
\end{eqnarray}

\subsection{Equilibrium relations}
\label{sec_er}

At equilibrium ($\Theta_c=0$), the free energy (\ref{ef1bol}) reduces to
\begin{eqnarray}
\label{ef8b}
F=\Theta_Q+W-T S.
\end{eqnarray}
On the other hand, multiplying Eq. (\ref{tigp3}) by $\rho$ and integrating over
${\bf r}$, we obtain
\begin{equation}
\label{tigp4bb}
E=\Theta_Q+W+k_B T-TS.
\end{equation}
Comparing Eqs. (\ref{ef8b}) and
(\ref{tigp4bb}), we get
\begin{equation}
\label{tigp4c}
F=E-k_B T.
\end{equation}
Equations (\ref{ef8b})-(\ref{tigp4c}) are the equilibrium forms of Eqs.
(\ref{ef1bol}), (\ref{dif1}) and (\ref{dif2}). We note that, at equilibrium, 
$\langle
E\rangle=E$.

\section{The virial theorem for a harmonic potential}
\label{sec_virial}

In this section, we provide the form of the virial theorem associated with the
generalized Schr\"odinger equation (\ref{qb10}) or, equivalently, with the
quantum damped isothermal Euler equations (\ref{tmad4})-(\ref{tmad7}). To make
the results more explicit, we restrict ourselves to the case of a harmonic
potential (see \cite{dgp} for generalizations). We write the  harmonic potential
as
\begin{equation}
\label{vhp1}
\Phi=\frac{1}{2}\omega_0^2 r^2.
\end{equation}
When $\omega_0^2=-\Omega^2<0$, this potential 
mimics the effect of a solid-body rotation of the system (this analogy is exact
in $d=2$). When $\omega_0^2>0$, this potential mimics the effect of a confining
trap. For the harmonic potential (\ref{vhp1}), the force (by unit of mass)
exerted on the particle is ${\bf F}=-\nabla\Phi=-\omega_0^2{\bf r}$. On the
other hand, the potential energy (\ref{ef6}) can be written as
\begin{eqnarray}
\label{vhp2}
W=\frac{1}{2}\omega_0^2 I,
\end{eqnarray}
where
\begin{eqnarray}
\label{vhp3}
I=\int m \rho r^2\, d{\bf r}
\end{eqnarray}
is the moment of inertia. We note that $I=m\langle r^2\rangle$, where $\langle r^2\rangle$ measures the quantum spread of the position of the particle.

For the harmonic potential (\ref{vhp1}), the time-dependent virial theorem
writes (see Appendix G of \cite{dgp}):
\begin{equation}
\label{vhp4}
\frac{1}{2}\ddot I+\frac{1}{2}\xi\dot I+\omega_0^2 I=2(\Theta_c+\Theta_Q)+dk_B T.
\end{equation}
In the strong friction limit $\xi\rightarrow +\infty$, corresponding to the
quantum Smoluchowski equation (\ref{tmad11}),  it reduces to
\begin{equation}
\label{vhp5}
\frac{1}{2}\xi\dot I+\omega_0^2 I=2\Theta_Q+d k_B T.
\end{equation}
At equilibrium ($\ddot I=\dot I=\Theta_c=0$), we obtain
\begin{equation}
\label{vhp6}
2\Theta_Q+d k_B T-\omega_0^2I=0.
\end{equation}
On the other hand, for the harmonic potential (\ref{vhp1}), the relations
(\ref{ef8b}) and
(\ref{tigp4bb})  become
\begin{eqnarray}
\label{quad3}
F=\Theta_Q+\frac{1}{2}\omega_0^2 I-T S
\end{eqnarray}
and
\begin{equation}
\label{tigp4bc}
E=\Theta_Q+\frac{1}{2}\omega_0^2 I+k_B T-TS.
\end{equation}

\section{Comparison with other works}
\label{sec_conn}

The generalized Schr\"odinger equation (\ref{qb10}) includes two new terms 
with respect to the Schr\"odinger equation (\ref{s9}): a friction term
$\xi$ and a temperature term $T$. These terms correspond to the real and
imaginary parts of the complex friction coefficient (\ref{qb11}). These two
terms have a common origin and they satisfy a
sort of fluctuation-dissipation theorem. The Schr\"odinger equation
(\ref{s9}) is recovered when $T=\xi=0$. In this section, we mention the
connection between the generalized Schr\"odinger equation (\ref{qb10}) and
nonlinear Schr\"odinger equations that have been introduced in the past with
different arguments.

\subsection{The damped Schr\"odinger equation}
\label{sec_kostin}

A Schr\"odinger equation including a damping term similar to the one present
in Eq. (\ref{qb10}), namely
\begin{equation}
\label{koss}
i\hbar\frac{\partial\psi}{\partial t}=-\frac{\hbar^2}{2m}\Delta\psi+m\Phi\psi-i\frac{\hbar}{2}\xi\left\lbrack \ln\left (\frac{\psi}{\psi^*}\right )-\left\langle \ln\left (\frac{\psi}{\psi^*}\right )\right\rangle\right\rbrack \psi,
\end{equation}
has been introduced by Kostin \cite{kostin}. He derived 
it from the Heisenberg-Langevin equation describing a quantum Brownian particle
interacting with a thermal bath environment. In Ref. \cite{sspcosmo}, we
independently obtained this equation by looking for the nonlinear Schr\"odinger
equation
that leads, through the Madelung transformation, to a quantum Euler equation
with a linear friction force proportional and opposite to the velocity. Our
approach gives another interpretation of the damped Schr\"odinger equation
(\ref{koss}) from a hydrodynamic representation of the wave equation
(see Sec. \ref{sec_tmad}) associated with an effective thermodynamic formalism
(see Sec. \ref{sec_effth}).

\subsection{The logarithmic Schr\"odinger equation}
\label{sec_bbm}

A nonlinear Schr\"odinger equation with a logarithmic potential $-b\ln|\psi|^2$
similar to the one present in Eq. (\ref{qb10}), namely
\begin{equation}
\label{bial}
i\hbar\frac{\partial\psi}{\partial t}=-\frac{\hbar^2}{2m}\Delta\psi+m\Phi\psi-2b\ln|\psi|\, \psi,
\end{equation}
has been introduced by Bialynicki-Birula and Mycielski \cite{bbm} 
as a possible generalization of the Schr\"odinger equation in quantum mechanics.
Their original motivation was the following. If we consider a free particle
described by the Schr\"odinger equation (\ref{s9}) with $\Phi=0$, the
kinetic term in the Schr\"odinger equation, accounting for the Heisenberg
uncertainty principle, leads to the spreading of the wavepacket. As a result, 
the position of the particle becomes more and more uncertain and the
Schr\"odinger equation does not have localized stationary solutions. When
applied to macroscopic objects, this leads to a problem because, according to
classical mechanics, a macroscopic object has a well-defined position. This
problem can be solved if we
assume that the Schr\"odinger equation is replaced by a nonlinear Schr\"odinger
equation with a logarithmic potential $-b\ln|\psi|^2$.\footnote{Other forms of
nonlinearity in the Schr\"odinger equation can prevent the spreading of the
wave packet. However, nonlinear Schr\"odinger equations have in general the
undesirable feature to generate  correlation between non-interacting particles
unless the nonlinearity is logarithmic. Therefore, the motivation of
Bialynicki-Birula and Mycielski \cite{bbm} to introduce a logarithmic
nonlinearity in the Schr\"odinger equation is that this term still satisfies the
additivity property for noninteracting subsystems while solving the spreading of
the wave packet problem. Therefore, the logarithmic term was selected by
assuming the factorization of the wave function for the composed
system. This is what they called the property of
separability of noninteracting subsystems. They suggested
that to maintain the lack of correlation between noninteracting particle
subsystems (separation property).
As we have seen in our effective thermodynamic
formalism, this logarithmic potential is associated with the Boltzmann entropy.
We note that a power-law nonlinearity $-|\psi|^{2(\gamma-1)}$, which
corresponds to a polytropic equation of state (see Ref. \cite{dgp}),
also solves the spreading of the wave packet problem but does not satisfy the
additivity property for noninteracting subsystems. A polytropic equation of
state is associated with the
Tsallis entropy (see Ref. \cite{dgp}) that has been introduced precisely in
order to deal with non-extensive and non-additive systems \cite{tsallisbook}.
Therefore, the Tsallis entropy may find applications in the context of nonlinear
Schr\"odinger equations with a power-law nonlinearity.}
The coefficient $b$ can be positive or negative.
When it is positive, the nonlinearity $-b\ln|\psi|^2$ counteracts the spreading
of the wave packet, thereby allowing solutions which behave macroscopically as
classical particles. The balance between the dispersion of the wave packet and
the logarithmic nonlinearity leads to a stationary solution of the wave equation
with a Gaussian profile and a finite width called a gausson. The radius of a
gausson is
$R=\hbar/(2mb)^{1/2}$ (see Appendix \ref{sec_exact}). 
In the interpretation of Bialynicki-Birula and Mycielski \cite{bbm},
the Schr\"odinger equation is an {\it approximation} of this nonlinear wave
equation. Therefore, the logarithmic nonlinearity has a fundamental origin and
the
coefficient $b<0$ is interpreted as a fundamental constant of physics. Of
course, it must be sufficiently small in order to satisfy the constraints set by
laboratory experiments. Bialynicki-Birula and Mycielski \cite{bbm} obtained
$b<4\times 10^{-10}\, {\rm eV}$ which implies a bound to the electron soliton
spatial width of $10\, \mu{\rm m}$.  Shull {\it et al.} \cite{shull} obtained
$b<3.4\times 10^{-13}\,  {\rm eV}$. Finally, an upper limit $b<3.3\times
10^{-15}\,
{\rm eV}$ was obtained by G\"ahler {\it et al.} \cite{gkz} from precise
measurements of Fresnel diffraction with slow neutrons. This implies a bound to
the electron soliton spatial width of $3\, {\rm mm}$. For $m\rightarrow
+\infty$, the density probability becomes a delta-function and the particle is
localized. It means that there exists the classical motion of a particle with a
sufficient big mass.

In Ref. \cite{sspcosmo}, we independently obtained the logarithmic
Schr\"odinger equation
\begin{equation}
\label{bialb}
i\hbar\frac{\partial\psi}{\partial
t}=-\frac{\hbar^2}{2m}\Delta\psi+m\Phi\psi+2k_B T\ln|\psi|\, \psi,
\end{equation}
by looking for the nonlinear Schr\"odinger
equation that leads, through the Madelung transformation, to an Euler equation
including a pressure term with an isothermal equation of state (we also obtained
more general nonlinear Schr\"odinger equations associated with
arbitrary barotropic equations of state). This leads to a logarithmic potential
of the form $k_B T \ln|\psi|^2$ where $T$ is an effective temperature that can
be positive or negative. Comparing Eqs. (\ref{bial}) and (\ref{bialb}), we find 
that 
\begin{equation}
b=-k_B T
\end{equation}
so that a positive coefficient $b$ corresponds to a negative
effective temperature. Our approach gives another
interpretation of the logarithmic Schr\"odinger equation (\ref{bial}) from a
hydrodynamic representation of the wave equation (see Sec. \ref{sec_tmad})
associated with an effective thermodynamic formalism
(see Sec. \ref{sec_effth}). In this hydrodynamic representation, the
evolution of a free quantum particle described by the Schr\"odinger
equation (\ref{s9}) with $\Phi=0$ is governed by the quantum Euler equation
(\ref{ntmad6}). The evolution of the fluid particle is due to the quantum
potential which has a repulsive effect and tends to delocalize the particle.
This accounts for the spreading of the wavepacket in the wave function
representation. As a result, the quantum Euler equation (\ref{ntmad6}) with
$\Phi=0$ has no steady state. However, in the presence of a logarithmic
nonlinearity, we
get the quantum Euler equation (\ref{tmad6}) which involves a pressure term with
an isothermal equation of state  (\ref{tmad7}). When the effective temperature
is negative ($T<0$), the pressure term produces an attractive force
corresponding to a negative pressure that  can balance the repulsion due to the
quantum potential and leads to a stationary solution given by the Boltzmann
distribution (\ref{ef14bb}) resulting from the condition of hydrostatic
equilibrium (\ref{he1}). This is the hydrodynamic representation of
the gausson. We note that the radius of the gausson can be written as
$R=\hbar/\sqrt{2m|k_B T|}$ which can be interpreted as a generalized de Broglie
relation.

\subsection{The unified description}

In this paper, we have derived a generalized Schr\"odinger equation
(\ref{qb10}) that unifies the damped Schr\"odinger equation (\ref{koss}) and
the logarithmic Schr\"odinger equation (\ref{bial}) or (\ref{bialb}). We have
shown that 
the friction and temperature terms in the generalized Schr\"odinger equation
(\ref{qb10}) have a common origin and that they can be obtained from a unified
description based on Nottale's theory of scale relativity. They satisfy a sort
of fluctuation-dissipation theorem. In our
formalism
\begin{equation}
\label{uni}
\xi=\gamma_R\qquad {\rm and}\qquad b=-k_B T=-\frac{1}{2}\hbar\gamma_I.
\end{equation}
If we assume that $\gamma\sim 1$, we
find that $\xi\sim 1$ and $b=-k_B T\sim
\hbar$. This establishes the fact that $b=-k_B T\sim \hbar$ has its origin in
quantum mechanics (it vanishes in the classical limit $\hbar\rightarrow 0$)
while $\xi\sim 1$ survives in the classical limit. We can also reach this
conclusion by noticing that the term ${\rm Re}(\gamma {\bf U})$ in Eq.
(\ref{qb1}) can be written as ${\rm Re}(\gamma {\bf U})=\gamma_R {\bf
u}+\gamma_I {\bf u}_Q$ (see Appendix \ref{sec_gne}). Therefore, the real
friction coefficient $\gamma_R=\xi$
(friction term) acts on the classical velocity ${\bf u}$  and the
imaginary friction coefficient $\gamma_I=-2b/\hbar=2k_B T/\hbar$ (temperature
term) acts on the quantum velocity ${\bf u}_Q=(\hbar/2m)\nabla\ln\rho$
[see Eqs. (\ref{ri3}) and (\ref{ri2})]. On the other hand, the temperature term
in Eq. (\ref{qb10}) can be written as $\frac{1}{2}\hbar\gamma_I\ln\rho\, \psi$
(disappearing when $\hbar\rightarrow 0$) and the friction term can be written as
$\gamma_R(S-\langle S\rangle)\psi$ (surviving when $\hbar\rightarrow 0$). The
fact that $b=-k_B T$ is proportional to the Planck constant $\hbar$ may explain
its
small value and why it is not detectable in earth experiments. We also note,
parenthetically, that the scaling  $T\sim\hbar$ is similar to the one arising in
the Hawking temperature $k_B T= \hbar c^3/8\pi GM$ of a black
hole. This may suggest a connection to quantum
gravity.

{\it Remark:} The gausson is a black hole if its radius
$R=\hbar/\sqrt{2mb}$ is of the order of the Schwarzschild radius $R=2Gm/c^2$.
This corresponds to a mass $m\sim (\hbar^2c^4/8G^2b)^{1/3}$.
Since $b=-k_B T$, this relation can be rewritten as $|k_B
T|=\hbar^2c^4/8G^2m^3=G\hbar^2/c^2R^3$. This relation may be compared with the
Hawking formula $k_B T= \hbar c^3/8\pi Gm=\hbar c/4\pi R$ for the temperature of
a Schwarzschild black hole. If we identify the two, we obtain $m\sim m_P$ where
$m_P=(\hbar c/G)^{1/2}$ is the Planck mass, $R\sim l_P$ where $l_P=(\hbar
G/c^3)^{1/2}$ is the Planck length, and $|k_B T|\sim m_P c^2\sim k_B T_P$ where
$T_P=(\hbar c^5/Gk_B^2)^{1/2}$ is the Planck temperature. This corresponds to a 
Planck black hole. We also note that $R\sim \lambda_C$ where $\lambda_C=\hbar/m
c$ is the Compton wavelength. These results are of course only qualitative and
suggestive. It is not clear whether they have a deeper meaning than is apparent
at first sight from, essentially, dimensional analysis.

\section{Another generalized Schr\"odinger equation}
\label{sec_ags}

In this section, we consider another form of dissipative 
Schr\"odinger equation. We write it as
\begin{equation}
\label{ag0}
i\hbar\frac{\partial\psi}{\partial t}=(1-i\gamma)\left
(-\frac{\hbar^2}{2m}\Delta\psi+m\Phi\psi-\langle E\rangle\psi\right ),
\end{equation}
where $\gamma$ is a dimensionless dissipation coefficient and
$\langle
E\rangle (t)$ is the time-dependent average energy defined by Eq.
(\ref{ae4}).
This equation is inspired by the so-called stochastic GP
equation introduced in the context of BECs \cite{gardiner}.\footnote{The
dissipative term $-i\gamma$ was introduced by Pitaevskii \cite{pita} (see also
\cite{choi,tsubota}).} However,
for the
simplicity of the presentation, we have ignored the stochastic term and the
nonlinearity (they can be introduced straightforwardly).
More importantly, we have replaced the constant chemical potential $\mu$ by the
time-dependent average energy $\langle E\rangle (t)$.  Physically, this means
that the generalized Schr\"odinger
equation (\ref{ag0}) is written in a frame that rotates with an angular
velocity $\omega(t)=\langle E\rangle(t)/\hbar$ that adapts itself at each time.
As will be shown below, this
allows us to satisfy the conservation of the normalization condition
globally.\footnote{Note that the number of bosons is not conserved
in the stochastic GP equation \cite{gardiner}.}

Making the Madelung transformation (\ref{ntmad2}) and proceeding as in Sec.
\ref{sec_mag}, we obtain the hydrodynamic equations
\begin{equation}
\label{ag1}
\frac{\partial\rho}{\partial t}+\nabla\cdot (\rho {\bf
u})=-\frac{2\gamma}{\hbar}\rho\left (\frac{1}{2}m{\bf u}^2+m\Phi+Q-\langle
E\rangle\right ),
\end{equation}
\begin{equation}
\label{ag2}
\frac{\partial S}{\partial t}+\frac{(\nabla S)^2}{2m}+m\Phi+Q-\langle
E\rangle-\frac{\gamma\hbar}{2m}\left (\frac{\nabla\rho\cdot\nabla
S}{\rho}+\Delta S\right )=0,
\end{equation}
\begin{equation}
\label{ag3}
\frac{\partial {\bf u}}{\partial t}+({\bf u}\cdot \nabla){\bf
u}=-\nabla\Phi-\frac{\nabla Q}{m}+\frac{\gamma\hbar}{2m}\nabla \left
(\frac{{\bf u}\cdot \nabla\rho}{\rho}\right )+ \frac{\gamma\hbar}{2m}{
\Delta} {\bf u},
\end{equation}
where $\langle E\rangle (t)$ is given by Eq. (\ref{dif1q}).
Equation (\ref{ag1}) is a generalized continuity equation. We note that
the normalization condition is not conserved locally. However, it is conserved
globally thanks to the time-dependent chemical potential $\langle E\rangle(t)$
introduced in Eq. (\ref{ag0}). On the other hand, we see that the dissipation
coefficient $\gamma$ introduces a viscous term $\nu\Delta{\bf u}$ in the
momentum
equation (\ref{ag3}) with a quantum viscosity 
\begin{equation}
\label{ag4}
\nu=\frac{\gamma\hbar}{2m}.
\end{equation}
In this sense, Eq. (\ref{ag3}) can be interpreted as a quantum Navier-Stokes
equation. The
hydrodynamic equations (\ref{ag1})-(\ref{ag3}) can
be written as
\begin{equation}
\label{ag1b}
\frac{\partial\rho}{\partial t}+\nabla\cdot (\rho {\bf
u})=-\frac{2\gamma}{\hbar}\rho\left (E-\langle
E\rangle\right ),
\end{equation}
\begin{equation}
\label{ag2v}
\frac{\partial S}{\partial t}+E-\langle
E\rangle-\frac{\gamma\hbar}{2m}\frac{\nabla(\rho\nabla S)}{\rho}=0,
\end{equation}
\begin{equation}
\label{ag3b}
\frac{\partial {\bf u}}{\partial
t}=-\frac{\nabla E}{m}+\frac{\gamma\hbar}{2m}\nabla \left
(\frac{\nabla\cdot (\rho {\bf u})}{\rho}\right ),
\end{equation}
where $E({\bf r},t)$ is defined by Eq.
(\ref{new1}). We note that $E\neq -\partial S/\partial
t$ in the present situation.

Making the WKB transformation (\ref{s4}) in Eq. (\ref{ag0}), we obtain
\begin{equation}
\label{ag5}
\frac{1}{1-i\gamma}\frac{\partial {\cal S}}{\partial t}+\frac{(\nabla
{\cal S})^2}{2m}-i\frac{\hbar}{2m}\Delta {\cal S}+m\Phi-\langle
E\rangle=0.
\end{equation}
Defining the complex velocity by Eq. (\ref{s2}), we find that the  generalized
Schr\"odinger equation (\ref{ag0}) is equivalent to the complex hydrodynamic
equation
\begin{equation}
\label{ag6}
\frac{1}{1-i\gamma}\frac{\partial {\bf U}}{\partial t}+({\bf U}\cdot \nabla){\bf
U}=\frac{i\hbar}{2m}\Delta {\bf U}-\nabla\Phi.
\end{equation}
We can interpret this equation as a viscous Navier-Stokes equation with a
complex
time $(1-i\gamma)t$. It would be interesting to see if this equation can be
derived from Nottale's theory of scale relativity. We note that the
average energy $\langle E\rangle(t)$ does not appear in this equation.
Therefore, Eq. (\ref{ag6}) can be viewed as the most basic equation of the
theory. If we start from this equation and derive the generalized  
Schr\"odinger equation (\ref{ag0}) from this equation by proceeding backwards
(following a presentation similar to the one given in Secs.
\ref{sec_derivation} and \ref{sec_diss}), we find that the
average energy  $\langle E\rangle(t)$  (chemical potential) appears in Eqs.
(\ref{ag0}) and (\ref{ag5}) as a ``constant'' of
integration
possibly depending on time. We are then free to choose it so as to assure
the global conservation of the normalization condition as done above.

If we consider a static ($\partial_t\rho=0$, ${\bf u}={\bf 0}$) solution of Eqs.
(\ref{ag1})-(\ref{ag3}), we get
\begin{equation}
\label{ag7}
m\Phi+Q=E\qquad {\rm and}\qquad S=0,
\end{equation}
where $E$ is a constant. We then have $\langle E\rangle =E$. The wavefunction
$\psi({\bf
r})=\sqrt{\rho({\bf r})}=\phi({\bf r})$ is real and independent on time because
we are already in
the frame rotating with angular velocity $\omega=E/\hbar$. It satisfies the
eigenvalue problem
\begin{equation}
\label{ag0aq}
-\frac{\hbar^2}{2m}\Delta\phi+m\Phi\phi=E\phi
\end{equation}
identical to the time-independent solution (\ref{sta4}) of the  conservative
($\gamma=0$) Schr\"odinger equation.

In the presence of dissipation ($\gamma\neq 0$), the average
energy $\langle E\rangle$ defined by Eq. (\ref{dif1q}) is not conserved.
Taking
its time derivative and proceeding as in Appendix \ref{sec_cote} by using Eqs.
(\ref{ag1b}) and (\ref{ag3b}), we obtain
\begin{equation}
\label{ag8}
\dot{\langle E\rangle}=-\frac{2\gamma}{\hbar}(\langle E^2\rangle-\langle
E\rangle^2)-\frac{\gamma\hbar}{2}\int\frac{\nabla(\rho {\bf u})^2}{\rho}\,
d{\bf r}\le 0,
\end{equation}
where $\langle E\rangle=\int\rho E\, d{\bf r}$ and $\langle E^2\rangle=\int\rho
E^2\, d{\bf r}$. 
At equilibrium, where $\dot{\langle E\rangle}=0$, Eq. (\ref{ag8}) implies that
${\bf u}({\bf r})={\bf 0}$ and $E({\bf r})=\langle E\rangle=E$, leading to Eq.
(\ref{ag7}). Therefore, Eq. (\ref{ag8}) forms an
$H$-theorem. We can obtain this
$H$-theorem directly from the generalized
Schr\"odinger equation (\ref{ag0}). Using the results of Appendix
\ref{sec_aep1}, this equation can be written as
\begin{equation}
\label{ag9}
i\hbar\frac{\partial\psi}{\partial t}=(1-i\gamma)({\hat H}-\langle
E\rangle)\psi=(1-i\gamma)\left \lbrack\frac{\delta H}{\delta\psi^*}-\langle
E\rangle\psi\right \rbrack.
\end{equation}
From Eqs. (\ref{ae11}) and (\ref{ag9}), we obtain
\begin{equation}
\label{ag10}
\dot H=-\frac{2\gamma}{\hbar}\int  |({\hat H}-\langle E\rangle)\psi
|^2\, d{\bf r},
\end{equation}
which is equivalent to Eq. (\ref{ag8}). Therefore, the generalized
Schr\"odinger equation  (\ref{ag0}) relaxes towards an equilibrium state that
minimizes the energy $H=\langle E\rangle$ defined by
Eqs. (\ref{dif1q}) and (\ref{ae4}) under the normalization condition 
(\ref{s11}). This
leads to the variational principle (\ref{lh5b}) returning the static state
(\ref{ag7}). Therefore, the  generalized Schr\"odinger equation  (\ref{ag0}) is
a relaxation equation that can serve as a numerical algorithm to compute
time-independent solutions of the conservative ($\gamma=0$) Schr\"odinger
equation (see the discussion in Sec. \ref{sec_ht}).

{\it Remark:} In situations of physical interest $\gamma\ll 1$. Inversely, if we
consider formally the limit $\gamma\rightarrow
+\infty$, we find that Eqs. (\ref{ag1b}) and (\ref{ag3b}) reduce to
\begin{equation}
\label{ag1bqq}
\frac{\partial\rho}{\partial t}=-\frac{2\gamma}{\hbar}\rho\left
(E-\langle
E\rangle\right ),
\end{equation}
where $E$ and $\langle E\rangle$ are defined by Eqs. (\ref{new1})
and (\ref{dif1q}) with ${\bf u}={\bf 0}$. Equation (\ref{ag1bqq}) provides an
even simpler numerical algorithm relaxing towards the static state (\ref{ag7}).
Indeed, taking the time derivative of $\langle E\rangle$ and using Eq.
(\ref{ag1bqq}), we obtain the $H$-theorem
\begin{equation}
\label{ag8q}
\dot{\langle E\rangle}=-\frac{2\gamma}{\hbar}(\langle E^2\rangle-\langle
E\rangle^2)\le 0.
\end{equation}
It can also be obtained from Eq. (\ref{ag8}) with ${\bf u}={\bf 0}$.

\section{Conclusion}

In this paper, we have derived the generalized Schr\"odinger 
equation (\ref{qb10}) from Nottale's theory of scale relativity. This equation
is nonlinear and irreversible. It takes into account the interaction of the
system with an external environment. We started from the covariant form of the
equation of dynamics (\ref{qb1}) including a friction force. This equation
can be interpreted as a damped viscous Burgers equation (\ref{qb2}) for a
complex velocity field ${\bf U}$ with an imaginary viscosity $\nu=i\hbar/2m$ and
a complex friction coefficient $\gamma$. The real part of the friction
coefficient gives rise to ordinary frictional effects while the imaginary part
of the friction
coefficient gives rise to thermal effects [see Eq. (\ref{qb11})]. The
friction and the temperature are connected to each other by a form of
fluctuation-dissipation theorem. Using Madelung's transformation, we have
developed a hydrodynamic representation of the generalized  Schr\"odinger
equation (\ref{qb10}) and obtained the quantum  damped isothermal Euler
equations
(\ref{tmad4})-(\ref{tmad7}). We have also shown that the generalized 
Schr\"odinger equation (\ref{qb10}) is consistent with a generalized
thermodynamic formalism based on the Boltzmann entropy (\ref{ef7}). In
particular, we have shown that the generalized Schr\"odinger equation
satisfies an $H$-theorem (\ref{ef11}) for the Boltzmann free energy
(\ref{ef1bol}).

The hydrodynamic representation of Madelung 
may lack a clear physical interpretation when the Schr\"odinger equation
describes just {\it one} particle. However, it takes more sense when the
Schr\"odinger equation describes a Bose-Einstein condensate (BEC) made of many
particles in the same quantum state \cite{revuebec}. In that case, the BEC can
be interpreted as a real fluid described by the quantum Euler equations. For
self-interacting BECs, there is an additional pressure force coming from the
self-interaction potential. It depends on the scattering length $a_s$ of the
bosons.

BECs found unexpected applications in the context of dark matter
(see the Introduction of \cite{prd1} for a review). Indeed, it has been
proposed that
dark matter halos are self-gravitating BECs at $T=0$ described by the
Gross-Pitaevskii-Poisson (GPP) equations
\begin{eqnarray}
\label{mi1qq}
i\hbar \frac{\partial\psi}{\partial
t}=-\frac{\hbar^2}{2m}\Delta\psi+m\Phi\psi+\frac{1}{2}m\omega_0^2r^2\psi
+\frac{4\pi a_s\hbar^2}{m^2}|\psi|^{2}\psi,
\end{eqnarray}
\begin{equation}
\label{mi2qq}
\Delta\Phi=4\pi G |\psi|^2.
\end{equation}
For the sake of generality,
we have included a harmonic potential that may mimic a confinement due to tidal
effects when $\omega_0^2>0$ or a solid rotation when $\omega_0^2<0$. The
quantum nature of the BECs may solve the problems of the cold dark matter (CDM)
model such as the
cusp problem and the missing satellite problem. However, the standard BEC model
based on the GPP equations at $T=0$ faces apparent difficulties. First of all,
these
equations are conservative so their spontaneous relaxation towards an
equilibrium state, corresponding to a dark matter halo, is not obvious at first
sight. On the other hand, the
mass-radius relation obtained by determining the stable steady state of the GPP
equations \cite{prd1} is not consistent
with the observations of large dark matter halos. When the bosons are
self-interacting, one
finds that all the halos should have the same radius independently of
their mass and, when the bosons
are non-interacting, one finds that the radius of the halos should decrease
with their
mass as $R\propto M^{-1}$, in contradiction with the observations that reveal
that the radius of the
halos increases with their mass as $R\propto M^{1/2}$ correponding to a
constant surface density (see
\cite{dgp}
for details). This
apparent contradiction is usually solved by considering that the stable steady
state of
the GPP equations only
describes the {\it core} of the halos that is completely condensed (ground
state).
This
solitonic/BEC core must be surrounded by an approximately isothermal
atmosphere. This isothermal atmosphere is necessary to account for the flat
rotation curves of the galaxies. It also explains why the size of dark matter
halos  increases with
their mass. Indeed, for large halos, it is the isothermal atmosphere that fixes
their size, not their solitonic core. Therefore, the flat rotation
curves of the galaxies and the fact that their size increases with their mass
suggest that dark matter halos have an almost isothermal envelope with a 
relatively large temperature. This seems to be paradoxical because one
expects from general considerations
that the thermodynamic temperature of dark matter is very low, much lower
than the condensation temperature $T_c$, so that taking $T=0$ is an excellent
approximation. Therefore, the  origin of an atmosphere with a nonzero
temperature $T\neq 0$ is not obvious at first sight. As a result, in order to
agree with the observations, we must find a source of dissipation and understand
the formation an approximately isothermal envelope surrounding the core.

In Ref. \cite{dgp}, we have proposed to describe dark matter halos by the
generalized GPP equations 
\begin{eqnarray}
\label{mi1}
i\hbar \frac{\partial\psi}{\partial t}=-\frac{\hbar^2}{2m}\Delta\psi+m\Phi\psi+\frac{1}{2}m\omega_0^2r^2\psi
+\frac{4\pi a_s\hbar^2}{m^2}|\psi|^{2}\psi+2k_B T\ln|\psi|\psi
-i\frac{\hbar}{2}\xi\left\lbrack \ln\left (\frac{\psi}{\psi^*}\right )-\left\langle \ln\left (\frac{\psi}{\psi^*}\right )\right\rangle\right\rbrack\psi,
\end{eqnarray}
\begin{equation}
\label{mi2}
\Delta\Phi=4\pi G |\psi|^2,
\end{equation}
or by the corresponding  quantum damped isothermal Euler equations
\begin{equation}
\label{mi11}
\frac{\partial\rho}{\partial t}+\nabla\cdot (\rho {\bf u})=0,
\end{equation}
\begin{eqnarray}
\label{mi12}
\frac{\partial {\bf u}}{\partial t}+({\bf u}\cdot \nabla){\bf u}=-\frac{k_B T}{m}\nabla \ln\rho
-\frac{4\pi a_s\hbar^2}{m^3}\nabla\rho
-\nabla\Phi-\omega_0^2{\bf r}-\frac{1}{m}\nabla Q-\xi{\bf u},
\end{eqnarray}
\begin{equation}
\label{mi13}
\Delta\Phi=4\pi G\rho,
\end{equation}
that include a friction term and a temperature term. The
original GPP equations (\ref{mi1qq}) and (\ref{mi2qq}) are recovered for
$\xi=T=0$. The friction term accounts for the
relaxation of the system towards a steady state. The temperature term accounts
for their
isothermal atmosphere leading to flat
rotation curves and to a mass-radius relation consistent with the
observations. The quantum pressure (due to the Heisenberg exclusion principle
or to the scattering of the bosons) accounts for
the formation of a solitonic core. The system described by the
generalized GPP equations (\ref{mi1}) and (\ref{mi2}) undergoes damped
oscillations and
finally
reaches an equilibrium state with a core-halo structure described by an equation
of state 
\begin{equation}
P=\rho\frac{k_B T}{m}+\frac{2\pi a_s\hbar^2\rho^2}{m^3}.
\end{equation}
The polytropic equation of state $P=2\pi
a_s\hbar^2\rho^{2}/m^3$ dominates in the core where the density is high and the
isothermal equation of state $P=\rho k_B T/m$ dominates in the halo where the
density is low. The fluid equations (\ref{mi11})-(\ref{mi13}) associated with
the damped GPP
equations (\ref{mi1}) and (\ref{mi2}) generalize the hydrodynamic equations of
the CDM model by accounting for a quantum force due to the Heisenberg
uncertainty principle, a pressure force due to the self-interaction of the
bosons, a harmonic potential, a temperature, and a friction. The hydrodynamic
equations of the CDM
model are recovered for $\hbar=a_s=\omega_0=T=\xi=0$.

We have proposed in \cite{dgp} three possible justifications of the generalized
GPP equations
(\ref{mi1}) and (\ref{mi2}):

(i) In a first interpretation, the generalized GP equation (\ref{mi1}) may 
be justified by physical processes. The dissipation could be due to non ideal
effects or to the  interaction of the system with an external environment and
the
nonlinear potential $2k_B T\ln|\psi|$ may account for the self-interaction
of the bosons. In that case, $T$ would be a formal measure of the
collisionless interactions inside the BEC at zero (thermal) temperature.

(ii) In a second interpretation, the generalized GPP
equations (\ref{mi1}) and   (\ref{mi2}) may provide a heuristic parametrization
of the process of gravitational cooling \cite{seidel94}. 
Gravitational cooling is a relaxation mechanism by which a self-gravitating
BEC, initially out-of-equilibrium, ejects some mass and energy in order to reach
a stable steady state (ground state) which is a stable stationary solution of
the GPP equations (\ref{mi1qq}) and   (\ref{mi2qq}). As a
result of this process, accompanied by damped oscillations (virialization), the
system takes a
core-halo structure in which the core is a soliton/BEC (ground state) and the
halo is made of
scalar radiation. The halo behaves approximately as an isothermal
atmosphere like in the process of violent collisionless relaxation
\cite{lb,csr}. However, it cannot be exactly isothermal otherwise it would have
an infinite mass \cite{bt}. In reality, the halo has a density profile
decreasing at large distances 
as $r^{-3}$ similarly to the Navarro-Frenk-White (NFW) \cite{nfw} and Burkert
\cite{burkert} profiles instead of decreasing as
$r^{-2}$ as implied by the self-gravitating isothermal
sphere \cite{bt}.\footnote{We note that the exponent
$\alpha=3$ is the closest exponent to $\alpha=2$ that yields a halo with a
(marginally) finite mass.} The
difference between these two profiles may be due to complicated physical
processes such as incomplete violent relaxation \cite{lb,csr}, tidal effects
\cite{kingclassique,kingfermionic}, stochastic
perturbations... In our model, an additional source of confinement
(modeling, e.g., tidal effects) can be taken into account by the harmonic
potential $\omega_0^2r^2/2$.  In this interpretation, the fundamental
equations of the problem are the GPP
equations (\ref{mi1qq}) and (\ref{mi2qq}) at $T=0$  but
the generalized GPP equations (\ref{mi1}) and (\ref{mi2}) with a bar on $\psi$ 
may provide an
effective parametrization of the GPP equations on the coarse-grained
scale\footnote{In particular, they are able to
account for the
damped oscillations of a system experiencing gravitational cooling
\cite{guzman,guzmanapj}. This damping is apparent on the simplified equation of
motion
(\ref{exact13}).} in the same manner that
the relaxation equations introduced in \cite{csr} provide a
parametrization of the Vlasov-Poisson equations on
the coarse-grained scale. The damping term  can heuristically explain {\it how}
a
system of self-gravitating bosons rapidly reaches an equilibrium state due to
gravitational cooling. Furthermore, this equilibrium state is characterized by a
nonzero effective temperature $T$ (the one that appears in Eq.
(\ref{mi1})) even if $T=0$ fundamentally. The
equilibrium structure of the dark matter halos result from the balance between
the repulsive
quantum force, the repulsive pressure due to the scattering of the bosons, the
repulsive pressure due to the effective temperature, and the gravitational
attraction. The equilibrium state of the generalized GPP equations (\ref{mi1})
and (\ref{mi2})  has a
core-halo structure. It is made of a compact condensed core (BEC/soliton) which
is a stable stationary solution of the GPP equation at $T=0$ (ground state)
surrounded by a halo of scalar radiation similar to an isothermal atmosphere at
temperature $T$. Therefore, Eqs. (\ref{mi1}) and (\ref{mi2}) could
provide a relevant model of dark matter halos experiencing gravitational
cooling.  In that interpretation, the friction coefficient and the temperature
are effective parameters modeling the process of gravitational cooling. Their
values
may differ from halo to halo depending on the efficiency of the
relaxation
process.

(iii) Another, more speculative, interpretation of the generalized GP equation
(\ref{mi1}) is
possible. We may argue that dark matter halos are fundamentally described by the
generalized GP/Schr\"odinger equation (\ref{mi1}) for which the standard
GP/Schr\"odinger equation (\ref{mi1qq}) is an approximation. In
this
interpretation, the
nonlinear terms of friction and temperature in the generalized GP equation
(\ref{mi1}) have a fundamental origin. They are due to the existence of an
hypothetical aether in which the quantum particles move.\footnote{This aether
may
be the manifestation of the fractal nature of the space time in Nottale's
theory.}  This aether is responsible for  the stochastic motion of the particles
due to quantum fluctuations. In this interpretation, $\xi$ and $T$ are
fundamental constants that represent the friction with the aether and the
temperature of the aether. For bosons with an ultra small mass, one can show
that the temperature of the halos is so small (see \cite{dgp}
for
details)
that the term $2k_B T\ln|\psi|\psi$ cannot be detected in laboratory
experiments. However, this term could manifest itself at large scales, in
astrophysics and cosmology, and account for the nonzero effective temperature
of the dark matter halos and their flat rotation curves. In that interpretation,
the flat rotation curves of the galaxies would have a fundamental origin
related to the
nonlinear term $2k_B T\ln|\psi|\psi$ in the generalized Schr\"odinger equation.
However, if $T$ is a fundamental constant (the temperature of the aether), all
the dark matter halos should have the same temperature while it can be shown
that their temperature increases linearly with their size as $T\propto R$ (see
\cite{dgp} for
details). At that stage, we do not know how to solve
this difficulty. Therefore, the interpretations (i) and (ii) should be
privileged over
the interpretation (iii).

\appendix

\section{Connection between the Smoluchowski equation, the Schr\"odinger
equation, and the viscous Burgers equation}
\label{sec_c}

\subsection{Connection between the Smoluchowski equation and the Schr\"odinger equation}
\label{sec_ssb}

We consider a Brownian particle of mass $m$ moving in an external
potential $U({\bf r})$. In the strong friction limit $\xi\rightarrow +\infty$,
where $\xi$ is the friction coefficient, the evolution of the density
probability $\rho({\bf r},t)$ of finding the Brownian particle at position ${\bf
r}$ at time $t$ is governed by a Fokker-Planck equation of the form
\begin{eqnarray}
\label{ssb1}
\frac{\partial \rho}{\partial t}=D\nabla\cdot (\nabla \rho+\beta\rho\nabla U)
\end{eqnarray}
called the Smoluchowski equation. Here, $D$ is the diffusion coefficient and
$\beta=m/k_B T$ is the inverse temperature (for commodity, we have included the
mass of the particle in the definition of $\beta$). The diffusion coefficient,
the friction coefficient and the temperature are related to each other by the
Einstein relation $D=k_B T/\xi m$. The equilibrium solution of the Smoluchowski
equation (\ref{ssb1}) is the Boltzmann distribution
\begin{eqnarray}
\label{ssb2}
\rho_{\rm eq}({\bf r})=A e^{-\beta U({\bf r})}.
\end{eqnarray}
If we make the change of variables
\begin{eqnarray}
\label{ssb3}
\rho({\bf r},t)=\psi({\bf r},t) e^{-\frac{1}{2}\beta U({\bf r})},
\end{eqnarray}
we can transform the Smoluchowski equation (\ref{ssb1}) into a Schr\"odinger
equation in imaginary time
\begin{eqnarray}
\label{ssb4}
\frac{\partial \psi}{\partial t}=D\Delta \psi+\frac{1}{2D}V\psi
\end{eqnarray}
with a potential
\begin{eqnarray}
\label{ssb5}
V({\bf r})=D^2\beta\Delta U-\frac{1}{2}D^2\beta^2(\nabla U)^2.
\end{eqnarray}
In order to make the analogy with the Schr\"odinger equation closer, we write the diffusion coefficient as
\begin{eqnarray}
\label{ssb6}
D=\frac{\hbar}{2m},
\end{eqnarray}
where $\hbar$, which has the dimension of an action, is analogous to the Planck constant (in the present case, we have $\hbar=2k_BT/\xi$ according to the Einstein relation). We also introduce the potential
\begin{eqnarray}
\label{ssb7}
\Phi({\bf r})=-V({\bf r}).
\end{eqnarray}
Then, we can rewrite Eq. (\ref{ssb4}) as
\begin{eqnarray}
\label{ssb8}
-\hbar\frac{\partial \psi}{\partial t}=-\frac{\hbar^2}{2m}\Delta \psi+m \Phi \psi.
\end{eqnarray}
Equation (\ref{ssb8}) can be viewed as a Schr\"odinger equation in imaginary
time with a potential $\Phi$ opposite to the potential $V$ appearing in Eq.
(\ref{ssb4}). Therefore, in all the cases where we can solve the Schr\"odinger
equation (\ref{ssb8}) for the potential $\Phi$, we can solve the Smoluchowski
equation (\ref{ssb1}) for the potential $U$ related to $\Phi$ by Eqs. 
(\ref{ssb5}) and  (\ref{ssb7}). Inversely, if we known an analytical solution of
the Smoluchowski equation (\ref{ssb1}) for the potential $U$, we can obtain an
analytical solution of the Schr\"odinger equation (\ref{ssb8}) with the
corresponding potential $\Phi$.

If we consider a solution of the Smoluchowski equation (\ref{ssb1}) of the form
(\ref{ssb3}) with
\begin{eqnarray}
\label{ssb9}
\psi({\bf r},t)=\phi({\bf r})e^{-\lambda t},
\end{eqnarray}
we obtain from Eq. (\ref{ssb8}) the eigenvalue equation
\begin{eqnarray}
\label{ssb10}
D\Delta \phi+\frac{1}{2D} V \phi=-\lambda\phi,
\end{eqnarray}
where the eigenvalue $\lambda$ gives the growth or damping rate of the
eigenmode (\ref{ssb9}).
In comparison, if we consider a stationary solution of the  Schr\"odinger
equation (\ref{s9}) of the form
\begin{eqnarray}
\label{ssb12}
\psi({\bf r},t)=\phi({\bf r})e^{-iE t/\hbar},
\end{eqnarray}
we obtain the eigenvalue equation
\begin{eqnarray}
\label{ssb13}
-\frac{\hbar^2}{2m}\Delta \phi+m \Phi \phi=E\phi,
\end{eqnarray}
where $E$ is the eigenenergy so that $\omega=E/\hbar$ gives 
the pulsation of the wave. We note that the two eigenvalue equations
(\ref{ssb10}) and (\ref{ssb13}) coincide provided that we make the
correspondences of Eqs. (\ref{ssb6}), (\ref{ssb7}) and set
$E/\hbar=\omega=\lambda$. Therefore, if we know analytically the
eigenvalues of the Schr\"odinger equation for the potential $\Phi$ we can obtain
analytically the eigenvalues of the Smoluchowski equation for the potential $U$
that is related to $\Phi$ according to Eqs. (\ref{ssb5}) and (\ref{ssb7}).

{\it Remark:} For $\lambda=0$ the solution of Eq. (\ref{ssb10})  is given by
[see Eq. (\ref{ssb2})]:
\begin{eqnarray}
\label{ssb18}
\psi_{\rm eq}({\bf r})=A e^{-\frac{1}{2}\beta U({\bf r})},
\end{eqnarray}
where
$U$ is related to $V$ according to Eq. (\ref{ssb5}). This is also the 
solution of the stationary Schr\"odinger equation (\ref{ssb13}) for
$E=0$. However, to make the solution explicit, we need to solve Eq. (\ref{ssb5})
to express $U$ as a function of $\Phi=-V$. We note that  Eq. (\ref{ssb5}) has
the form of a Riccati equation for $\nabla U$.

\subsection{Connection between the viscous Burgers equation and the Schr\"odinger equation: The Cole-Hopf transformation}
\label{sec_coh}

The Navier-Stokes equation without pressure
\begin{eqnarray}
\label{coh1}
\frac{\partial {\bf u}}{\partial t}+({\bf u}\cdot \nabla){\bf u}=\nu\Delta {\bf u}-\nabla V
\end{eqnarray}
is called the viscous Burgers equation. For the sake of generality, we have
introduced an external potential $V({\bf r})$. For a potential flow, the
velocity field can be written as ${\bf u}=\nabla \theta$. In that case, the flow
is irrotational: $\nabla\times {\bf u}={\bf 0}$.  Using the 
identity $({\bf u}\cdot \nabla){\bf u}=\nabla ({{\bf u}^2}/{2})-{\bf u}\times
(\nabla\times {\bf u})$ which reduces to $({\bf u}\cdot \nabla){\bf u}=\nabla
({{\bf u}^2}/{2})$ for an irrotational flow,  Eq. (\ref{coh1}) can be
integrated. This yields the viscous Bernoulli equation
\begin{eqnarray}
\label{coh2}
\frac{\partial \theta}{\partial t}+\frac{(\nabla \theta)^2}{2}=\nu\Delta \theta-V,
\end{eqnarray}
where the constant of integration $C(t)$ has been set equal to zero. If we make
the change of variables
\begin{eqnarray}
\label{sam}
\theta=-2\nu\ln\psi
\end{eqnarray}
in Eq. (\ref{coh2}), we obtain
\begin{eqnarray}
\label{coh3}
\frac{\partial \psi}{\partial t}=\nu\Delta \psi+\frac{1}{2\nu}V\psi.
\end{eqnarray}
For $V=0$, Eq. (\ref{coh3}) reduces to the classical diffusion equation.
Therefore, the viscous Burgers equation (\ref{coh1}) for a potential flow is
equivalent to the diffusion equation. The transformation (\ref{sam}) leading
from Eq.
(\ref{coh1}) to Eq. (\ref{coh3}) is called the Cole-Hopf transformation
\cite{cole,hopf}. With
the presence of the external potential $V$, Eq. (\ref{coh3}) is similar to the
Schr\"odinger equation in imaginary time. To make the analogy closer, we
write the viscosity as
\begin{eqnarray}
\label{coh4}
\nu=\frac{\hbar}{2m},
\end{eqnarray}
where $\hbar$, which has the dimension of an action, is analogous to the Planck constant (this is a pure notation here since the mass $m$ has no physical meaning). We also introduce the potential
\begin{eqnarray}
\label{coh5}
\Phi({\bf r})=-V({\bf r}).
\end{eqnarray}
Then, we can rewrite Eq. (\ref{coh3}) as
\begin{eqnarray}
\label{coh6}
-\hbar\frac{\partial \psi}{\partial t}=-\frac{\hbar^2}{2m}\Delta \psi+m \Phi \psi.
\end{eqnarray}
Equation (\ref{coh6}) can be viewed as a Schr\"odinger equation in imaginary
time with a potential $\Phi$ opposite to the potential $V$ appearing in the
viscous Burgers equation (\ref{coh1}).
Therefore, in all the cases where we can solve the Schr\"odinger equation (\ref{coh6}) for the potential $\Phi$ we can solve the Burgers equation (\ref{coh1}) for the potential $V=-\Phi$, and conversely.

We look for the general stationary solution (${\bf u}({\bf r})$ independent of
time) of the viscous Burgers equation
(\ref{coh1}). The condition $\partial_t{\bf u}={\bf 0}$ is equivalent to
$\partial_t\nabla\theta={\bf 0}$ implying $\theta({\bf r},t)=\theta_1({\bf
r})+\theta_2(t)$. Substituting this decomposition into Eq. (\ref{coh2}), we
obtain
\begin{eqnarray}
\label{coh9}
\frac{d\theta_2}{dt}=-\frac{(\nabla \theta_1)^2}{2}+\nu\Delta \theta_1-V\equiv 2\nu\lambda.
\end{eqnarray}
Since the first term depends only on $t$ and the second term depends only on ${\bf r}$, they must be individually equal to a constant that we note  $2\nu\lambda$. We then obtain $\theta_2=2\nu\lambda t$. On the other hand, making the transformation $\theta_1=-2\nu\ln\phi$, we find that the general stationary solution of the viscous Burgers equation (\ref{coh1}) is
\begin{eqnarray}
\label{coh10}
\theta({\bf r},t)=-2\nu (\ln\phi-\lambda t),
\end{eqnarray}
\begin{eqnarray}
\label{coh11}
{\bf u}({\bf r})=\nabla\theta=-2\nu \frac{\nabla\phi}{\phi},
\end{eqnarray}
where $\phi$ is the solution of the eigenvalue equation
\begin{eqnarray}
\label{coh12}
-2\nu^2\Delta\phi-V\phi=2\nu\lambda\phi.
\end{eqnarray}
Using Eqs. (\ref{coh4}) and (\ref{coh5}) and defining
$E/\hbar=\omega=\lambda$, we
can rewrite Eq. (\ref{coh12}) into the form of a stationary Schr\"odinger
equation
\begin{eqnarray}
\label{coh13}
-\frac{\hbar^2}{2m}\Delta \phi+m \Phi \phi=E\phi.
\end{eqnarray}
The stationary solution of the viscous Burgers equation with a potential $V$ 
corresponds to the stationary solution of the  Schr\"odinger equation with a
potential $\Phi=-V$. Therefore, in all the cases where we can solve the
stationary  Schr\"odinger equation analytically, we can obtain an analytical
solution of the stationary Burgers equation and conversely.

\subsection{Connection between the Smoluchowski equation and the viscous Burgers equation}

According to  the results of Appendix \ref{sec_ssb}, the Smoluchowski 
equation (\ref{ssb1}) can be transformed into the Schr\"odinger equation in
imaginary time (\ref{ssb8}) with $D=\hbar/2m$ and $\Phi=-V$, where $V$ is
related to $U$ by Eq. (\ref{ssb5}).  According to  the results of Appendix
\ref{sec_coh}, the viscous Burgers equation (\ref{coh1}) can be transformed
into the Schr\"odinger equation in imaginary time (\ref{coh6}) with
$\nu=\hbar/2m$ and $\Phi=-V$. As a result, the Smoluchowski equation 
(\ref{ssb1}) can be be transformed into the viscous Burgers equation 
(\ref{coh1}) with $\nu=D$ and $V$ given by Eq. (\ref{ssb5}). The velocity field
${\bf u}$ in the Burgers equation (\ref{coh1}) is related to the density $\rho$
and to the potential $U$ in the Smoluchowski equation (\ref{ssb1}) by means of
the Cole-Hopf transformation giving
\begin{eqnarray}
\label{ssb14}
\theta=-2\nu\ln\psi=-2D\ln\psi=-2D\left (\ln\rho+\frac{1}{2}\beta U\right ),
\end{eqnarray}
\begin{eqnarray}
\label{ssb15}
{\bf u}=\nabla\theta=-2D\left (\frac{\nabla\rho}{\rho}+\frac{1}{2}\beta\nabla U\right ).
\end{eqnarray}
For some simple potentials $U$, we can solve the Smoluchowski equation
(\ref{ssb1}) analytically. Using Eqs. (\ref{ssb14}) and (\ref{ssb15}), we can
then obtain an analytical solution of the viscous Burgers equation for the
corresponding potential $V$ defined by Eq. (\ref{ssb5}).

Let us consider a solution of the Smoluchowski equation of the form of Eq.
(\ref{ssb3}) with Eq. (\ref{ssb9}). 
The rate $\lambda$ of this eigenmode is the eigenvalue of the stationary 
Schr\"odinger equation (\ref{ssb10}).  Using the Cole-Hopf transformation, we
find that
\begin{eqnarray}
\label{ssb16}
\theta({\bf r},t)=-2\nu\ln\psi=-2D\ln\psi=-2D (\ln\phi-\lambda t),
\end{eqnarray}
\begin{eqnarray}
\label{ssb17}
{\bf u}({\bf r})=\nabla\theta=-2D \frac{\nabla\phi}{\phi}.
\end{eqnarray}
This returns the general stationary solution (\ref{coh10})-(\ref{coh12}) of the viscous Burgers equation with an external potential (see Appendix \ref{sec_coh}). More specifically, we consider the equilibrium solution of the Smoluchowski equation corresponding to $\lambda=0$. Using Eq. (\ref{ssb18}) and making the Cole-Hopf transformation, we find that
\begin{eqnarray}
\label{ssb19}
\theta({\bf r})=-2\nu\ln\psi=-2D\ln\psi=-2D\left (\ln A-\frac{1}{2}\beta U\right ),
\end{eqnarray}
\begin{eqnarray}
\label{ssb20}
{\bf u}({\bf r})=\nabla\theta=D\beta\nabla U.
\end{eqnarray}
This provides a particular stationary solution of the viscous Burgers equation with a potential $V$ related to $U$ by Eq. (\ref{ssb5}).

\section{Necessity of the Nelson relation (\ref{s8})}
\label{sec_qdc}

If we define the wave function by Eq. (\ref{s5}) without imposing the relation
(\ref{s8}), we obtain an equation of the form
\begin{equation}
\label{am1}
i\hbar\frac{\partial\psi}{\partial t}=\hbar\left ({\cal
D}-\frac{\hbar}{2m}\right )\frac{(\nabla\psi)^2}{\psi}-{\cal
D}\hbar\Delta\psi+m\Phi\psi.
\end{equation}
Applying the Madelung transformation to this equation, we obtain the
hydrodynamic equations
\begin{equation}
\label{ntmad3b}
\frac{\partial\rho}{\partial t}+\nabla\cdot (\rho {\bf u})=\left
(1-\frac{2m}{\hbar}{\cal D}\right )\rho \nabla\cdot {\bf u},
\end{equation}
\begin{equation}
\label{am3}
\frac{\partial S}{\partial t}+\frac{\cal D}{\hbar}(\nabla
S)^2+m\Phi-\frac{1}{2}{\cal D}\hbar \left
\lbrack\frac{\Delta\rho}{\rho}-\frac{1}{2\rho^2}(\nabla\rho)^2\right
\rbrack+\hbar\left ({\cal D}-\frac{\hbar}{2m}\right )\left\lbrack
\frac{1}{4\rho^2}(\nabla\rho)^2-\frac{1}{\hbar^2}(\nabla S)^2\right\rbrack=0,
\end{equation}
\begin{equation}
\label{am4}
\frac{\partial {\bf u}}{\partial t}+\frac{2m}{\hbar}{\cal D}\nabla\left
(\frac{{\bf u}^2}{2}\right )=-\nabla\Phi+{\cal D}\frac{\hbar}{2m}\nabla \left
\lbrack\frac{\Delta\rho}{\rho}-\frac{1}{2\rho^2}(\nabla\rho)^2\right
\rbrack-\frac{\hbar}{m}\left ({\cal D}-\frac{\hbar}{2m}\right
)\nabla\left\lbrack \frac{1}{4\rho^2}(\nabla\rho)^2-\frac{1}{\hbar^2}(\nabla
S)^2\right\rbrack.
\end{equation}
We see from  Eq. (\ref{ntmad3b}) that Eq.
(\ref{am1}) conserves
the integral $\int |\psi|^2\, d{\bf r}$ if, and only if, the quantum diffusion
coefficient ${\cal D}$ satisfies the Nelson relation (\ref{s8}).
In that case, Eqs. (\ref{am1})-(\ref{am4}) reduce to  the Schr\"odinger equation
(\ref{s9})
and to the hydrodynamic equations (\ref{tmad4c})-(\ref{ntmad8}).

\section{Generalized Ehrenfest theorem}
\label{sec_eh}

In this Appendix, using the fluid representation (\ref{tmad4})-(\ref{tmad7}) of
the 
generalized Schr\"odinger equation (\ref{qb10}), we derive a generalization of
the Ehrenfest \cite{ehrenfest} theorem that takes dissipative effects 
into account.

In the hydrodynamic representation, the average values of the position
and velocity of the quantum particle are given by
\begin{equation}
\label{eh1}
\langle {\bf r}\rangle (t)=\int \rho {\bf r}\, d{\bf r},\qquad \langle {\bf
u}\rangle (t) =\int \rho {\bf u}\, d{\bf r}.
\end{equation}
Taking the time derivative of the first relation in Eq. (\ref{eh1}), using the
continuity equation (\ref{tmad4}), integrating by parts, and comparing with the
second relation in Eq. (\ref{eh1}), we obtain
\begin{equation}
\label{eh2}
\frac{d}{dt}\langle {\bf r}\rangle=\langle {\bf u}\rangle.
\end{equation}
Taking the time derivative of the second relation in Eq. (\ref{eh1}), using the 
Euler equation (\ref{tmad9}), and recalling that the average value of the
quantum force vanishes [see Eq. (\ref{zqf})], we obtain
\begin{equation}
\label{eh3}
\frac{d}{dt}\langle {\bf u}\rangle=-\langle \nabla \Phi\rangle-\xi \langle {\bf
u}\rangle.
\end{equation}
Therefore, the average values of ${\bf r}$ and ${\bf u}$ 
follow ordinary equations of motion as in classical mechanics. This is the
meaning of the Ehrenfest theorem. We see that this theorem can be generalized
to the case of dissipative systems.\footnote{It can also be generalized to the
class of nonlinear Schr\"odinger equations considered in \cite{dgp} that are
equivalent to hydrodynamic equations with an arbitrary barotropic equation of
state $P(\rho)$. Indeed, in the derivation of Eq. (\ref{eh3}), the term
$\int\nabla P\, d{\bf r}$ vanishes for all functions $P({\bf r},t)$ that tend
to zero at infinity sufficiently rapidly.} We note, however, that the Ehrenfest
theorem involves the average value of the force $\langle \nabla\Phi\rangle
(t)=\int \rho \nabla\Phi\, d{\bf r}$, not the force $\nabla\Phi(\langle {\bf
r}\rangle (t),t)$ taken at the average position of the particle. In general,
they are not the same. 

In order to establish that Eqs. (\ref{eh2}) and (\ref{eh3}) are really 
equivalent to the Ehrenfest equations, we need to discuss the relation between
the hydrodynamic variables and the quantum operators. According to the
correspondence principle, the impulse operator is defined by
$\hat{\bf p}=-i\hbar\nabla$ [see Eq. (\ref{corrpple})]. The average value of the
impulse is
\begin{equation}
\label{eh5}
\langle {\bf p}\rangle=\langle \psi|\hat{\bf p}|\psi\rangle =-i\hbar\int
\psi^*\nabla\psi\, d{\bf r}.
\end{equation}
In Nottale's theory, the complex impulse is defined by ${\bf
P}=m{\bf U}=\nabla{\cal S}=-i\hbar\nabla\ln\psi$ [see Eq. (\ref{cqv2})]. Its
average value is
\begin{equation}
\label{eh7}
\langle {\bf P}\rangle=\int {\bf P}|\psi|^2\, d{\bf r}=-i\hbar\int
\psi^*\nabla\psi\, d{\bf r}.
\end{equation}
It coincides with the average value of the impulse: $\langle {\bf
P}\rangle = \langle {\bf p}\rangle$. On the other hand, we note that
$\langle {\bf u}_Q\rangle ={\bf 0}$ according to Eq. (\ref{cqv7}). Therefore,
$\langle {\bf U}\rangle =\langle {\bf u}\rangle$. This first shows that the
average value of the complex velocity is real and that it satisfies Eq.
(\ref{eh3}). Furthermore, $\langle {\bf P}\rangle=m\langle {\bf U}\rangle
=m\langle {\bf u}\rangle$ so that $\langle {\bf p}\rangle=m\langle {\bf
u}\rangle$. Therefore,
Eqs. (\ref{eh2}) and (\ref{eh3}) are equivalent to the Ehrenfest equations
appropriately generalized to the case of dissipative systems.

\section{Conservation of the average energy in the hydrodynamic representation}
\label{sec_cote}

In the hydrodynamic representation, the average energy (\ref{dif1q}) associated
with the
Schr\"odinger equation (\ref{s9}) can be written as
\begin{equation}
\label{cote1}
\langle E\rangle=\Theta_c+\Theta_Q+W,
\end{equation}
where $\Theta_c$ is the classical kinetic energy (\ref{ef3}), $\Theta_Q$ is the
quantum kinetic energy  (\ref{ef4}), and $W$ is the potential energy
(\ref{ef6}).
Their
first variations are
\begin{equation}
\label{var1}
\delta\Theta_c=\int m\frac{{\bf u}^2}{2}\delta\rho \, d{\bf r}+\int \rho m {\bf
u}\cdot\delta {\bf u} \, d{\bf r},
\end{equation}
\begin{equation}
\label{var2}
\delta\Theta_Q=\int Q\delta\rho \, d{\bf r},
\end{equation}
\begin{equation}
\label{var5}
\delta W=\int m \Phi\delta\rho \, d{\bf r}.
\end{equation}
Taking the time derivative of the
average
energy (\ref{cote1}), and using Eqs. (\ref{var1})-(\ref{var5}), we get
\begin{eqnarray}
\label{cons5}
\dot{\langle E\rangle}=\int \left (m \frac{{\bf u}^2}{2}+Q+m\Phi\right )
\frac{\partial\rho}{\partial t}\, d{\bf r}+\int \rho m {\bf u}\cdot
\frac{\partial {\bf u}}{\partial t}\, d{\bf r}.
\end{eqnarray}
Substituting the continuity equation  (\ref{tmad4c}) into Eq. (\ref{cons5}),
integrating by parts, and using the Euler equation (\ref{ntmad6}) or
(\ref{new1b}), we
obtain $\dot{\langle E\rangle}=0$
showing the conservation of the average energy.

{\it Remark:} The calculations leading to relations (\ref{var1})-(\ref{var5})
are straightforward
except, maybe, the ones leading Eq. (\ref{var2}). We give below three different
derivations of this relation:

(i) From Eqs.  (\ref{ntmad8}), (\ref{ef5b}) and
(\ref{ef4}) we directly obtain
\begin{equation}
\label{ide4}
\delta\Theta_Q=\frac{\hbar^2}{m}\int \nabla\sqrt{\rho}\cdot
\nabla\delta\sqrt{\rho}\, d{\bf r}=-\frac{\hbar^2}{m}\int
\delta\sqrt{\rho}\Delta\sqrt{\rho}\, d{\bf r}=-\frac{\hbar^2}{2m}\int
\delta\rho\frac{\Delta\sqrt{\rho}}{\sqrt{\rho}}\, d{\bf r}=\int Q\delta\rho\,
d{\bf r}.
\end{equation}

(ii) From the first equality of Eq. (\ref{ntmad8}), we
find that
\begin{equation}
\label{var6}
\delta Q=-\frac{\hbar^2}{2m\rho}\nabla\cdot
(\sqrt{\rho}\nabla\delta\sqrt{\rho}-\delta\sqrt{\rho}\nabla\sqrt{\rho}
)=-\frac{\hbar^2}{4m\rho}\nabla\cdot
(\rho\delta\nabla\ln\rho). 
\end{equation}
This implies the identity
\begin{equation}
\label{var7}
\int \rho \delta Q\, d{\bf r}=0,
\end{equation}
from which we get $\delta\Theta_Q=\int \rho \delta Q\, d{\bf r}+\int Q\delta\rho
\, d{\bf r}=\int Q\delta\rho \, d{\bf r}$. We note
that Eq. (\ref{var6}) is the equivalent to the tensorial equation (\ref{mad19})
with Eq. (\ref{mad20}) except that it applies to a perturbation $\delta$ instead
of a space derivative $\partial_i$. 

(iii) Taking the first order variations of Eq.
(\ref{cqq7}) and using Eq. (\ref{cqv7}), we obtain
\begin{equation}
\label{cqq7b}
\rho\delta Q=-\frac{\hbar}{2}\nabla\cdot (\rho\, \delta {\bf u}_Q),
\end{equation}
implying Eq. (\ref{var7}), then Eq.  (\ref{var2}).

\section{Derivation of the generalized Nelson equations}
\label{sec_gne}

\subsection{The complex impulse}
\label{sec_cqv}

The wave function can be written in the WKB form as
\begin{equation}
\label{cqv1}
\psi=e^{i{\cal S}/\hbar},
\end{equation}
where ${\cal S}$ is the complex action. Since ${\cal S}=-i\hbar\ln\psi$, the
complex velocity field ${\bf U}=\nabla {\cal S}/m$ and the complex impulse
${\bf P}=m{\bf U}=\nabla {\cal S}$ are given by
\begin{equation}
\label{cqv2}
{\bf U}=-i\frac{\hbar}{m}\nabla\ln\psi\qquad {\rm and}\qquad {\bf P}=-i
\hbar\nabla\ln\psi.
\end{equation}
We note that
\begin{equation}
{\bf P}\psi=-i \hbar\nabla\psi
\end{equation}
which is similar to the correspondance principle (\ref{corrpple}). The total
kinetic
energy is defined by Eq. (\ref{ef2}). Using Eq. (\ref{cqv2})
it can be written as
\begin{equation}
\label{cqv3}
\Theta=\frac{1}{2}\int\rho m |{\bf U}|^2\, d{\bf r}.
\end{equation}
It has the form of an ordinary kinetic energy for a complex velocity field.

The wave function can also be written in the Madelung form as
\begin{equation}
\label{cqv4}
\psi=\sqrt{\rho}e^{i S/\hbar},
\end{equation}
where $S$ is the real action and $\rho=|\psi|^2$ is the probability density.
Comparing the
two expressions (\ref{cqv1}) and (\ref{cqv4}) of the wave function, we find that
the complex action is related to the real action and to the probability density
by
\begin{equation}
\label{cqv5}
{\cal S}=S-i\frac{\hbar}{2}\ln\rho.
\end{equation}
As a result, the complex velocity field ${\bf U}=\nabla {\cal S}/m$  can be
rewritten as
\begin{equation}
\label{cqv5b}
{\bf U}={\bf u}-i{\bf u}_Q,
\end{equation}
where
\begin{equation}
\label{cqv6}
{\bf u}=\frac{\nabla S}{m}
\end{equation}
is the classical velocity and
\begin{equation}
\label{cqv7}
{\bf u}_Q=\frac{\hbar}{2m}{\nabla\ln\rho}
\end{equation}
is the quantum velocity.\footnote{The complex velocity 
(\ref{cqv5b}) combining 
the classical velocity (\ref{cqv6}) and the quantum velocity (\ref{cqv7}) was
first introduced by
Madelung in a not well-known paper \cite{madelungearly}. It also
appears in Refs.
\cite{hcp,km}.}  Using Eq. (\ref{cqv5b}),
the total kinetic energy (\ref{cqv3}) can be written as
$\Theta=\Theta_c+\Theta_Q$ where
\begin{equation}
\label{cqv8}
\Theta_c=\frac{1}{2}\int\rho m {\bf u}^2\, d{\bf r}
\end{equation}
is the classical kinetic energy and
\begin{equation}
\label{cqv9}
\Theta_Q=\frac{1}{2}\int\rho m {\bf u}_Q^2\, d{\bf r}
\end{equation}
is the quantum kinetic energy. Using Eq. (\ref{cqv7}) we can check that the
quantum kinetic energy $\Theta_Q$ is equivalent to Eq. (\ref{ef4}) which has the
form of a quantum potential
energy. Alternatively, Eq. (\ref{cqv9}) indicates that the quantum kinetic
energy can be written as an ordinary kinetic energy for a quantum velocity
field.

{\it Remark:} Using Eq. (\ref{cqv7}), the quantum potential (\ref{ntmad8}) can
be
expressed in terms of the quantum velocity as
\begin{equation}
\label{cqq7}
Q=-\frac{\hbar}{2}\nabla\cdot {\bf u}_Q-\frac{1}{2}m{\bf u}_Q^2.
\end{equation}
If we consider a static state of the Schr\"odinger equation (\ref{s9}), we have
$Q+m\Phi=E$ [see Eq. (\ref{sta5})]. Using Eq. (\ref{cqq7}), we obtain
\begin{equation}
\label{cqq7briccati}
-\frac{\hbar}{2}\nabla\cdot {\bf u}_Q-\frac{1}{2}m{\bf u}_Q^2=E-m\Phi({\bf r}).
\end{equation}
This equation can be viewed as a Riccati equation that may be easier to solve
than the Schr\"odinger
equation itself.

\subsection{The complex quantum potential}
\label{sec_cqq}

Using Eq. (\ref{s8}), the
scale-covariant equation of dynamics (\ref{s1}) can be
written as
\begin{equation}
\label{cqq1}
\frac{\partial {\bf U}}{\partial t}+({\bf U}\cdot \nabla){\bf U}=i
\frac{\hbar}{2m}\Delta{\bf U}-\nabla\Phi.
\end{equation}
In Sec. \ref{sec_s}, we have interpreted the term $i (\hbar/2m)\Delta{\bf U}$ as
a viscous 
term with an imaginary viscosity $\nu=i\hbar/2m$. Alternatively, using
the identity $\Delta{\bf U}=\nabla (\nabla\cdot {\bf U})=\nabla (\Delta {\cal
S}/m)$ valid for a potential flow, the term $i (\hbar/2m)\Delta{\bf U}$ can be
interpreted as a complex quantum force by unit of mass
\begin{equation}
\label{cqq1b}
{\cal F}_Q=-\frac{1}{m}\nabla q
\end{equation}
deriving from a complex quantum potential
\begin{equation}
\label{cqq2}
q=-i\frac{\hbar}{2}\nabla\cdot {\bf U}=-i\frac{\hbar}{2m}\Delta{\cal
S}=-\frac{\hbar^2}{2m}\Delta\ln\psi.
\end{equation}
Using Eq. (\ref{cqv5b}), the quantum potential can be written in terms of the
classical and quantum velocities as
\begin{equation}
\label{cqq5}
q=-\frac{\hbar}{2}\nabla\cdot {\bf u}_Q-i\frac{\hbar}{2}\nabla\cdot {\bf u}.
\end{equation}
On the other hand, using Eq. (\ref{ntmad2}) and the identity (\ref{s6}), we find
that
\begin{equation}
\label{cqq3}
q=-\frac{\hbar^2}{4m}\Delta(\ln\rho)-i\frac{\hbar}{2m}\Delta
S=-\frac{\hbar^2}{4m}\left\lbrack \frac{\Delta
\rho}{\rho}-\frac{(\nabla\rho)^2}{\rho^2}\right\rbrack-i\frac{\hbar}{2m}\Delta
S.
\end{equation}
Comparing Eq. (\ref{cqq3}) with Eq. (\ref{ntmad8}), or Eq. (\ref{cqq5}) with Eq.
(\ref{cqq7}), we note that the real part of the
complex quantum potential $q$ is {\it not} the quantum potential $Q$.

\subsection{The complex energy}
\label{sec_hj}

According to Eq. (\ref{cqv1}), the complex energy ${\cal E}=-{\partial
{\cal S}}/{\partial t}$ is given by
\begin{equation}
{\cal E}=i\hbar \frac{\partial\ln\psi}{\partial t}.
\end{equation}
We note that
\begin{equation}
{\cal E}\psi=i\hbar \frac{\partial\psi}{\partial t}
\label{eno}
\end{equation}
which is similar to the correspondance principle (\ref{corrpple}). Using
Eq.
(\ref{cqv5}), we find that
${\cal E}=E+iE_Q$, where
\begin{equation}
\label{hj5qq}
E=-\frac{\partial S}{\partial t}
\end{equation}
and
\begin{equation}
\label{hj5qqb}
E_Q=\frac{\hbar}{2\rho}\frac{\partial\rho}{\partial t}.
\end{equation}
The complex Hamilton-Jacobi equation (\ref{s3}) can be written as
\begin{equation}
\label{hj2}
{\cal E}=\frac{1}{2m}(\nabla {\cal S})^2-i{\cal D}\Delta{\cal S}+m\Phi.
\end{equation}
Using Eq. (\ref{s2}), we obtain
\begin{equation}
\label{hj3}
{\cal E}=\frac{1}{2}m{\bf U}^2-i\frac{\hbar}{2}\nabla\cdot {\bf U}+m\Phi.
\end{equation}
Introducing the complex quantum potential (\ref{cqq2}), the complex
energy takes the form 
\begin{equation}
\label{hj4}
{\cal E}=\frac{1}{2}m{\bf U}^2+q+m\Phi.
\end{equation}
It is the sum of the complex kinetic energy, the complex quantum potential, and
the external potential. As noted by Nottale \cite{nottale}, the theory of scale
relativity yields a new contribution $q$ to the energy that comes from the very
geometry of space-time. This is similar to the new contribution $mc^2$ to the
energy in Einstein's theory of relativity. Using Eqs. (\ref{cqv5b}) and
(\ref{cqq5}), we find that
${\cal E}=E+iE_Q$ with
\begin{equation}
\label{hj5}
E=\frac{1}{2}m{\bf u}^2-\frac{1}{2}m{\bf u}_Q^2-\frac{\hbar}{2}\nabla\cdot {\bf
u}_Q+m\Phi
\end{equation}
and
\begin{equation}
\label{hj6}
E_Q=-m{\bf u}\cdot {\bf u}_Q-\frac{\hbar}{2}\nabla\cdot {\bf u}.
\end{equation}
Using Eq. (\ref{cqq7}), we can rewrite Eq. (\ref{hj5}) as
\begin{equation}
\label{hj7}
E=\frac{1}{2}m{\bf u}^2+Q+m\Phi.
\end{equation}
Equations (\ref{cqv6}),  (\ref{hj5qq}) and (\ref{hj7}) return the quantum
Hamilton-Jacobi equations (\ref{ntmad5}) and (\ref{new1}). On
the
other hand, using Eq. (\ref{cqv7}), we can rewrite Eq. (\ref{hj6}) as
\begin{equation}
\label{hj7b}
E_Q=-\frac{\hbar}{2\rho}\nabla(\rho {\bf u}).
\end{equation}
Equations (\ref{hj5qqb}) and (\ref{hj7b}) return the equation of continuity
(\ref{tmad4c}).  When $\gamma=0$, the average energy $\langle E\rangle$ is
conserved (see Appendix \ref{sec_cote}). On the other hand,
from Eq.
(\ref{hj7b}), we see that $\langle E_Q\rangle$ is trivially conserved since 
$\langle E_Q\rangle =0$. As a result, $\langle
{\cal E}\rangle=\langle E\rangle$ is conserved.

{\it Remark:} Using Eqs. (\ref{s9}) and (\ref{eno}), we find that
\begin{equation}
\label{hj11}
{\cal E}({\bf r},t)=-\frac{\hbar^2}{2m}\frac{\Delta\psi}{\psi}+m\Phi.
\end{equation}
This equation can be rewritten as
\begin{equation}
\label{hj12}
-\frac{\hbar^2}{2m}\Delta\psi+m\Phi\psi={\cal E}({\bf r},t)\psi.
\end{equation}
For a wave function of the form (\ref{sta3}), we find that ${\cal E}({\bf
r},t)=E$
where $E$ is a constant. In that case, Eq. (\ref{hj12}) reduces to the
time-independent Schr\"odinger equation (\ref{sta4}).

\subsection{The real and imaginary parts of the generalized complex viscous
Burgers equation}
\label{sec_ri}

Writing the complex velocity field as ${\bf U}={\bf u}-i{\bf u}_Q$ and taking
the real and imaginary parts of the generalized complex viscous Burgers
equation
(\ref{qb2}), we obtain the two real coupled equations
\begin{equation}
\label{ri3}
\frac{\partial {\bf u}_Q}{\partial t}+({\bf u}_Q\cdot \nabla){\bf u}+({\bf
u}\cdot \nabla){\bf u}_Q= -\frac{\hbar}{2m}\Delta{\bf u},
\end{equation}
\begin{equation}
\label{ri2}
\frac{\partial {\bf u}}{\partial t}+({\bf u}\cdot \nabla){\bf u}-({\bf u}_Q\cdot
\nabla){\bf u}_Q=\frac{\hbar}{2m}\Delta{\bf u}_Q-\nabla\Phi-\gamma_R {\bf
u}-\gamma_I {\bf u}_Q.
\end{equation}
When $\gamma=0$ these equations coincide 
with those derived by Nelson \cite{nelson} in his stochastic interpretation of
quantum mechanics. In his theory, ${\bf u}=\nabla S/m$ is called the ``current''
velocity
and $-{\bf u}_Q={\cal D}\nabla\ln\rho$ the ``osmotic''
velocity.\footnote{According to Einstein's theory of
Brownian motion, the terminal velocity ${\bf u}=-({1}/{\xi})\nabla\Phi$
acquired by a Brownian particle submitted to an external force
must balance the process of diffusion produced by the thermal molecular motion.
The first process is described by a current ${\bf j}=\rho {\bf
u}=-({1}/{\xi})\rho\nabla\Phi$. The second process is described by a current
${\bf J}=-D\nabla\rho$ given by Fick's law, leading to a diffusion equation
$\partial_t\rho=-\nabla\cdot {\bf J}=D\Delta\rho$. Therefore,
$-D\nabla\rho-({1}/{\xi})\rho\nabla\Phi={\bf 0}$. When compared to the
condition of equilibrium $\nabla P+\rho\nabla\Phi={\bf 0}$ where $P=\rho k_B
T/m$ is the osmotic pressure, one gets the Einstein relation $D=k_B T/\xi m$.
Furthermore, the velocity can be written as ${\bf u}=D\nabla\ln\rho$, hence
the name ``osmotic velocity'' given by Nelson \cite{nelson} by analogy with
Brownian motion.} Therefore, the Nelson equations are
mathematically
equivalent to
the complex viscous Burgers equation. This is the reason why they lead to
the Schr\"odinger equation. Similarly, the generalized Nelson equations
(\ref{ri3}) and (\ref{ri2}) lead to the generalized Schr\"odinger
equation (\ref{qb10}). We note that the friction term occurs only in Eq.
(\ref{ri2}). It appears through the combination $\gamma_R {\bf
u}+\gamma_I {\bf u}_Q$ where $\gamma_R$ acts on the classical velocity and
$\gamma_I$ on the quantum velocity. Let us check that Eqs. (\ref{ri3}) and
(\ref{ri2}) lead
to the damped
isothermal Euler equations (\ref{tmad4})-(\ref{tmad7}).

We first consider Eq. (\ref{ri3}). Using the identities $\Delta {\bf
u}=\nabla(\nabla\cdot {\bf u})-\nabla\times(\nabla\times {\bf u})$ and $\nabla
({\bf u}\cdot {\bf u}_Q)=({\bf u}\cdot \nabla){\bf u}_Q+({\bf u}_Q\cdot
\nabla){\bf u}+{\bf u}\times(\nabla\times {\bf u}_Q)+{\bf
u}_Q\times(\nabla\times {\bf u})$ which reduce to $\Delta {\bf
u}=\nabla(\nabla\cdot {\bf u})$ and $\nabla ({\bf u}\cdot {\bf u}_Q)=({\bf
u}\cdot \nabla){\bf u}_Q+({\bf u}_Q\cdot \nabla){\bf u}$ for an irrotational
flow  ($\nabla\times {\bf u}=\nabla\times {\bf u}_Q={\bf 0}$), Eq. (\ref{ri3})
can be rewritten as
\begin{equation}
\label{ri4}
\frac{\partial {\bf u}_Q}{\partial t}+\nabla\cdot ({\bf u}\cdot {\bf
u}_Q)=-\frac{\hbar}{2m}\nabla(\nabla\cdot {\bf u}).
\end{equation}
Using Eq.  (\ref{cqv7}), we recover the equation of continuity
(\ref{tmad4}). 

We now consider Eq. (\ref{ri2}). Using the identities $\Delta {\bf
u}_Q=\nabla(\nabla\cdot {\bf u}_Q)-\nabla\times(\nabla\times {\bf u}_Q)$ and 
$({\bf u}_Q\cdot \nabla){\bf u}_Q=\nabla ({{\bf u}_Q^2}/{2})-{\bf u}_Q\times
(\nabla\times {\bf u}_Q)$ which reduce to $\Delta {\bf u}_Q=\nabla(\nabla\cdot
{\bf u}_Q)$ and $({\bf u}_Q\cdot \nabla){\bf u}_Q=\nabla ({{\bf u}_Q^2}/{2})$
for an irrotational flow  ($\nabla\times {\bf u}_Q={\bf 0}$), Eq. (\ref{ri2})
can be rewritten as
\begin{equation}
\label{ri6}
\frac{\partial {\bf u}}{\partial t}+({\bf u}\cdot \nabla){\bf u}=
-\nabla\Phi-\gamma_R{\bf u}-\gamma_I{\bf u}_Q+\frac{\hbar}{2m}\nabla (\nabla
\cdot {\bf u}_Q)+\nabla \left (\frac{{\bf u}_Q^2}{2}\right ).
\end{equation}
Using Eqs. (\ref{cqv7}) and (\ref{cqq7}), and introducing the notations of Eq.
(\ref{qb8}), we recover 
the quantum damped isothermal Euler equation (\ref{tmad6}). We note that the
friction term corresponds to $-\gamma_R{\bf
u}$ and the pressure term corresponds to $-\gamma_I{\bf
u}_Q$.

\section{A damped Schr\"odinger equation that does not satisfy the local
conservation of the normalization condition}
\label{sec_pb}

If we attempt to take into account dissipative effects in the Schr\"odinger equation by writing  the scale-covariant equation of dynamics as
\begin{equation}
\label{pb1}
\frac{D{\bf U}}{Dt}=-\nabla\Phi-\gamma {\bf U},
\end{equation}
where $\gamma$ is a complex friction coefficient, we obtain a generalized complex viscous Burgers equation of the form
\begin{equation}
\label{pb2}
\frac{\partial {\bf U}}{\partial t}+({\bf U}\cdot \nabla){\bf U}=i {\cal D}\Delta{\bf U}-\nabla\Phi-\gamma {\bf U}.
\end{equation}
The corresponding complex Hamilton-Jacobi equation writes
\begin{equation}
\label{pb3}
\frac{\partial {\cal S}}{\partial t}+\frac{1}{2m}(\nabla {\cal S})^2-i{\cal
D}\Delta{\cal S}+m\Phi+V(t)+\gamma {\cal S}=0.
\end{equation}
Introducing the wave function (\ref{s4}) and repeating the calculations
of Sec. \ref{sec_s}, we obtain the generalized Schr\"odinger equation
\begin{equation}
\label{pb5}
i\hbar\frac{\partial\psi}{\partial t}=-\frac{\hbar^2}{2m}\Delta\psi+m\Phi\psi+V\psi-i\gamma\hbar (\ln\psi)\psi.
\end{equation}
As we shall see, this equation with $V=0$ does not conserve the normalization
condition. Therefore, it makes
sense to determine
$V(t)$ so that the average value of the friction term is equal to zero.
This yields
\begin{equation}
\label{pb6}
V(t)=i\gamma\hbar\langle \ln\psi\rangle.
\end{equation}
In that case, we obtain a  generalized Schr\"odinger equation of the form
\begin{equation}
\label{pb5q}
i\hbar\frac{\partial\psi}{\partial t}=-\frac{\hbar^2}{2m}\Delta\psi+m\Phi\psi-i\gamma\hbar (\ln\psi-\langle \ln\psi\rangle)\psi.
\end{equation}
Introducing the notations of Eq. (\ref{qb8}), it can be rewritten as
\begin{equation}
\label{pb5b}
i\hbar\frac{\partial\psi}{\partial t}=-\frac{\hbar^2}{2m}\Delta\psi+m\Phi\psi-i\xi\hbar (\ln\psi-\langle \ln\psi\rangle)\psi+2k_B T(\ln\psi-\langle \ln\psi\rangle)\psi.
\end{equation}
Making the Madelung transformation, we
obtain the hydrodynamic equations
\begin{equation}
\label{pb7}
\frac{\partial\rho}{\partial t}+\nabla\cdot (\rho {\bf u})=-\xi\rho(\ln\rho-\langle\ln\rho\rangle)+\frac{4k_B T}{\hbar^2}\rho (S-\langle S\rangle),
\end{equation}
\begin{equation}
\label{pb8}
\frac{\partial S}{\partial t}+\frac{1}{2m}(\nabla S)^2+m\Phi+Q+\xi (S-\langle S\rangle)+k_B T (\ln\rho-\langle\ln\rho\rangle)=0,
\end{equation}
\begin{equation}
\label{pb9}
\frac{\partial {\bf u}}{\partial t}+({\bf u}\cdot \nabla){\bf u}=-\frac{1}{\rho}\nabla p-\nabla\Phi-\frac{1}{m}\nabla Q-\xi {\bf u}.
\end{equation}
Since Eq. (\ref{pb7}) differs from the continuity equation, we conclude that Eq.
(\ref{pb5q}) does not conserve the normalization condition locally. We note,
however, that it conserves the
normalization condition globally. Indeed \begin{equation}
\label{pb10}
\frac{d}{dt}\int \rho\, d{\bf r}=0.
\end{equation}
This is not the case of Eq. (\ref{pb5}) with $V=0$ which leads to Eq.
(\ref{pb7}) without the terms in bracket. In that case, we get
\begin{equation}
\label{pb11}
\frac{d}{dt}\int \rho\, d{\bf r}=-\xi\int\rho\ln\rho\, d{\bf r}+\frac{4k_B
T}{\hbar^2}\int\rho S\, d{\bf r},
\end{equation}
which is usually different from zero.\footnote{We
note that the source terms in Eqs. (\ref{pb7}) and (\ref{pb11}) involve the
Boltzmann entropy 
$S_B=-k_B \langle\ln\rho\rangle$.}

{\it Remark:} Using the identity
\begin{equation}
\label{pb12}
\ln\psi=\frac{1}{2}\ln\left (\frac{\psi}{\psi^*}\right )+\ln|\psi|,
\end{equation}
we can rewrite Eq. (\ref{pb5}) as
\begin{equation}
\label{pb13}
i\hbar\frac{\partial\psi}{\partial
t}=-\frac{\hbar^2}{2m}\Delta\psi+m\Phi\psi+V\psi-i\gamma\hbar
\ln|\psi|\psi-i\frac{\gamma}{2}\hbar \ln\left (\frac{\psi}{\psi^*}\right )\psi.
\end{equation}
This is similar to Eq. (\ref{qb7}) with the crucial difference that the
coefficients in front of $\ln|\psi|$ and $\ln(\psi/\psi^*)$ are complex while
they are real and imaginary in Eq. (\ref{qb7}).

\section{Einstein-like relation}
\label{sec_el}

The quantum diffusion coefficient that appears in the theory of scale
relativity, in particular in the Fokker-Planck-like equation (\ref{b9}), is
related to the Planck constant by the Nelson relation (\ref{s8}). Combining Eqs.
(\ref{s8}) and
(\ref{qb8}), we obtain the identity
\begin{equation}
\label{el1}
{\cal D}=\frac{k_B T}{m\gamma_I},
\end{equation}
which is similar to the Einstein relation in Brownian theory. 
This can be interpreted as a sort of fluctuation-dissipation theorem. 

On the other hand, in the
strong friction limit $\xi\rightarrow +\infty$, the generalized Schr\"odinger
equation (\ref{qb10}) is equivalent, through the Madelung transformation,
to the quantum Smoluchowski equation
\begin{equation}
\label{el2}
\xi\frac{\partial\rho}{\partial t}=\nabla\cdot\left (\frac{k_B T}{m}\nabla \rho+\rho\nabla\Phi+\frac{\rho}{m}\nabla Q\right ).
\end{equation}
It involves a classical diffusion coefficient
\begin{equation}
\label{el3}
D=\frac{k_B T}{m \xi}=\frac{k_B T}{m \gamma_R}
\end{equation}
which is given by the standard Einstein relation. 

The classical and quantum
diffusion coefficients satisfy the relation
\begin{equation}
\label{el4}
\frac{D}{\cal D}=\frac{\gamma_I}{\gamma_R}.
\end{equation}
We note that they coincide ($D={\cal D}$ or, equivalently,
$\gamma_R=\gamma_I$) when
\begin{equation}
\label{el5}
\frac{2k_B T}{\xi}=\hbar.
\end{equation}

\section{The energy operator and the Hamiltonian}
\label{sec_ae}

\subsection{The Schr\"odinger equation}
\label{sec_aep1}

The Hamiltonian of a classical particle moving in a potential $\Phi$ is
\begin{eqnarray}
\label{ae1}
H=\frac{p^2}{2m}+m\Phi.
\end{eqnarray}
Using the correspondence principle 
\begin{eqnarray}
\label{corrpple}
\hat{\bf p}=-i\hbar\nabla\qquad{\rm and}\qquad  \hat{H}=i\hbar\partial_t,
\end{eqnarray}
we obtain the Schr\"odinger equation
\begin{equation}
\label{ae3}
i\hbar\frac{\partial\psi}{\partial t}=\hat H\psi
\end{equation}
with the Hamiltonian operator
\begin{eqnarray}
\label{ae2}
\hat H=-\frac{\hbar^2}{2m}\Delta+m\Phi.
\end{eqnarray}
The average value of the Hamiltonian is
\begin{eqnarray}
\label{ae4}
H = \left \langle \psi\left |\hat H\right |\psi\right
\rangle=-\frac{\hbar^2}{2m}\int \psi^*\Delta\psi\, d{\bf r}+m\int
\psi^*\Phi\psi\, d{\bf r}=\frac{\hbar^2}{2m}\int |\nabla\psi|^2\, d{\bf r}+m\int
|\psi|^2 \Phi\, d{\bf r}.
\end{eqnarray}
This is the sum of the kinetic energy 
\begin{eqnarray}
\label{ef2}
\Theta=\left \langle \psi\left |-\frac{\hbar^2}{2m}\Delta\right |\psi\right
\rangle=\frac{\hbar^2}{2m}\int |\nabla\psi|^2 \, d{\bf r}
\end{eqnarray}
and the potential energy
\begin{eqnarray}
W =\langle \psi |m\Phi|\psi\rangle=m\int
|\psi|^2 \Phi\, d{\bf r}.
\end{eqnarray}
Using the Madelung transformation, the kinetic energy (\ref{ef2}) can be
decomposed
into  the  classical kinetic energy (\ref{ef3}) and the quantum kinetic energy
(\ref{ef5}). The potential energy can be rewritten as in Eq. (\ref{ef6}). Using
the results of Sec. \ref{sec_ef}, we see that the average value of the
Hamiltonian (\ref{ae4}) concides with the average energy given by Eq.
(\ref{new1}), namely
\begin{eqnarray}
H=\langle E\rangle.
\end{eqnarray}

Taking the first variations of the energy functional (\ref{ae4}),
we get
\begin{eqnarray}
\label{ae11}
\delta H=\int\left (
-\frac{\hbar^2}{2m}
\Delta\psi^*+m\Phi\psi^*\right )\delta\psi\, d{\bf
r}+{\rm c.c.}
\end{eqnarray}
The term in parenthesis coincides with the Hamiltonian operator (\ref{ae2})
applied on $\psi^*$. Therefore
\begin{equation}
\label{ag9qq}
\frac{\delta H}{\delta\psi^*}={\hat H}\psi.
\end{equation}
As a result, the Schr\"odinger equation (\ref{ae3}) can be
rewritten as
\begin{eqnarray}
\label{ae12q}
i\hbar \frac{\partial\psi}{\partial
t}=\frac{\delta H}{\delta\psi^*}.
\end{eqnarray}
This expression shows that $H$ represents the true
Hamiltonian of the particle. Indeed, in terms of the wavefunction $\psi({\bf
r},t)$ and its canonical momentum $\pi({\bf r},t)=i\hbar\psi^*({\bf r},t)$, the
 Schr\"odinger equation is exactly reproduced by the Hamilton equations
\begin{eqnarray}
\label{lh5w}
\frac{\partial\psi}{\partial t}=\frac{\delta H}{\delta\pi},\qquad
\frac{\partial\pi}{\partial t}=-\frac{\delta H}{\delta\psi}.
\end{eqnarray}
This formulation directly implies the
conservation of the total energy
$H$ since
\begin{eqnarray}
\label{lh5wb}
{\dot H}=\int \frac{\delta H}{\delta\psi}\frac{\partial\psi}{\partial t}\,
d{\bf r}+\int \frac{\delta H}{\delta\pi}\frac{\partial\pi}{\partial t}\, d{\bf
r}=0.
\end{eqnarray}
From general arguments \cite{holm}, a minimum of the energy
functional $H$  given by
Eq. (\ref{ae4}) under
the normalization condition  (\ref{s11}) is a stationary
solution of the Schr\"odinger equation   that is formally nonlinearly
dynamically stable. Writing the variational principle as
\begin{eqnarray}
\label{lh5b}
\delta H-\mu\delta\int |\psi|^2\, d{\bf r}=0,
\end{eqnarray}
where $\mu$
is a Lagrange
multiplier (chemical potential), we obtain the time-independent Schr\"odinger
equation
(\ref{sta4}) with $\mu=E$.  This shows that the chemical potential $\mu$ can be
identified with the eigenenergy $E$. The
variational
principle (\ref{lh5b}) was introduced by Schr\"odinger
\cite{schrodinger1} in his first
paper on wave mechanics. This is actually how he derived the fundamental 
eigenvalue equation (\ref{sta4}) (see Appendix F of \cite{chavmatos}).

\subsection{The generalized Schr\"odinger equation}

The generalized Schr\"odinger equation (\ref{qb10}) can be written as
\begin{equation}
\label{qb10e}
i\hbar\frac{\partial\psi}{\partial
t}={\hat E}\psi-i\frac{\hbar}{2}\xi\left\lbrack
\ln\left (\frac{\psi}{\psi^*}\right )-\left\langle \ln\left
(\frac{\psi}{\psi^*}\right )\right\rangle\right\rbrack \psi,
\end{equation}
where the energy operator is given by
\begin{equation}
\label{ae5}
{\hat E}=-\frac{\hbar^2}{2m}\Delta+m\Phi+2k_B
T\ln|\psi|.
\end{equation}
Its average value is
\begin{eqnarray}
\label{ae6}
\langle E\rangle =\frac{\hbar^2}{2m}\int
|\nabla\psi|^2\, d{\bf r}+m\int |\psi|^2 \Phi\, d{\bf r}+k_B T\int
|\psi|^2\ln|\psi|^2\, d{\bf r}.
\end{eqnarray}
Using the results of Sec. \ref{sec_ef}, we see that the average value of the
energy operator  concides with the average value of the 
energy given by Eq. (\ref{dif1}). It differs from the free energy (\ref{ef8})
which can be rewritten as
\begin{eqnarray}
\label{ae10}
F=\frac{\hbar^2}{2m}\int |\nabla\psi|^2\, d{\bf r}+m\int |\psi|^2 \Phi\, d{\bf
r}+k_B T\int |\psi|^2(\ln|\psi|^2-1)\, d{\bf r}.
\end{eqnarray}
We have
\begin{equation}
\label{ae8b}
F=\langle E\rangle-k_B T.
\end{equation}
From Eq. (\ref{qb10e}), we easily obtain the
identity 
\begin{eqnarray}
\label{ae13}
\dot F=-\xi\int \frac{\hbar^2}{4m}|\psi|^2\left |\nabla\ln\left
(\frac{\psi}{\psi^*}\right )\right |^2\, d{\bf r},
\end{eqnarray}
which coincides with the $H$-theorem (\ref{ef11}).

{\it Remark:} In the context of nonlinear Schr\"odinger
equations \cite{dgp}, it can be shown that the true Hamiltonian $H$ of the
system is $F$, not $\langle E\rangle$. This is because the
nonlinear Schr\"odinger
equations considered in \cite{dgp}, and the corresponding fluid equations, can
be expressed in terms of the functional derivative of the free energy $F$ (see
Sec. III.F and Appendix B of \cite{dgp}). For the standard Schr\"odinger
equation (\ref{s9}), we have $F=H=\langle E\rangle$. For the logarithmic
Schr\"odinger equation (\ref{qb10}), since $F$ and $\langle E\rangle$ only
differ by
a constant, we can also interpret $\langle E\rangle$ as the
Hamiltonian of the
system. However, this identification is not true anymore for other nonlinear
Schr\"odinger equations \cite{dgp}. In general, $F=H\neq \langle E\rangle$.

\section{Exact solution of the generalized Schr\"odinger equations (\ref{qb10})
and (\ref{ag0}) with a
harmonic potential}
\label{sec_exact}

When the external potential is harmonic [see Eq. (\ref{vhp1})], leading to a
force $-\nabla\Phi=-\omega_0^2{\bf r}$ depending linearly on the distance, we
can obtain an exact analytical solution of the generalized Schr\"odinger
equations (\ref{qb10}) and (\ref{ag0}) or of the equivalent hydrodynamic
equations (\ref{tmad4})-(\ref{tmad7}) and (\ref{ag1})-(\ref{ag3}). We consider a
wave
function of the form
\begin{eqnarray}
\label{exact1}
\psi({\bf r},t)=\left \lbrack \frac{2}{S_d \Gamma(d/2)R(t)^d}\right
\rbrack^{1/2}e^{-\frac{r^2}{2R(t)^2}}e^{imH(t)r^2/2\hbar}e^{iS_0(t)/\hbar},
\end{eqnarray}
where $R(t)$ measures the size of the wave packet. Comparing Eq. (\ref{exact1})
with
Eq. (\ref{ntmad2}), we find that the density and the action are given by
\begin{eqnarray}
\label{exact2}
\rho({\bf r},t)=\frac{2}{S_d \Gamma(d/2)R(t)^d}e^{-\frac{r^2}{R(t)^2}} \qquad
{\rm and}\qquad 
S({\bf r},t)=\frac{1}{2} m H(t) r^2+S_0(t).
\end{eqnarray}
The velocity defined by Eq. (\ref{ntmad3}) is then given by
\begin{eqnarray}
\label{exact3}
{\bf u}({\bf r},t)=H(t){\bf r}.
\end{eqnarray}
Using Eqs. (\ref{exact2}) and (\ref{exact3}), we easily obtain
\begin{eqnarray}
\label{exact10}
\frac{\partial\rho}{\partial t}+\nabla\cdot (\rho {\bf u})=-\rho\left
(\frac{\dot R}{R}-H\right )\left (d-2\frac{r^2}{R^2}\right ),
\end{eqnarray}
\begin{eqnarray}
\label{exact7}
\frac{\partial {\bf u}}{\partial t}+({\bf u}\cdot \nabla){\bf u}=(\dot
H+H^2){\bf r},
\end{eqnarray}
\begin{eqnarray}
\label{exact4}
\frac{\nabla\rho}{\rho}=-\frac{2{\bf r}}{R^2},\qquad \nabla
Q=-\frac{\hbar^2}{m}\frac{\bf r}{R^4},
\end{eqnarray}
\begin{eqnarray}
\label{exact11}
E-\langle E\rangle=\left
(\frac{1}{2}mH^2+\frac{1}{2}m\omega_0^2-\frac{\hbar^2}{2mR^4}\right
)\left (r^2-\frac{d}{2}R^2\right ),
\end{eqnarray}
\begin{eqnarray}
\label{exact8}
\nabla\left (\frac{{\bf u}\cdot\nabla\rho}{\rho}\right )=-4H\frac{\bf
r}{R^2},\qquad \Delta{\bf u}={\bf 0}.
\end{eqnarray}
We also have
\begin{eqnarray}
\label{a5}
I=\alpha MR^2,\qquad \Theta_c=\frac{1}{2}\alpha M\left (\frac{dR}{dt}\right
)^2,\qquad \Theta_Q=\sigma \frac{\hbar^2M}{m^2R^2},
\end{eqnarray}
\begin{eqnarray}
\label{a6}
U=-d\frac{k_B T}{m}M\ln R+C,\qquad W=\frac{1}{2}\omega_0^2 \alpha M
R^2,
\end{eqnarray}
with the coefficients
\begin{eqnarray}
\label{a8}
\alpha=\frac{d}{2}, \qquad \sigma=\frac{d}{4},\qquad C=\frac{k_B
T}{m}M\left\lbrack \ln\left (\frac{2M}{S_d\Gamma(d/2)}\right
)-1-\alpha\right\rbrack.
\end{eqnarray}

\subsection{The generalized Schr\"odinger equation  (\ref{qb10})}

Substituting Eq. (\ref{exact10}) into the continuity equation (\ref{tmad4}), we
obtain the relation
\begin{eqnarray}
\label{exact12}
H=\frac{\dot R}{R}.
\end{eqnarray}
The function $H$ is similar to the Hubble parameter in cosmology, where $R$
plays the role of the scale factor. Using Eqs. (\ref{exact3}), (\ref{exact7}),
(\ref{exact4}) and (\ref{exact12}), the quantum damped isothermal Euler
equation (\ref{tmad6}) becomes
\begin{eqnarray}
\label{exact13}
\ddot R+\xi\dot R+\omega_0^2
R=\frac{2k_B T}{mR}+\frac{\hbar^2}{m^2R^3}.
\end{eqnarray}
In the strong friction limit $\xi\rightarrow +\infty$, corresponding to the
quantum Smoluchowski equation (\ref{tmad11}), it reduces to
\begin{eqnarray}
\label{exact13bw}
\xi\dot R+\omega_0^2
R=\frac{2k_B T}{mR}+\frac{\hbar^2}{m^2R^3}.
\end{eqnarray}
Equations (\ref{exact13}) and (\ref{exact13bw}) are closed
ordinary differential equation determining $R(t)$. They are studied in our
companion paper
\cite{companion}. Together with Eq. (\ref{exact1}), they determine a particular
exact solution of
the generalized Schr\"odinger equation (\ref{qb10}) with a harmonic potential.
This solution exhibits damped oscillations about the equilibrium state because
of the friction term
$\xi$.

The free energy (\ref{ef1}) can be written as
\begin{eqnarray}
\label{a10}
F=\frac{1}{2}\alpha M\left (\frac{dR}{dt}\right )^2+V(R),
\end{eqnarray}
where the first term is the classical kinetic energy and the second term is an
effective potential energy 
\begin{eqnarray}
\label{a11}
V(R)=\sigma \frac{\hbar^2M}{m^2R^2}+\frac{1}{2}\omega_0^2\alpha
MR^2-d\frac{M k_B T}{m}\ln
R+C
\end{eqnarray}
including the quantum kinetic energy, the potential energy, and the internal
energy. The virial theorem (\ref{vhp4}) becomes
\begin{eqnarray}
\label{exact13b}
\alpha M\frac{d^2R}{dt^2}+\xi\alpha M\frac{dR}{dt}=-\frac{dV}{dR},
\end{eqnarray}
which is equivalent to Eq. (\ref{exact13}). The $H$-theorem (\ref{ef11}) takes
the form
\begin{eqnarray}
\label{a20}
\frac{dF}{dt}=-\xi\alpha M\left (\frac{dR}{dt}\right )^2\le 0.
\end{eqnarray}
It can also be obtained from Eqs. (\ref{a10}) and (\ref{exact13b}). In the
strong friction limit $\xi\rightarrow +\infty$, where $F=V$, Eqs.
(\ref{exact13b}) and (\ref{a20}) reduce to 
\begin{eqnarray}
\label{a19d}
\xi\alpha M\frac{dR}{dt}=-\frac{d{V}}{dR} \qquad {\rm and}\qquad 
\frac{dF}{dt}=-\frac{1}{\xi\alpha M}\left
(\frac{dV}{dR}\right )^2\le 0.
\end{eqnarray}
They can also be obtained from Eqs. (\ref{vhp5}) and (\ref{ef13}).

{\it Remark:} When $T<0$ and $\omega_0=0$, the equilibrium state of the
generalized Schr\"odinger equation  (\ref{qb10}) is the gausson. Its radius,
given by Eq. (\ref{exact13}), is $R=\hbar/\sqrt{2mk_B |T|}=\hbar/\sqrt{2mb}$. 

\subsection{The generalized Schr\"odinger equation 
(\ref{ag0})}

Substituting Eqs. (\ref{exact10}) and  (\ref{exact11}) into the
generalized continuity equation (\ref{ag1}), we
obtain
\begin{eqnarray}
\label{exact14}
\frac{m\gamma R^2}{2\hbar}H^2-H+\frac{\gamma
m\omega_0^2R^2}{2\hbar}-\frac{\gamma\hbar}{2mR^2}+\frac{\dot R}{R}=0.
\end{eqnarray}
Substituting Eqs. (\ref{exact7}), (\ref{exact4}) and
(\ref{exact8}) into the quantum Navier-Stokes equation (\ref{ag3}), we get 
\begin{eqnarray}
\label{exact15}
\dot H+H^2=-\omega_0^2+\frac{\hbar^2}{m^2R^4}-\frac{2\gamma\hbar H}{mR^2}.
\end{eqnarray}
Equation (\ref{exact14}) is a second degree equation giving $H$ as a function of
$\dot R$ and $R$. When $H(R,\dot R)$ is inserted into Eq (\ref{exact15}) we
obtain a closed differential equation determining $R(t)$. It is studied
in our companion paper \cite{companion}. Together with Eq. (\ref{exact1}), it
determines a particular
exact solution of the generalized Schr\"odinger equation (\ref{ag0}) with a
harmonic potential. 

For $\gamma\ll 1$, we can expand Eq. (\ref{exact14}) to first order in $\gamma$,
and obtain
\begin{eqnarray}
\label{exact18}
H=\frac{\dot R}{R}+\frac{\gamma
m}{2\hbar}\omega_0^2R^2-\frac{\gamma\hbar}{2mR^2}+\frac{m\gamma R\dot
R}{2\hbar}.
\end{eqnarray} 
Taking its time derivative and combining the result with Eq. (\ref{exact15}) we
obtain
\begin{eqnarray}
\label{exact19}
\ddot R+\frac{3\gamma\hbar}{mR^2}\dot R+\frac{\gamma
m}{\hbar}\omega_0^2 R^2\dot R+\frac{m\gamma R^2\ddot
R}{2\hbar}+\frac{m\gamma R\dot R^2}{2\hbar}+\omega_0^2 R=\frac{\hbar^2}{m^2R^3}.
\end{eqnarray}

If we neglect the term in the r.h.s. of Eq. (\ref{ag1}), accounting for the
fact that $\gamma\ll 1$, it reduces
to the ordinary continuity equation (\ref{tmad4}) and
Eq. (\ref{exact14}) reduces to Eq. (\ref{exact12}). Substituting this relation
into Eq. (\ref{exact15}), we obtain the simpler differential equation
\begin{eqnarray}
\label{exact16}
\ddot R+\frac{2\gamma\hbar}{mR^2}\dot R+\omega_0^2 R=\frac{\hbar^2}{m^2R^3}.
\end{eqnarray}
The solution of this equation exhibits damped oscillations about the equilibrium
state  because of the friction term
$\gamma$. We stress that the damping is not due to the viscous term
$({\gamma\hbar}/{2m})\Delta {\bf u}$ which vanishes for a velocity field ${\bf
u}=H{\bf r}$, but to the term $({\gamma\hbar}/{2m})\nabla
({\bf u}\cdot \nabla\rho/{\rho})$.

In the case $\gamma\gg 1$, we have ${\bf
u}\simeq 0$ and $H\simeq 0$ (see the Remark in Sec. \ref{sec_ags}), and
Eq. (\ref{exact14})
reduces to
\begin{eqnarray}
\label{exact17}
\dot R=\frac{\gamma\hbar}{2mR}-\frac{\gamma m \omega_0^2R^3}{2\hbar}.
\end{eqnarray} 
We note Eq. (\ref{exact17})  does not correspond to the strong friction limit
of Eq. (\ref{exact16}) which has been established under the assumption
$\gamma\ll 1$.

\section{Dispersion relation for the generalized Schr\"odinger equations
(\ref{qb10}) and (\ref{ag0}) without
external potential}
\label{sec_dispersion}

\subsection{The generalized Schr\"odinger equation  (\ref{qb10})}

We consider the  generalized Schr\"odinger equation  (\ref{qb10}) without
external potential ($\Phi=0$). We use its hydrodynamic representation 
(\ref{tmad4})-(\ref{tmad7}). We consider a homogeneous equilibrium state with
$\rho({\bf r},t)=\rho$, ${\bf u}({\bf r},t)={\bf 0}$ and $S({\bf r},t)=-Et$ with
$E=k_B T\ln\rho$. Considering a small perturbation about this equilibrium state
and writing the linearized hydrodynamic equations, we find after simple
calculations that the equation for the density contrast
$\delta({\bf r},t)=\delta\rho({\bf r},t)/\rho$ writes
\begin{eqnarray}
\label{dis1}
\frac{\partial^2\delta}{\partial t^2}+\xi\frac{\partial\delta}{\partial
t}=-\frac{\hbar^2}{4m^2}\Delta^2\delta+c_s^2\Delta\delta,
\end{eqnarray} 
where $c_s^2=P'(\rho)=k_B T/m$ is the square of the speed of sound.\footnote{We
use the notation $c_s^2$ instead of $k_B T/m$ so that our results can be
immediately generalized to the
class of nonlinear Schr\"odinger equations considered in \cite{dgp}.}
Considering plane waves of the form $\delta\propto e^{i({\bf k}\cdot {\bf
r}-\omega t)}$, we obtain the dispersion relation
\begin{eqnarray}
\label{dis2}
\omega^2+i\xi \omega=\frac{\hbar^2k^4}{4m^2}+c_s^2k^2 \qquad \Rightarrow\qquad 
\omega=-i\frac{\xi}{2}\pm\sqrt{-\frac{\xi^2}{4}+c_s^2k^2+\frac{\hbar^2k^4}{
4m^2}}.
\end{eqnarray} 
This
dispersion relation is studied in detail in our companion  paper
\cite{companion}. When $c_s^2>0$, it involves the characteristic wavenumber 
\begin{eqnarray}
\label{dis3}
k_*^2=-\frac{2m^2c_s^2}{\hbar^2}+\sqrt{\frac{4m^4c_s^4}{\hbar^4}+\frac{m^2\xi^2}
{\hbar^2}}
\end{eqnarray} 
that separates the regime where the perturbations are purely damped ($k<k_*$)
from the regime where the perturbations have damped oscillations  ($k>k_*$).
When  $c_s^2<0$, there is an additional critical wavenumber 
\begin{eqnarray}
\label{dis4}
k_c^2=\frac{4m^2|c_s^2|}{\hbar^2}.
\end{eqnarray} 
When $k<k_c$ the perturbations grow exponentially rapidly, when  $k_c<k<k_*$
they are purely damped, and when $k>k_*$ they have
damped oscillations. We maximum growth rate and the most unstable
wavenumber are
\begin{eqnarray}
\label{dis5}
\gamma_{\rm
max}=-\frac{\xi}{2}+\sqrt{\frac{\xi^2}{4}+\frac{c_s^4m^2}{\hbar^2}},\qquad
k_m^2=\frac{2m^2|c_s^2|}{\hbar^2}.
\end{eqnarray}

{\it Remark:} For $\xi=0$, we recover the results obtained in
Sec. V of \cite{prd1} in the nongravitational case. For the generalized
Schr\"odinger equation (\ref{qb10}) for which
$c_s^2=P'(\rho)=k_B T/m$, the critical length $\lambda_c=\pi\hbar/\sqrt{mk_B
|T|}=\pi\hbar/(mb)^{1/2}$ obtained in the linear regime
of the dynamics [see Eq. (\ref{dis4})] 
coincides with the radius $R=\hbar/(2mb)^{1/2}$ of the gausson obtained in the
strongly nonlinear (clustered) regime of
the dynamics (see Sec.
\ref{sec_bbm} and Appendix \ref{sec_exact}).

\subsection{The generalized Schr\"odinger equation 
(\ref{ag0})}

We consider the  generalized Schr\"odinger equation  (\ref{ag0}) without
external potential ($\Phi=0$). We use its hydrodynamic representation 
(\ref{ag1})-(\ref{ag3}). We consider a homogeneous equilibrium state with
$\rho({\bf r},t)=\rho$, ${\bf u}({\bf r},t)={\bf 0}$ and $S({\bf r},t)=0$.
Considering a small perturbation about this equilibrium state
and writing the linearized hydrodynamic equations, we find after simple
calculations that the equation for the density contrast
$\delta({\bf r},t)=\delta\rho({\bf r},t)/\rho$ writes
\begin{eqnarray}
\label{dis6}
\frac{\partial^2\delta}{\partial t^2}-2\nu\Delta\frac{\partial\delta}{\partial
t}=-\left (\frac{\hbar^2}{4m^2}+\nu^2\right )\Delta^2\delta,
\end{eqnarray} 
where $\nu$ is the quantum viscosity defined by Eq. (\ref{ag4}). Considering
plane waves of the form $\delta\propto e^{i({\bf k}\cdot {\bf
r}-\omega t)}$, we obtain the dispersion relation
\begin{eqnarray}
\label{dis7}
\omega^2+2i\nu k^2 \omega=\left (\frac{\hbar^2}{4m^2}+\nu^2\right )k^4\qquad
\Rightarrow \qquad \omega=\pm\frac{\hbar k^2}{2m}-i\nu k^2.
\end{eqnarray} 
The perturbation oscillates with a pulsation $\Omega=\hbar k^2/2m$ and is
damped with a damping rate $\gamma=-\nu k^2$.

\section{Uncertainty principle}
\label{sec_up}

In this Appendix, we present a simple derivation of the Heisenberg uncertainty
principle. According to the correspondance
principle (\ref{corrpple}), the mean square value of the impulse is
\begin{equation}
\label{up1}
\langle {\bf p}^2\rangle=\langle \psi|\hat{\bf p}^2|\psi\rangle =-\hbar^2\int
\psi^*\Delta\psi\, d{\bf r}=\hbar^2\int |\nabla\psi|^2\, d{\bf r}.
\end{equation}
Using Eqs. (\ref{cqv3}), (\ref{cqv5b}) and (\ref{ef2}),  we obtain the
identities
\begin{equation}
\label{up2}
\left\langle {\bf p}^2\right\rangle=2m\Theta=\langle |{\bf
P}|^2\rangle=m^2\langle |{\bf U}|^2\rangle=m^2\langle {\bf u}^2\rangle
+m^2\langle {\bf u}_Q^2\rangle.
\end{equation}
Therefore,  the mean square value of the impulse is equal to the mean 
square value of the modulus of the complex impulse, which is the sum of the mean
square value of the classical impulse and the mean square value
of the quantum impulse. This shows that the mean square value of the
impulse is always larger than the mean square value of the quantum impulse
(they become equal in a steady state where ${\bf u}={\bf 0}$). For the
quantum harmonic oscillator, using the exact Gaussian solution of Appendix
\ref{sec_exact}, we have
\begin{equation}
\label{up3}
\langle {\bf
u}_Q^2\rangle=\frac{2\Theta_Q}{m}=\frac{d}{2}\frac{\hbar^2}{m^2R^2}\qquad {\rm
and}\qquad
\langle {\bf r}^2\rangle=\frac{I}{m}=\frac{d}{2}R^2.
\end{equation}
This implies 
\begin{equation}
\label{up4}
m^2 \langle {\bf u}_Q^2\rangle\langle {\bf r}^2\rangle =\frac{d^2}{4}\hbar^2.
\end{equation}
From Eqs. (\ref{up2}) and (\ref{up4}), we obtain the inequality
\begin{equation}
\label{up5}
\langle (\Delta  {p}_i)^2\rangle \langle (\Delta {x}_j)^2\rangle \ge
\frac{1}{4}\hbar^2\delta_{ij},
\end{equation}
corresponding to the Heisenberg uncertainty principle.

\end{document}